\documentclass[nofootinbib,reprint]{revtex4-1}
\usepackage{amsmath,amssymb}
\usepackage{graphicx,color}
\usepackage[pdftex,colorlinks]{hyperref}

\begin{document}


\title{Exact analytical solutions to the master equation of quantum Brownian motion for a general environment}

\author{C.~H.~Fleming}
\address{Joint Quantum Institute and Department of Physics, University of Maryland, College Park, Maryland 20742}
\author{Albert Roura}
\address{Max-Planck-Institut f\"ur Gravitationsphysik (Albert-Einstein-Institut), Am M\"uhlenberg 1, 14476 Golm, Germany}
\author{B.~L.~Hu}
\address{Joint Quantum Institute and Department of Physics, University of Maryland, College Park, Maryland 20742}

\begin{abstract}
We revisit the model of a  quantum Brownian oscillator linearly
coupled to an environment of  quantum oscillators at finite
temperature. By introducing a compact and particularly well-suited
formulation, we give a rather quick and direct derivation of the
master equation and its solutions for general spectral functions and
arbitrary temperatures. The flexibility of our approach allows for an
immediate generalization to cases with an external force and with an
arbitrary number of Brownian oscillators. More importantly, we point
out an important mathematical subtlety concerning boundary-value
problems for integro-differential equations which led to incorrect
master equation coefficients and impacts on the description of
nonlocal dissipation effects in all earlier derivations. Furthermore,
we provide explicit, exact analytical results for the master equation
coefficients and its solutions in a wide variety of cases, including
ohmic, sub-ohmic and supra-ohmic environments with a finite cut-off.
\end{abstract}



\maketitle

\tableofcontents

\section{Introduction}

\subsection{New Results placed in Background Context}

An open quantum systems (OQS) \cite{Breuer02} refers to a quantum system
interacting with an environment, which could be multi-partite,
possessing many more degrees of freedom (it could also be identified
as the remaining ``irrelevant'' degrees of freedom of the system
itself). An environment in some simplified modeling can be described
in terms of its spectral density and parametrized by its temperature.
Its influence  on the open system can be expressed in terms of
fluctuations (vacuum and thermal) and noises (the most general form
can be colored and multiplicative). A theory of OQS describes the
nature and dynamics of this system as a result of such interactions,
which manifest in quantum dissipation and diffusion and can alter
significantly the quantum coherence, entanglement and correlation
properties of the otherwise closed quantum system. The familiar
quantum statistical mechanics is the extreme limiting case when the
system remains in equilibrium through interaction with a thermal or
chemical reservoir.

Open quantum system is the theoretical construct suitable for the
investigation of the properties and dynamics of nonequilibrium
quantum systems in the Langevin vein (as distinguished from the
Boltzmann vein, which considers closed systems albeit often with a
hierarchical structure; see, e.g., Ref.~\cite{CalzettaHu08}). It plays
an important role in addressing the fundamental issues such as the
quantum-to-classical transition through the environment-induced
decoherence mechanism \cite{Giulini96,Zurek03}. For practical
purposes it has been effectively applied to exciting phenomena in
many new directions of micro- and meso-physics in the last two
decades, made possible by innovative experiments aided by
technological advances in high-precision instrumentation. These
include the areas of superconductivity such as quantum dissipative
tunneling in SQUIDs \cite{CaldeiraLeggett83b,Leggett87,Weiss99},
atomic and quantum optical systems using ultrafast lasers with atoms
in cavities and optical lattices \cite{Scully07,Meystre01,Wiseman10}, as
well as nanoelectromechanical devices \cite{Naik06,Lahaye04} which
have great potential in physical, chemical and bioscience
applications. For an accurate description of the system's properties
and evolution in these processes, the effects of its interaction with
the environment are essential.

Quantum Brownian motion (QBM) of an oscillator coupled to a thermal
bath of quantum oscillators has been extensively studied as a
canonical model for open quantum systems because there is a
considerable amount of insight that one can learn from it while being
treatable analytically to a significant degree. In this paper we
continue the lineage of work on QBM via the influence functional
path-integral method of Feynman and Vernon \cite{Feynman63} used by
Caldeira and Leggett \cite{CaldeiraLeggett83} to derive a master equation for a
high-temperature ohmic environment, which corresponds to the
Markovian regime. Following this, Caldeira, Cerdeira and Ramaswamy
\cite{Caldeira89} derived the Markovian master equation for the system
with weak coupling to an ohmic bath, which was claimed to be valid at
arbitrary temperature (see Sec.~\ref{sec-caldeira} for a critique of
this claim). At the same time Unruh and Zurek \cite{UnruhZurek89} derived a
more complete and general master equation that incorporated a colored
noise at finite temperature, but there is a problem with their
fluctuation-dissipation relation (see Ref.~\cite{HPZ92}). Finally, in
a path-integral calculation from first principles, Hu, Paz and Zhang
(HPZ) \cite{HPZ92} derived a master equation for a general environment
(arbitrary temperature and spectral density), barring certain subtle
errors in the coefficients, which lead to inaccurate treatment of the
nonlocal dissipation cases, as we will discuss.
After that, this equation has been rederived by a number of authors.
Halliwell and Yu \cite{HalliwellYu96} exploited the phase-space
transformation properties of the Wigner function for the full system
plus environment and derived a Fokker-Planck equation corresponding
to the HPZ equation. Calzetta, Roura and Verdaguer (CRV)
\cite{CRV03,CRV01} derived it using a stochastic description for
open quantum systems based on Langevin equations, whereas Ford and
O'Connell \cite{FordOconnell01} employed a somewhat related method via the
quantum Langevin equation \cite{FordOconnell88} and obtained also the
solution to the HPZ equation for a Gaussian wave-packet.

The present paper's contribution to this legacy is threefold:
\begin{enumerate}
\item We have completely determined the precise form of the HPZ master
equation coefficients and pointed out a problem with earlier derivations for nonlocal dissipation (Sec.~\ref{sec:MED}).
\item We have found concise and efficient solutions to the master equation with a number of exact nonpertrubative analytical results (Sec.~\ref{sec:solutions}).
\item
We have extended the theory to that of a system of multiple
oscillators bilinearly coupled amongst themselves and to the bath in
an arbitrary fashion while acted upon by classical forces
(Sec.~\ref{sec:generalization}).
\end{enumerate}

In this paper we will follow the approach of CRV in
Refs.~\cite{CRV03,CRV01} and make use of a stochastic description
whose central element is a Langevin equation for the dynamics of the
open quantum system. This offers an efficient mathematical tool for
obtaining all the quantum properties of the system. An important
feature of the present approach is the reformulation in phase-space
(rather than position space) together with the use of vector and
matrix notation. The combination of all these elements makes this new
approach far more flexible and compact. For example, we are able to
derive the general expression for the solution of the master equation
in essentially two short lines [see Eq.~\eqref{eq:sigma1}]. The
flexibility of our formalism is also illustrated by the
straightforward generalizations to the cases of an external force
(this is nontrivial for nonlocal dissipation) and an arbitrary number
of system oscillators that will be presented.
This goes far beyond previous generalizations of the theory \cite{Chou08}
which assume specific forms of coupling.

One of our key contributions, however, is uncovering a significant
shortcoming of earlier results for the master equation coefficients.
We point out a subtlety involving boundary conditions for solutions
of integro-differential equations and explain how certain properties
that hold for ordinary differential equations are not true for
nonlocal dissipation. These properties had always been employed
erroneously, in one way or another, when deriving the expressions for
the master equation coefficients, even those which were then
evaluated numerically. This long-standing error could have deep
implications for regimes where the effects of nonlocal dissipation
are significant and one should be cautious with all results for those
cases reported in the literature.

Taking into account the aspect mentioned in the previous paragraph,
and using our compact formulation, we have provided a relatively
simplified expression for the correct master equation. Moreover, one
can also obtain the general solution to the master equation in terms
of the matrix propagator of a linear integro-differential equation,
and see that at late times it tends to a Gaussian state completely
characterized by a constant covariance matrix. For odd meromorphic
spectral functions, and many others,  we are able to reduce the
calculation of this covariance matrix to a simple contour integral
and obtain exact nonperturbative results for finite cut-off and
arbitrarily strong coupling. This includes examples of ohmic,
sub-ohmic and supra-ohmic environments; and from this late-time
covariance one can immediately obtain the late-time diffusion
coefficients as well. Our results generalize the work of Anastopoulos
and Halliwell \cite{Anastopoulos95} as well as Ford and O'Connell
\cite{FordOconnell01}, who already found the late time state to be a
Gaussian, and the earlier work of Hu and Zhang
\cite{HuZhang93,HuZhang95} on the generalized uncertainty function
for Gaussian states.

In addition, working with Laplace transforms and then transforming
back to time domain, we manage to find the exact solutions for the
propagators associated with the integro-differential equations
corresponding to ohmic, sub-ohmic and supra-ohmic environments with a
finite cut-off. This enables us to gain very valuable information on
the dynamics of the system. For instance, for an ohmic environment
one can show that using the local approximation for the propagator is
a valid approximation in the large cut-off limit, which makes it
possible to obtain relatively manageable analytic results for the
diffusion coefficients at all times. Furthermore, the exact solution
of a specific sub-ohmic environment reveals that long-time
correlations (due to excessive coupling with IR modes of the
environment) give rise to contributions to the propagator that decay
at late times like power laws. This invalidates the use of an
effectively local description at late times, whose contributions
decay exponentially, and provides a clear example of a situation
where nonlocal dissipation needs to be properly dealt with. Finally,
studying the exact solutions for some particular supra-ohmic
environment we also find significant nonlocal effects which are due
in this case to the UV regulator function. This leads to a marked
cut-off sensitivity of the momentum covariance that had not been
noticed before.

\subsection{Key Points and Organization}

Those readers who want to find out quickly the problem with earlier
derivations of the master equation can simply read Sec.~\ref{sec:TLE}
to get acquainted with our notation and formalism and go to
Sec.~\ref{sec:MED}, where the master equation is derived, aided
perhaps by \ref{sec:nonlocal_prop}, which explains in detail
the key mathematical subtlety concerning integro-differential
equations and its implications for the existing derivations. They may
also find Sec.~\ref{sec:non-ohmic} valuable since it contains
specific examples where nonlocal dissipation effects give dominant
contributions and can lead to significant discrepancies from previous
results.

The other useful results are mentioned below alongside a description
of how this paper is organized.

The key framework providing the stochastic description for an open quantum
system in terms of a Langevin equation and its compact phase-space
formulation is introduced in Sec.~\ref{sec:TLE}, where a very simple
derivation of the general solution for the state evolution of the
system is given. The problems with previous derivations are
pointed out and the correct derivation of the master equation is
given in Sec.~\ref{sec:master}. The master equation is then solved
using the method of characteristic curves and the solution is shown
to be equivalent to that obtained in a more straightforward manner
from the Langevin equation.

The general  solution of the master equation is employed in
Sec.~\ref{sec:solutions} to discuss general properties of the state
evolution of the QBM subsystem, tending to a Gaussian stationary
state at late times.
A very simple and intuitive picture of environment-induced
decoherence in terms of the reduced Wigner function can be directly
extracted, which could easily be made quantitative and precise. In
addition, a generic discussion of late-time dynamics is provided.

In Sec.~\ref{sec:ohmic} we find the exact nonlocal propagator for an
ohmic environment with finite cut-off and identify a new regime at
ultra-strong coupling. We provide exact nonpertrubative results for
the late-time thermal covariance and full-time results for the
diffusion coefficients in the large cut-off limit.

Explicit examples of sub-ohmic and supra-ohmic spectral functions are
considered in Sec.~\ref{sec:non-ohmic} for which the exact propagator
is computed and dominant contributions from nonlocal dissipation
effects are found (of IR origin in one case and UV in the other).

The generalization to a system of multiple oscillators bilinearly coupled to themselves and the bath in arbitrary fashion and acted upon by classical forces is presented in Sec.~\ref{sec:generalization}. Finally, in Sec.~\ref{sec:discussion} we summarize our results and discuss their main implications as well as possible applications.

In addition to a couple of appendices on special functions and properties of Laplace transforms for reference purposes, \ref{sec:divergences} contains technical aspects concerning divergences of the dissipation kernel and frequency renormalization, as well as initial kicks and a discussion of divergences associated with uncorrelated initial states.

\ref{sec:nonlocal_prop} contains a detailed explanation of the mathematical subtlety involving boundary-value problems for integro-differential equations and a discussion of how it affected different classes of earlier derivations of the master equations. The important formula for the late-time covariance in terms of a single frequency integral is derived in \ref{sec:late_cov_der}, and the explicit analytic results for the diffusion coefficients of an ohmic environment at all times in the large cut-off limit are computed in \ref{sec:mod-time_app}.

Throughout the paper we use units with $\hbar=k_\mathrm{B}=1$.

\section{The Langevin Equation}
\label{sec:TLE}
\subsection{General Theory}
\label{sec:GT1}
The Lagrangian of a closed system consisting of a quantum Brownian oscillator with mass $M$, natural frequency $\Omega$ and coordinate $x$, bilinearly coupled with coupling constants $c_n$ to an environment 
consisting of oscillators with mass $m_n$, natural frequency $\omega_n$ and coordinates $x_n$, is most straightforwardly given by
\begin{align}
\mathsf{L} =&\; \frac{1}{2} M \left( \dot{x}^2 - \Omega_\mathrm{bare}^2 x^2 \right)  \label{eq:lagrangian1} \\
& + \sum_n \frac{1}{2} m_n \left( \dot{x}_n^2 - \omega_n^2 x_n^2 \right) \nonumber
- \sum_n c_n x x_n \, .
\end{align}
One introduces a ``bare'' frequency $\Omega_\mathrm{bare}$ because the interaction with the environment shifts the coefficient of the potential term by a certain amount $\delta \Omega^2$, given by Eq.~\eqref{eq:freq_ren}, so that the square of the actual frequency characterizing the subsystem of interest is given by $\Omega^2_\mathrm{bare} - \delta \Omega^2$.
Alternatively, one can consider the following Lagrangian:
\begin{align}
\mathsf{L}  =&\; \frac{1}{2} M \left( \dot{x}^2 - \Omega^2 x^2 \right) \label{eq:switch} \\
&+ \sum_n \frac{1}{2} m_n \left( \dot{x}_n^2 - \omega_n^2 \left(x_n - \frac{c_n \theta_\mathrm{s}(t)}{m_n \omega_n^2} x \right)^2 \right) \, , \nonumber
\end{align}
where $\Omega$ corresponds to the actual frequency of the Brownian oscillator.
For $\theta_\mathrm{s}(t) = 1$ and provided that one identifies $\Omega^2$ with $\Omega^2_\mathrm{bare} - \delta \Omega^2$, this new Lagrangian is equivalent to that of Eq.~\eqref{eq:lagrangian1} (further details on frequency renormalization and related issues are provided in \ref{sec:divergences}).
In addition, we included a switch-on function $\theta_\mathrm{s}(t)$ which vanishes at the initial time and smoothly increases to reach a constant unit value after a characteristic time-scale $t_\mathrm{s}$.
While we consider initially uncorrelated states for the Brownian oscillator and the environment throughout the paper, which can sometimes lead to certain unphysical results, introducing a smooth switch-on function provides a way of effectively generating well-behaved initial states with the high-frequency modes of the environment properly correlated with the Brownian oscillator.
Further discussion on this point can be found in \ref{sec:initial_coupling}, but throughout the rest of the paper we will take $\theta_\mathrm{s}(t) = 1$ (or, equivalently, $t_\mathrm{s} = 0$) unless stated otherwise, and will only occasionally describe how the results would differ for a non-vanishing switch-on time.

The subsystem corresponding to the quantum Brownian oscillator constitutes an open quantum system: while the evolution of the whole closed system is unitary, the Brownian oscillator (referred to as the ``system'' from now on) evolves non-unitarily due to the entanglement generated by the interaction with the environment.
An important object characterizing the open system is the reduced density matrix, which results from taking the density matrix of the closed system and tracing out the environment: $\rho_\mathrm{r} = \mathrm{Tr}_\mathcal{E} \rho$.
The expectation value of observables $\mathcal{O}$ that only depend on the system variables and are local in time can be directly obtained from it: $\langle \mathcal{O} \rangle (t) = \mathrm{Tr} \left[\mathcal{O}\, \rho_\mathrm{r}(t)\right]$.
Given the density matrix for a continuous degree of freedom in position representation, one can always define the corresponding Wigner function:
\begin{equation}
W_{\!\mathrm{r}} (X,p,t) = \frac{1}{2\pi }\int_{\!-\infty }^{+\infty } \!\!\! d\Delta \, e^{ip\Delta} \, \rho_\mathrm{r}\!\left(X\!-\!\frac{\Delta}{2},X\!+\!\frac{\Delta}{2},t\right) \label{eq:wigner_def1},
\end{equation}
which contains the same amount of information.
See for instance Ref.~\cite{HilleryOconnell86} for a detailed description of the main properties of Wigner functions.
In addition, the so-called dissipation and noise kernels (which involve respectively the commutator and anticommutator of the environment position operators in interaction picture) play an important role when studying the open system dynamics \cite{Gardiner85,roura99}.
The case of a time-dependent coupling has been considered by Hu and Matacz \cite{HuMatacz94}, wherein all parameters of the system and bath oscillators and their couplings were allowed to be time-dependent.
When only the system-environment coupling is time-dependent, as in our case, and the initial state of the environment is a thermal state with temperature $T$, the dissipation and noise kernels are given respectively by
\begin{align}
\label{eq:dissipation} \mu(t,\tau) & = - \! \int_0^\infty \!\!\!d\omega \sin\!\left[\omega(t\!-\!\tau)\right] \, I(\omega) \, \theta_\mathrm{s}(t) \theta_\mathrm{s}(\tau) , \\
\label{eq:noise} \nu(t,\tau) & = + \! \int_0^\infty \!\!\!d\omega \coth\!\left( \frac{\omega}{2 T} \right) \cos\!\left[\omega(t\!-\!\tau)\right] \, I(\omega) \, \theta_\mathrm{s}(t) \theta_\mathrm{s}(\tau) ,
\end{align}
where $I(\omega)$ is the spectral density function defined by
\begin{eqnarray}
I(\omega) &=& \sum_n \frac{c_n^2}{2 m_n \omega_n} \delta(\omega\!-\!\omega_n) .
\end{eqnarray}
It is often taken to be ohmic, i.e.\ $I(\omega) = (2/\pi) M \gamma_0 \, \omega$, but with a cut-off regulator so that it vanishes (or decays sufficiently fast) above some high-frequency scale $\Lambda$. However, more general spectral functions have been considered before and will be considered here as well.

It was shown in Ref.~\cite{CRV03} that the quantum properties of this kind of open systems can be entirely studied using a stochastic description whose central element is a Langevin equation of the form $(L \cdot x)(t) = \xi(t)$, where $\xi(t)$ is a Gaussian stochastic source with a vanishing mean and correlation function equal to the noise kernel, i.e. $\left\langle \xi(t) \right\rangle_\xi = 0$ and  $\left\langle \xi(t) \xi(\tau) \right\rangle_\xi = \nu(t,\tau)$. The dissipation kernel in turn appears in the Langevin integro-differential operator $L$, which is defined by
\begin{align}
(L \cdot x)(t) =&\; M \ddot{x}(t) + M \Omega^2 x(t) + 2 \! \int_0^t \!\!d\tau \, \mu(t,\tau) \, x(\tau) \nonumber \\
& + M \delta\Omega^2 \, \theta^2_\mathrm{s}(t) \, x(t) \label{eq:integro-diff1} \, ,
\end{align}
and where $\delta\Omega^2$ is given by Eq.~\eqref{eq:freq_ren}.
One can then express the time-evolving reduced Wigner function in terms of solutions of the Langevin equation and a double average over their initial conditions, weighed with the reduced Wigner function at the initial time, and over the realizations of the stochastic source [see Eq.~\eqref{eq:average0} below]. Furthermore, one can also obtain the quantum correlation functions for system observables at multiple times (which in general cannot be obtained from the reduced Wigner function and its evolution via the master equation) in terms of the solutions of the Langevin equation \cite{CRV03}, as briefly illustrated in Sec.~\ref{sec:correlations}. See also Ref.~\cite{FordOconnell88} for a similar formulation involving a Langevin equation for operators in the Heisenberg picture.

If we take a vanishing switch-on time, which amounts to discarding the switch-on function entirely, both the noise and dissipation kernels become time-translation invariant.
Moreover, it is convenient to introduce a \emph{damping kernel} $\gamma(t\!-\!\tau)$ which is related to the dissipation kernel by $\mu(t,\tau) = \mu(t\!-\!\tau) = M (\partial/\partial t) \gamma(t\!-\!\tau)$ and is hence given by
\begin{equation}
\gamma(t,\tau) = \gamma(t \!-\! \tau) = \frac{1}{M} \int_0^\infty \!\!\! d\omega \,
\cos[\omega(t\!-\!\tau)] \,\frac{I(\omega)}{\omega} \label{damping2} .
\end{equation}
Note that this kernel is symmetric and positive definite like the noise kernel.
Integrating by parts, the left-hand side of the Langevin equation can be written as follows (see \ref{sec:divergences} for further details):
\begin{align}
(L \cdot x)(t) =&\; M \ddot{x}(t) + 2 M\! \int_0^t \!\!d\tau \, \gamma(t\!-\!\tau)\, \dot{x}(\tau) +  M\Omega^2 x(t) \nonumber \\
& + 2 M \gamma(t) \, x(0) \label{eq:integro-diff2} ,
\end{align}
The damping-kernel representation provides a cancelation of the frequency renormalization while introducing a \emph{slip} in the initial conditions.
This is caused by the last term on the right-hand side of Eq.~\eqref{eq:integro-diff2}, which corresponds to a transient driving term proportional to the position of the system at the initial time.
Leaving the slip term aside, one can show that all the (accumulated) energy dissipated through the nonlocal damping kernel term will be strictly positive (no amplification) as a consequence of the damping kernel being positive-definite.
\subsection{Solutions of the Langevin Equation}
\label{sec:MSSL}
The Langevin equation can be written as
\begin{equation}
L \cdot x = M \left( \ddot{x} + 2 \, \gamma * \dot{x} +  \Omega^2 x \right) + 2 M x_0 \gamma = \xi , \label{eq:integro-diff3}
\end{equation}
where $*$ denotes the Laplace convolution, i.e. $(A*B) (t) = \int_0^t d\tau A(t\!-\!\tau)B(\tau)$,
and $x_0$ is the initial condition at $t=0$.
It is, thus, convenient to perform a Laplace transform
\begin{equation}
\hat{f}(s) = \mathcal{L}\{f\}(s) = \int_0^\infty \!\!\! dt \, e^{-s t} f(t) ,
\end{equation}
under which the equation becomes purely algebraic.
The Laplace transform of Eq.~\eqref{eq:integro-diff3} is given by
\begin{equation}
M \left( s^2 + 2 s \hat{\gamma}(s)+ \Omega^2 \right) \hat{x}(s) = M \left( s x_0 + \dot{x}_0 \right) + \hat{\xi}(s) ,
\end{equation}
whose solution is
\begin{eqnarray}
\hat{x}(s) &=& M \left( s x_0 + \dot{x}_0 \right)\hat{G}(s) + \hat{G}(s) \hat{\xi}(s) ,
\label{eq:laplace0}\\
\hat{G}(s) &=& \frac{1/M}{s^2 + 2 s \hat{\gamma}(s) + \Omega^2 } , \label{eq:green1}
\end{eqnarray}
where terms proportional to the initial conditions $x_0$ and $\dot{x}_0$ correspond to the homogeneous solution while the noise term corresponds to the driven solution.
$G(t)$ satisfies the initial boundary conditions $G(0)=0$, $\dot{G}(0)=\frac{1}{M}$ and fully determines the retarded Green function or propagator.
In the time domain, the solution can be expressed as
\begin{eqnarray}
x(t) &=& M \left( x_0 \dot{G}(t) + \dot{x}_0 G(t) \right) + (G * \xi)(t) .
\end{eqnarray}

\subsubsection{Meromorphic Spectra}
For an ohmic environment in the infinite cut-off limit one has $\hat{\gamma}(s) = \gamma_0$. More realistically, $\hat{\gamma}(s)$ will decay sufficiently fast at high $s$, implying a certain degree of nonlocal dissipation (non-polynomial behavior in Laplace space).
Thus, as illustrated by this example, one will generally need to deal with non-polynomial damping kernels $\hat{\gamma}(s)$.
If $\hat{\gamma}(s)$ is a meromorphic function (i.e. analytic except for an isolated set of poles), obtaining the inverse Laplace transform of $\hat{G}(s)$ amounts to calculating a simple contour integral.

On the other hand, given expression \eqref{damping2} for the damping kernel,  one can easily compute its Laplace transform:
\begin{equation}
\hat{\gamma}(s) = \frac{1}{M} \int_0^\infty \!\!\! d\omega \, \frac{I(\omega)}{\omega} \frac{s}{\omega^2 + s^2} \label{eq:dissip2}.
\end{equation}
If we take the odd extension of the spectral density for negative frequencies, i.e.\ $I(-|\omega|) \equiv -I(|\omega|)$, then the integral can be recast as
\begin{eqnarray}
\hat{\gamma}(s) &=& \frac{1}{2M} \int_{\!-\infty}^{+\infty} \!\!\! d\omega \, \frac{I(\omega)}{\omega} \frac{s}{\omega^2 + s^2} , \label{eq:mu-contour}
\end{eqnarray}
which can be easily evaluated if the odd extension of $I(\omega)$ is meromorphic, e.g.\ for $I(\omega) \sim \omega$ but not $I(\omega) \sim \omega^2$.
This is still less than ideal as the difficulty of solving the Langevin equation is more directly determined by the nature of the damping kernel.
One would rather make the choice of damping kernel first (preferably in the Laplace domain) than derive it from the spectral density. Nevertheless, since the spectral density is still required to compute the noise kernel, we need the inverse relationship.
Furthermore, as shown below, not every $\hat{\gamma}(s)$ (even sufficiently regular ones) can be obtained from a spectral function through  Eq.~\eqref{eq:mu-contour}.

Fortunately, Eq.~\eqref{damping2} implies a simple relation between the spectral density and the Fourier transform of the damping kernel: $I(\omega) = \frac{M}{\pi} \omega \tilde{\gamma}(\omega)$,
and using Eq.~\eqref{eq:laplace_fourier3} applied to $\tilde{\gamma}(\omega)$ we get the following result for $I(\omega)$ in terms of the Laplace transform of the damping kernel:%
\begin{equation}
I(\omega) = \frac{1}{\pi} M \omega \lim_{\epsilon \to 0}
\left[ \hat{\gamma}(\epsilon \!+\! \imath \omega)
+ \hat{\gamma}(\epsilon \!-\! \imath \omega) \right]
\label{eq:spectral_damp}.
\end{equation}
From this we see that meromorphic damping kernels result in spectral densities which are odd meromorphic functions. Conversely, we have also seen that odd meromorphic spectral densities lead to a meromorphic damping kernel in Laplace space that can be obtained via contour integration through Eq.~\eqref{eq:mu-contour}.
We will thus refer to this class of odd meromorphic spectral densities and corresponding damping kernels as \emph{meromorphic spectra}.
Moreover, as we will see in later sections, given that Bromwich's formula for the inverse Laplace transform can also be computed as a contour integral,
all the important quantities for these meromorphic spectra are calculable via contour integration.

Note that, as mentioned above, not every meromorphic function $\hat{\gamma}(s)$ corresponds to a damping kernel that can be obtained from a spectral function through  Eq.~\eqref{eq:mu-contour}.
This point can be seen by realizing that according to Eq.~\eqref{eq:spectral_damp} different $\hat{\gamma}(s)$ will give rise to the same spectral density as long as $\hat{\gamma}(\epsilon + \imath \omega) + \hat{\gamma}(\epsilon - \imath \omega)$ is the same.
Hence, if one wants to consider a candidate function $\hat{\gamma}(s)$, one should proceed as follows.
Eq.~\eqref{eq:spectral_damp} is first used to obtain the spectral density, which is then substituted into Eq.~\eqref{eq:mu-contour}.
If the initial candidate is recovered, it was a satisfactory one to begin with, otherwise it should be discarded, but the new damping kernel obtained in the last step is a valid one, which can be used instead.

\subsubsection{Phase-Space Representation}
\label{sec:PSR}
If we introduce the phase-space coordinates $\mathbf{z}^{\!\mathrm{T}}=(x,p)$, the Langevin equation~\eqref{eq:integro-diff3}, together with the relation $p = m \dot{x}$, can be recast as a first-order linear integro-differential system of equations:
\begin{equation}
\dot{\mathbf{z}} + \boldsymbol{\mathsf{H}} * \mathbf{z} = \boldsymbol{\xi} ,
\label{eq:langevin_bold}
\end{equation}
where we introduced the boldface notation for vectors and matrices, $\boldsymbol{\xi}^\mathrm{T}=(0,\xi)$ and the time-nonlocal pseudo-Hamiltonian $\boldsymbol{\mathsf{H}}(t,\tau)=\boldsymbol{\mathsf{H}}(t\!-\!\tau)$ is given by
\begin{equation}
\boldsymbol{\mathsf{H}}(\tau) = \left[ \begin{array}{cc} 0 & -\frac{1}{M} \delta(\tau) \\
M\Omega^2 \delta(\tau) & 2 \, \gamma(\tau) \end{array} \right] \label{eq:langevin_kernel} .
\end{equation}
Performing the Laplace transform of Eq.~\eqref{eq:langevin_bold}, which becomes a purely algebraic equation, and rearranging the terms to express the solution in terms of the initial conditions and the stochastic source, one gets
\begin{eqnarray}
\hat{\mathbf{z}}(s) &=& \hat{\boldsymbol{\Phi}}(s) \, \mathbf{z}_0 + \hat{\boldsymbol{\Phi}}(s) \, \hat{\boldsymbol{\xi}}(s) , \\
\hat{\boldsymbol{\Phi}}(s) &=& \left[ \begin{array}{cc} M s\, \hat{G}(s) & \hat{G}(s) \\ M^2 s^2 \hat{G}(s) - M & M s\, \hat{G}(s) \end{array} \right] , \label{eq:Phi}
\end{eqnarray}
where $\hat{G}(s)$ is the same propagator derived in the position representation and given by Eq.~\eqref{eq:green1}.
Transforming back to the time domain, we can express the initial-value solutions as
\begin{eqnarray}
\mathbf{z}(t) &=& \boldsymbol{\Phi}(t) \, \mathbf{z}_0 + (\boldsymbol{\Phi} * \boldsymbol{\xi})(t) , \label{eq:LEIVS} \\
\boldsymbol{\Phi}(t) &=& \left[ \begin{array}{cc} M \dot{G}(t) & G(t) \\ M^2 \ddot{G}(t) & M \dot{G}(t) \end{array} \right] , \label{eq:Phi(t)}
\end{eqnarray}
and $\boldsymbol{\Phi}(t)$ can be identified as the matrix propagator associated with the phase-space version of the Langevin equation, Eq.~\eqref{eq:langevin_bold}.

Combining the result for $\mathbf{z}(t)$ as given by Eq.~\eqref{eq:LEIVS} with an analogous expression for the solution $\mathbf{z}(\tau)$ evaluated at an earlier time $\tau < t$, one can write $\mathbf{z}(\tau)$ in terms of $\mathbf{z}(t)$ and the stochastic source as follows:
\begin{eqnarray}
\mathbf{z}(\tau) &=& \boldsymbol{\Phi}(\tau,t) \, \mathbf{z}(t)
- \int_\tau^t \!\! d\tau' \, \boldsymbol{\Phi}(\tau,t) \, \boldsymbol{\Phi}(t\!-\!\tau') \, \boldsymbol{\xi}(\tau') \nonumber \\
&& -\int_0^\tau \!\! d\tau' \left[ \boldsymbol{\Phi}(\tau,t) \, \boldsymbol{\Phi}(t\!-\!\tau') - \boldsymbol{\Phi}(\tau\!-\!\tau') \right] \boldsymbol{\xi}(\tau') , \label{eq:final}
\end{eqnarray}
where we introduced the transition matrix $\boldsymbol{\Phi}(t,\tau)$, which is defined as
\begin{equation}
\boldsymbol{\Phi}(t,\tau) = \boldsymbol{\Phi}(t) \, \boldsymbol{\Phi}^{\!-1}(\tau)
\label{eq:transition} .
\end{equation}
Note that $\boldsymbol{\Phi}(t,\tau) \neq \boldsymbol{\Phi}(t\!-\!\tau)$ unless one has local dissipation. Thus, in the general case of nonlocal dissipation the last term on the right-hand side of Eq.~\eqref{eq:final} does not vanish and $\mathbf{z}(\tau)$ also depends on $\boldsymbol{\xi}(\tau')$ with $\tau' < \tau$.
This means that, unlike with ordinary differential equations, when boundary conditions $\mathbf{z}(t)$ are specified at a final time $t$, there is no truly \emph{advanced} propagator for the inhomogeneous solutions of the integro-differential equation.
One can still express the solution of such a final-value problem in terms of a matrix propagator (or Green's function in position space) with the right boundary conditions:
\begin{equation}
\mathbf{z}(\tau) = \boldsymbol{\Phi}(\tau,t) \, \mathbf{z}(t) + \int_0^t \!\! d\tau' \, \boldsymbol{\Phi}_\mathrm{f}\left(\tau,\tau'\right) \, \boldsymbol{\xi}(\tau') ,\label{eq:final_solution}
\end{equation}
where
\begin{equation}
\boldsymbol{\Phi}_\mathrm{f}\left(\tau,\tau'\right)
= -\boldsymbol{\Phi}(\tau,t) \, \boldsymbol{\Phi}(t\!-\!\tau')
+ \theta\!\left( \tau \!-\! \tau' \right) \, \boldsymbol{\Phi}(\tau\!-\!\tau')
\label{eq:final_propag},
\end{equation}
but one only has $\boldsymbol{\Phi}_\mathrm{f}\left(\tau,\tau'\right)=0$ for $\tau > \tau'$ in the case of strictly local dissipation.

Such mathematical subtleties of final-value problems for integro-differential equations have been missed in the existing literature on the derivation of the master equation for QBM models and could lead to significant discrepancies whenever the nonlocal effects of dissipation are important. A detailed discussion of this and related points is provided in \ref{sec:nonlocal_prop}.

\subsection{Evolution of States}
\label{sec:MESL}

As found in Ref.~\cite{CRV03}, the reduced Wigner function can be expressed in terms of the solutions of the Langevin equation and a double average over their initial conditions and the realizations of the stochastic source. Using the vector notation for phase-space variables introduced in the previous subsection, the result can be written as
\begin{equation}
W_{\!\mathrm{r}}(\mathbf{z},t)
= \left\langle \left\langle \delta\!\left( \mathbf{z}(t) \!-\! \mathbf{z} \right) \right\rangle_{\boldsymbol{\xi}} \right\rangle_{\!\!\mathbf{z}_0}
\label{eq:average0} ,
\end{equation}
with the averages over the initial conditions and the stochastic source defined as follows:
\begin{eqnarray}
\langle \cdots \rangle_{\mathbf{z}_0} &=& \frac{1}{2 \pi} \int d\mathbf{z} \cdots
W_{\!\mathrm{r}}(\mathbf{z},0)
\label{eq:initial_average}, \\
\langle \cdots \rangle_{\boldsymbol{\xi}} &=& \frac{1}{\sqrt{2\pi \det(\nu)}} \int \! \mathcal{D}\xi \cdots e^{-\frac{1}{2} \xi \cdot \nu^{-1} \cdot \xi}
\label{eq:functional_measure},
\end{eqnarray}
where the right-hand side of Eq.~\eqref{eq:functional_measure} corresponds to the functional integral associated with the Gaussian stochastic source.
The characteristic function of the Wigner function, regarded as a phase-space distribution, is given by its Fourier transform and it can be shown to take a rather simple form:
\begin{eqnarray}
\mathcal{W}_\mathrm{r}(\mathbf{k},t) &=& \int \! d\mathbf{z} \, e^{-i \mathbf{k}^\mathrm{T} \mathbf{z}} \left\langle \left\langle \delta\!\left[\mathbf{z} \!-\! \mathbf{z}(t)\right] \right\rangle_{\mathbf{z}_0} \right\rangle_{\!\boldsymbol{\xi}} , \\
&=& \left\langle \left\langle e^{-i \mathbf{k}^\mathrm{T} \mathbf{z}(t)} \right\rangle_{\!\mathbf{z}_0} \right\rangle_{\!\!\boldsymbol{\xi}} , \\
&=& \left\langle  e^{-i \mathbf{k}^\mathrm{T} \boldsymbol{\Phi}(t) \mathbf{z}_0} \right\rangle_{\!\mathbf{z}_0} \left\langle e^{-i \mathbf{k}^\mathrm{T} (\boldsymbol{\Phi} * \boldsymbol{\xi})(t)} \right\rangle_{\!\boldsymbol{\xi}} \label{eq:MES0} , \\
&=& \mathcal{W}_\mathrm{r}\big(\boldsymbol{\Phi}^{\!\mathrm{T}}\!(t) \, \mathbf{k},0\big) \, e^{-\frac{1}{2} \mathbf{k}^\mathrm{T}\! \boldsymbol{\sigma}_T(t) \, \mathbf{k}} , \label{eq:MES1}
\end{eqnarray}
where the thermal covariance matrix $\boldsymbol{\sigma}_T(t)$ is given by
\begin{align}
\boldsymbol{\sigma}_T(t) &= \int_0^t \!\! d\tau \! \int_0^t \!\! d\tau' \, \boldsymbol{\Phi}(t\!-\!\tau) \, \boldsymbol{\nu}(\tau,\tau') \, \boldsymbol{\Phi}^{\!\mathrm{T}} (t\!-\!\tau') , \label{eq:sigma1} \\
\boldsymbol{\nu}(\tau,\tau') &= \left[ \begin{array}{cc} 0 & 0 \\ 0 & \nu(\tau,\tau') \end{array} \right].
\end{align}
In the third equality above we used the initial-value solution \eqref{eq:LEIVS} for $\mathbf{z}(t)$ to get Eq.~\eqref{eq:MES0},
and in the last step we completed the square to calculate the Gaussian functional integral corresponding to the noise average in order to obtain the final result in Eq.~\eqref{eq:MES1}.
Note that for our Lagrangian, the stochastic force $\boldsymbol{\xi}$ only has a momentum component and, therefore, all the components of its covariance matrix $\boldsymbol{\nu}$ vanish except for the momentum-momentum component, which coincides with the noise kernel.

The form of the solution is rather simple: all initial cumulants of the Wigner function undergo damped oscillations (for the underdamped case) while the thermal covariance starts from a vanishing value and evolves to the asymptotic values corresponding to the thermal equilibrium state for the system coupled to the environment.
We will discuss these solutions more thoroughly in Sec.~\ref{sec:solutions}.

\subsection{General Correlations}
\label{sec:correlations}
Using the initial-value solution of the Langevin equation given by Eq.~\eqref{eq:LEIVS} and following the same approach as in Ref.~\cite{CRV03}, it is straightforward to calculate quantum correlations between system observables at different times.
For instance, the symmetrized two-point quantum correlation function for position and momentum operators in the Heisenberg representation is given by:
\begin{align}
& \frac{1}{2} \left\langle \mathbf{z}\!\left(t_1\right) \mathbf{z}^{\!\mathrm{T}}\!\!\left(t_2\right) + \mathbf{z}\!\left(t_2\right) \mathbf{z}^{\!\mathrm{T}}\!\!\left(t_1\right) \right\rangle  = \nonumber\\
& \frac{1}{2} \left\langle \left\langle \mathbf{z}\!\left(t_1\right) \mathbf{z}^{\!\mathrm{T}}\!\!\left(t_2\right) + \mathbf{z}\!\left(t_2\right) \mathbf{z}^{\!\mathrm{T}}\!\!\left(t_1\right) \right\rangle_{\boldsymbol{\xi}} \right\rangle_{\!\!\mathbf{z}_0} ,
\end{align}
which with our solutions in Eq.~\eqref{eq:LEIVS} and some basic properties of the stochastic Gaussian source,
namely $\left\langle \boldsymbol{\xi}(t) \right\rangle_{\boldsymbol{\xi}} = 0$
and  $\left\langle \boldsymbol{\xi}(t) \boldsymbol{\xi}(\tau) \right\rangle_{\boldsymbol{\xi}} = \boldsymbol{\nu}(t,\tau)$,
will produce the two-time correlation
\begin{equation}
\left\langle \left\langle\mathbf{z}\!\left(t_1\right) \mathbf{z}^{\!\mathrm{T}}\!\!\left(t_2\right) \right\rangle_{\boldsymbol{\xi}} \right\rangle_{\!\!\mathbf{z}_0}
= \boldsymbol{\Phi}\!\left(t_1\right) \boldsymbol{\sigma}_0 \, \boldsymbol{\Phi}^{\!\mathrm{T}}\!\!\left(t_2\right) + \boldsymbol{\sigma}_T\!\left( t_1,t_2 \right) \label{eq:2point},
\end{equation}
in terms of the two-time thermal covariance
\begin{equation}
\boldsymbol{\sigma}_T\!\left( t_1,t_2 \right) = \int_0^{t_1} \!\!\! d\tau_1 \! \int_0^{t_2} \!\!\! d\tau_2 \, \boldsymbol{\Phi}\!\left(t_1\!-\!\tau_1\right) \, \boldsymbol{\nu}\!\left(\tau_1,\tau_2\right) \, \boldsymbol{\Phi}^{\!\mathrm{T}}\!\!\left(t_2\!-\!\tau_2\right) .
\end{equation}
The result for the coincidence-time limit, $t_1=t_2=t$, agrees with that of our master equation solution, Eqs.~\eqref{eq:MES1}-\eqref{eq:sigma1}, as discussed in Sec.~\ref{sec:wigner-sol}.
Higher-order correlations can be calculated in a similar manner,
but we can see from the form of our solution in Eq.~\eqref{eq:MES1} and the Gaussian character of the stochastic source and its vanishing mean that only the homogeneous part of the solution contributes to cumulants different from the second-order one,
which are therefore entirely characterized by the initial state of system and the homogeneous solutions of the Langevin equation.

\section{Master Equation}
\label{sec:master}

\subsection{General Theory}
Given the microscopic QBM model of Sec.~\ref{sec:GT1},
the HPZ master equation for the reduced density matrix operator $\rho_\mathrm{r}$ and for the reduced Wigner function are given respectively by
\begin{align}
\frac{\partial}{\partial t} \rho_\mathrm{r} =&\;  -\imath \left[ H_\mathrm{R}, \rho_\mathrm{r} \right] - \imath \Gamma \left[ x,\{ p, \rho_\mathrm{r}\}\right] \\
& - M D_{\!pp} \left[ x,\left[ x, \rho_\mathrm{r}\right]\right] - D_{\!xp} \left[ x,\left[ p, \rho_\mathrm{r}\right]\right] , \nonumber \\
\frac{\partial}{\partial t} W_{\!\mathrm{r}} =&\; \{ H_\mathrm{R},W_{\!\mathrm{r}} \} + 2\Gamma \frac{\partial}{\partial p} (p W_{\!\mathrm{r}}) \label{eq:wigner} \\
&+ M D_{\!pp} \frac{\partial^2}{\partial p^2} W_{\!\mathrm{r}} - D_{\!xp} \frac{\partial^2}{\partial x \partial p} W_{\!\mathrm{r}} , \nonumber
\end{align}
where $H_\mathrm{R}$ corresponds to the system Hamiltonian with $\Omega^2$ replaced by a time-dependent frequency $\Omega^2_\mathrm{R}(t) \sim \Omega^2$ whose detailed form, together with that of the time-dependent dissipation coefficient $\Gamma(t)$ and the diffusion coefficients $D_{\!xp}(t)$ and $D_{\!pp}(t)$, can be found in Ref.~\cite{HPZ92}.

However, as discussed in \ref{sec:nonlocal_prop}, previous derivations of this master equation missed a mathematical subtlety concerning the Green functions of integro-differential equations, which renders the existing results for the master equation coefficients invalid whenever the nonlocal aspects of dissipation become important. In the next subsection we provide a compact rederivation of the master equation where this issue is properly dealt with, and obtain the correct expressions for the coefficients in the general case (including the case of nonlocal dissipation). In addition, in Sec.~\ref{sec:MESM} we will provide an analytic expression for the solutions of the master equation and show its equivalence with the result for the state evolution obtained in the previous section using the Langevin equation.

\subsection{Derivation of the Master Equation}
\label{sec:MED} At this point, the quickest derivation of the QBM
master equation would merely consist of taking the time derivative of
Eq.~\eqref{eq:MES1} and calculating the inverse Fourier transform.
Nevertheless, in order to point out the differences with previous
derivations, which missed the subtleties of propagators associated
with integro-differential equations, we will now provide a more
traditional derivation involving the propagator associated with
final-value boundary conditions and show that, when done correctly,
the two are equivalent. We will follow the derivation by Calzetta,
Roura and Verdaguer (CRV) \cite{CRV03,CRV01} adapting it to our
compact notation in terms of phase-space vectors and matrices.

We start by considering the stochastic representation of the Wigner function
\begin{equation}
W_{\!\mathrm{r}}(\mathbf{z},t)
= \left\langle \left\langle \delta\!\left( \mathbf{z}(t) \!-\! \mathbf{z} \right) \right\rangle_{\boldsymbol{\xi}} \right\rangle_{\!\!\mathbf{z}_0},
\end{equation}
and differentiate with respect to time:
\begin{equation}
\frac{\partial}{\partial t} W_{\!\mathrm{r}}(\mathbf{z},t) = - \boldsymbol{\nabla}_{\!\!\mathbf{z}}^{\mathrm{T}} \left\langle \left\langle \dot{\mathbf{z}}(t) \, \delta\!\left( \mathbf{z}(t) \!-\! \mathbf{z} \right) \right\rangle_{\boldsymbol{\xi}} \right\rangle_{\!\!\mathbf{z}_0}  \label{eq:derivation1}.
\end{equation}
One can then use the Langevin equation $\dot{\mathbf{z}} + \boldsymbol{\mathsf{H}} * \mathbf{z} = \boldsymbol{\xi}$ to substitute $\dot{\mathbf{z}}(t)$ and rewrite Eq.~\eqref{eq:derivation1} as
\begin{align}
& \frac{\partial}{\partial t} W_{\!\mathrm{r}}(\mathbf{z},t) = \\
& \boldsymbol{\nabla}_{\!\!\mathbf{z}}^{\mathrm{T}} \left\langle \left\langle \left( \int_0^t \!\! d\tau \, \boldsymbol{\mathsf{H}}(t,\tau) \, \mathbf{z}(\tau) - \boldsymbol{\xi}(t) \right) \delta\!\left( \mathbf{z}(t) \!-\! \mathbf{z} \right) \right\rangle_{\!\!\boldsymbol{\xi}} \right\rangle_{\!\!\mathbf{z}_0} . \nonumber
\end{align}
Next, using Eq.~\eqref{eq:final_solution} one can express $\mathbf{z}(\tau)$ in terms of the final value $\mathbf{z}(t)=\mathbf{z}$ and the propagator $\boldsymbol{\Phi}_\mathrm{f}\!\left(\tau,\tau'\right)$ given by Eq.~\eqref{eq:final_propag}.
As already pointed out in Sec.~\ref{sec:MSSL} and discussed in detail in \ref{sec:nonlocal_prop}, $\boldsymbol{\Phi}_\mathrm{f}\!\left(\tau,\tau'\right)$ will only be a truly advanced propagator [with $\boldsymbol{\Phi}_\mathrm{f}\!\left(\tau,\tau'\right)=0$ for $\tau > \tau'$] when considering a strictly local damping kernel, contrary to what had been previously assumed.
After using Eq.~\eqref{eq:final_solution} we are left with a homogeneous term and two more terms involving the stochastic source:
\begin{align}
&\frac{\partial}{\partial t} W\!(\mathbf{z},t) = \boldsymbol{\nabla}_{\!\!\mathbf{z}}^{\mathrm{T}} \! \int_0^t \!\! d\tau \, \boldsymbol{\mathsf{H}}(t,\tau) \,  \boldsymbol{\Phi}(\tau,t) \, \mathbf{z} \, W\!(\mathbf{z},t) \nonumber \\
& + \boldsymbol{\nabla}_{\!\!\mathbf{z}}^{\mathrm{T}} \left\langle \left\langle \int_0^t \!\! d\tau \! \int_0^t \!\! d\tau' \, \boldsymbol{\mathsf{H}}(t,\tau) \, \boldsymbol{\Phi}_\mathrm{f}\!\left(\tau,\tau'\right) \, \boldsymbol{\xi}(\tau') \, \delta\!\left( \mathbf{z}(t) \!-\! \mathbf{z} \right) \right\rangle_{\!\!\boldsymbol{\xi}} \right\rangle_{\!\!\!\mathbf{z}_0} \nonumber \\
& - \boldsymbol{\nabla}_{\!\!\mathbf{z}}^{\mathrm{T}} \left\langle \left\langle \boldsymbol{\xi}(t) \, \delta\!\left( \mathbf{z}(t) \!-\! \mathbf{z} \right) \right\rangle_{\boldsymbol{\xi}} \right\rangle_{\!\!\!\mathbf{z}_0} .
\end{align}
The expectation value of the terms proportional to the stochastic source $\boldsymbol{\xi}$ can be evaluated with the help of Novikov's formula
\begin{align}
& \left\langle \boldsymbol{\xi}(\tau') \, \delta\!\left( \mathbf{z}(t) \!-\! \mathbf{z} \right) \right\rangle_{\boldsymbol{\xi}} = \label{eq:novikov} \\
& - \int_0^t \!\! d\tau'' \, \boldsymbol{\nu}\!\left( \tau' , \tau'' \right) \, \left\langle \left[ \frac{ \delta \mathbf{z}(t) }{ \delta \boldsymbol{\xi}(\tau'') } \right]^{\!\mathrm{T}} \! \boldsymbol{\nabla}_{\!\!\mathbf{z}} \, \delta\!\left( \mathbf{z}(t) \!-\! \mathbf{z} \right) \right\rangle_{\!\!\!\boldsymbol{\xi}} , \nonumber
\end{align}
which can be derived by using Eq.~\eqref{eq:functional_measure} and functionally integrating by parts with respect to $\xi$. The functional Jacobian matrix appearing in Eq.~\eqref{eq:novikov} can be easily obtained by functionally differentiating with respect to $\boldsymbol{\xi}(\tau'')$ the solution of the Langevin equation as given by Eq.~\eqref{eq:LEIVS}, and one gets
\begin{eqnarray}
\left[ \frac{ \delta \mathbf{z}(t) }{ \delta \boldsymbol{\xi}(\tau'') } \right] &=& \boldsymbol{\Phi}\!\left(t\!-\!\tau''\right) .
\end{eqnarray}
Putting these elements together we finally get the following result for the master equation:
\begin{equation}
\frac{\partial}{\partial t} W_{\!\mathrm{r}}(\mathbf{z},t) = \left\{ \boldsymbol{\nabla}_{\!\!\mathbf{z}}^{\mathrm{T}} \, \boldsymbol{\mathcal{H}}(t) \, \mathbf{z} + \boldsymbol{\nabla}_{\!\!\mathbf{z}}^{\mathrm{T}} \, \mathbf{D}(t) \, \boldsymbol{\nabla}_{\!\!\mathbf{z}} \right\} W_{\!\mathrm{r}}(\mathbf{z},t) ,
\end{equation}
with the time-local \emph{pseudo-Hamiltonian} and \emph{diffusion} matrices given respectively by
\begin{align}
& \boldsymbol{\mathcal{H}}(t) \equiv \int_0^t \!\! d\tau \, \boldsymbol{\mathsf{H}}(t,\tau) \, \boldsymbol{\Phi}(\tau,t) , \\
& \mathbf{D}(t) \equiv \mathrm{Sy} \! \int_0^t \!\! d\tau \, \boldsymbol{\nu}\!\left( t , \tau \right) \, \boldsymbol{\Phi}^{\!\mathrm{T}}\!(t\!-\!\tau) - \label{eq:diffusion0} \\
& \mathrm{Sy} \! \int_0^t \!\! d\tau \! \int_0^t \!\! d\tau' \!\! \int_0^t \!\! d\tau'' \, \boldsymbol{\mathsf{H}}(t,\tau) \, \boldsymbol{\Phi}_\mathrm{f}\!\left(\tau,\tau'\right) \, \boldsymbol{\nu}\!\left( \tau' , \tau'' \right) \, \boldsymbol{\Phi}^{\!\mathrm{T}}\!\left(t\!-\!\tau''\right) \nonumber,
\end{align}
and where $\boldsymbol{\Phi}_\mathrm{f}(\tau,\tau')$ was defined in Eq.~\eqref{eq:final_propag},
and only the symmetric part, $\mathrm{Sy} (\mathbf{M}) \equiv \big( \mathbf{M} + \mathbf{M}^\mathrm{T} \big)/2$, of the diffusion matrix contributes to the master equation.
These matrices relate to the conventional representation as follows:
\begin{align}
\boldsymbol{\mathcal{H}}(t) &= \left[ \begin{array}{cc} 0 & -\frac{1}{M} \\ M \Omega^2_\mathrm{R}(t) & 2\Gamma(t) \end{array} \right] , \\
\mathbf{D}(t) &= \left[ \begin{array}{cc} 0 & -\frac{1}{2} D_{\!xp}(t) \\ -\frac{1}{2} D_{\!xp}(t) & M D_{\!pp}(t) \end{array} \right] .
\end{align}
The result for the master equation coefficients is expressed here in a form analogous to that of previous derivations, but this is not the simplest representation.
We will next proceed to simplify them by eliminating the explicit dependence on the time-nonlocal pseudo-Hamiltonian $\boldsymbol{\mathsf{H}}(t,\tau)$.

\subsubsection{Simplification of the Master Equation Coefficients}

Let us start with the pseudo-Hamiltonian matrix
\begin{equation}
\boldsymbol{\mathcal{H}}(t) = (\boldsymbol{\mathsf{H}} \cdot  \boldsymbol{\Phi})(t) \, \boldsymbol{\Phi}^{\!-1}(t) .
\end{equation}
Taking into account that $\boldsymbol{\Phi}$ satisfies the integro-differential equation $\dot{\boldsymbol{\Phi}}(t) = -(\boldsymbol{\mathsf{H}} \cdot  \boldsymbol{\Phi})(t)$, the pseudo-Hamiltonian can be rewritten as
\begin{equation}
\boldsymbol{\mathcal{H}}(t) = -\dot{\boldsymbol{\Phi}}(t) \, \boldsymbol{\Phi}^{\!-1}(t)
\label{eq:simplifiedH}.
\end{equation}
This new expression for $\boldsymbol{\mathcal{H}}(t)$ immediately reveals that the homogenous solutions of the nonlocal Langevin equation can be equivalently related to the solutions of linear ordinary differential equation with time-dependent coefficients.
Indeed, the nonlocal propagator also satisfies the dual local equation
\begin{eqnarray}
\dot{\boldsymbol{\Phi}}(t) + \boldsymbol{\mathcal{H}}(t) \, \boldsymbol{\Phi}(t) &=& 0 .
\end{eqnarray}
Hence, for local dissipation one would simply have a time-independent $\boldsymbol{\mathcal{H}}$ and $\boldsymbol{\Phi}(t) = e^{-t \, \boldsymbol{\mathcal{H}}}$, whereas for nonlocal dissipation $\boldsymbol{\mathcal{H}}(t)$ would be time-dependent and $\boldsymbol{\Phi}(t)$ would be given by a time-ordered exponential.

One can proceed analogously for the diffusion matrix. In order to do so we need to simplify the following integral:
\begin{align}
& \int_0^t \!\! d\tau \, \boldsymbol{\mathsf{H}}(t,\tau) \, \boldsymbol{\Phi}(\tau\!-\!\tau') \, \theta\!\left( \tau \!-\! \tau' \right)
= \int_{\tau'}^t \!\! d\tau \, \boldsymbol{\mathsf{H}}(t\!-\!\tau) \, \boldsymbol{\Phi}(\tau\!-\!\tau') ,
\end{align}
which reduces to
\begin{align}
\int_{\tau'}^t \!\! d\tau \, \boldsymbol{\mathsf{H}}(t\!-\!\tau) \, \boldsymbol{\Phi}(\tau\!-\!\tau')
&= \int_{0}^{t-\tau'} \!\!\! d\tau \, \boldsymbol{\mathsf{H}}(t\!-\!\tau'\!-\!\tau) \, \boldsymbol{\Phi}(\tau) \nonumber \\
&= -\dot{\boldsymbol{\Phi}}(t\!-\!\tau') \label{eq:simplification1}.
\end{align}
where we made use of the stationary property of the dissipation kernel and introduced a simple change of variables.
Using Eqs.~\eqref{eq:simplifiedH} and \eqref{eq:simplification1}, Eq.~\eqref{eq:diffusion0} can be simplified to the following form, which involves terms with at most two time integrals:
\begin{align}
& \mathbf{D}(t) = \mathrm{Sy}\! \int_0^t \!\! d\tau \, \boldsymbol{\nu}\!\left( t , \tau \right) \, \boldsymbol{\Phi}^{\!\mathrm{T}}\!\!\left(t\!-\!\tau\right) + \label{eq:diff_simple} \\
& \mathrm{Sy}\! \int_0^t \!\! d\tau \! \int_0^t \!\! d\tau' \left\{ \left[ \frac{d}{dt} + \boldsymbol{\mathcal{H}}(t) \right] \boldsymbol{\Phi}(t\!-\!\tau) \right\} \boldsymbol{\nu}\!\left( \tau , \tau' \right) \, \boldsymbol{\Phi}^{\!\mathrm{T}}\!\left(t\!-\!\tau'\right) , \nonumber
\end{align}
where one can clearly see that the second term on the right-hand side vanishes for local dissipation, when the transition matrix is the exponential matrix $e^{-t \, \boldsymbol{\mathcal{H}}}$.
However, it can play a crucial role whenever the effects of nonlocal dissipation are important, as in the example of a sub-ohmic environment of Sec.~\ref{sec:sub-ohmic}.

From our new expression \eqref{eq:diff_simple} one can see that the diffusion matrix can be easily related to the thermal covariance, as given by Eq.~\eqref{eq:sigma1}, and its time derivative.
Our simplified representation of the master equation is then
\begin{align}
\frac{\partial}{\partial t} W_{\!\mathrm{r}}(\mathbf{z},t) &= \left\{ \boldsymbol{\nabla}_{\!\!\mathbf{z}}^{\mathrm{T}} \, \boldsymbol{\mathcal{H}}(t) \, \mathbf{z} + \boldsymbol{\nabla}_{\!\!\mathbf{z}}^{\mathrm{T}} \, \mathbf{D}(t) \, \boldsymbol{\nabla}_{\!\!\mathbf{z}} \right\} W_{\!\mathrm{r}}(\mathbf{z},t) , \\
\boldsymbol{\mathcal{H}}(t) &= -\dot{\boldsymbol{\Phi}}(t) \, \boldsymbol{\Phi}^{\!-1}(t) , \\
\mathbf{D}(t) &= \frac{1}{2} \left\{ \boldsymbol{\mathcal{H}}(t) \, \boldsymbol{\sigma}_T(t) + \boldsymbol{\sigma}_T(t) \, \boldsymbol{\mathcal{H}}^{\!\mathrm{T}}\!(t) + \dot{\boldsymbol{\sigma}}_T(t) \right\} , \label{eq:Ddynamic}
\end{align}
with the phase-space propagator $\boldsymbol{\Phi}(t)$ given by Eq.~\eqref{eq:Phi(t)} and the thermal covariance $\boldsymbol{\sigma}_T(t)$ given by Eq.~\eqref{eq:sigma1}.
This representation contains fewer integrals than the conventional representation and is completely determined in terms of $\boldsymbol{\Phi}(t)$ and the noise kernel.

\subsection{Master Equation Solutions}
\label{sec:MESM}
In this section we will show that the master equation itself can be solved to produce the same solution as derived in Sec.~\ref{sec:MESL}.
We consider the general master equation
\begin{equation}
\frac{\partial}{\partial t} W_{\!\mathrm{r}} = \left( \boldsymbol{\nabla}_{\!\!\mathbf{z}}^{\mathrm{T}} \, \mathbf{D}(t) \, \boldsymbol{\nabla}_{\!\!\mathbf{z}} + \boldsymbol{\nabla}_{\!\!\mathbf{z}}^{\mathrm{T}} \, \boldsymbol{\mathcal{H}}(t) \, \mathbf{z} \right) W_{\!\mathrm{r}} \, .
\end{equation}
This is a hyperbolic second-order partial differential equation (PDE).
The equation is not separable in time nor phase-space.
The nature of the PDE suggests taking a Fourier transform of the phase-space variables as the derivatives are of higher order than the algebraic parameters.
Furthermore, not only does a Fourier transform reduce the PDE to first order, but the computation of expectation values also becomes trivial since we are then working with the characteristic function of the distribution.

The Fourier transform is defined as
\begin{equation}
\mathcal{F}\{f\}(\mathbf{k}) = \int_{\!-\infty}^{+\infty} \!\!\! dx \! \int_{\!-\infty}^{+\infty} \!\!\! dp \, e^{- \imath \mathbf{k} \cdot \mathbf{z}} f(\mathbf{z}) ,
\end{equation}
and it exhibits the usual properties:
\begin{equation}
\imath^n \frac{\partial^n \mathcal{F}\{f\}}{\partial k_j^n}(\mathbf{0}) = \int_{\!-\infty}^{+\infty} \!\!\! dx \! \int_{\!-\infty}^{+\infty} \!\!\! dp \, q_j^n f(\mathbf{z}) \label{eq:correlations1}.
\end{equation}
The master equation becomes then
\begin{equation}
\left( \frac{\partial}{\partial t} + \mathbf{k}^{\!\mathrm{T}} \boldsymbol{\mathcal{H}} \boldsymbol{\nabla}_{\!\!\mathbf{k}} \right) \mathcal{W}_\mathrm{r} = - \mathbf{k}^{\!\mathrm{T}} \mathbf{D} \, \mathbf{k} \, \mathcal{W}_\mathrm{r} \label{eq:phase} .
\end{equation}
where $\mathcal{W}_\mathrm{r} = \mathcal{F}\{W_{\!\mathrm{r}}\}$ and the normalization of $W_{\!\mathrm{r}} (\mathbf{z},t)$ implies $\mathcal{W}_\mathrm{r}(\mathbf{0},t) = 1$.

From Eq.~\eqref{eq:phase} it is clear that if the master equation coefficients asymptote to constant values, then we will have a stationary Gaussian solution in the late-time limit given by
\begin{equation}
\mathcal{W}_T^\infty = e^{-\frac{1}{2} \mathbf{k}^{\!\mathrm{T}}\! \boldsymbol{\sigma}_T^\infty \, \mathbf{k} } \, ,
\end{equation}
with $\boldsymbol{\sigma}_T^\infty$ uniquely determined by the Lyapunov equation
\begin{equation}
\boldsymbol{\mathcal{H}}_\infty \, \boldsymbol{\sigma}_T^\infty + \boldsymbol{\sigma}_T^\infty \, \boldsymbol{\mathcal{H}}^{\!\mathrm{T}}_\infty = 2 \, \mathbf{D}_\infty \, \label{eq:thermal_relation}.
\end{equation}
To zeroth-order in the system-environment coupling, this corresponds to the free thermal state of the system.
It is also reasonable to believe that more generally this corresponds to the thermal state of our system coupled to the environment (i.e. the reduced density matrix of the thermal state of the whole system including the system-environment interaction).
For arbitrary systems this has been proven to second order in the system-environment coupling (here first order in damping, e.g. $\gamma_0$) \cite{FRHQOS}.

\subsubsection{Method of Characteristic Curves}
\label{sec:charcurv}
The method of characteristic curves involves looking for parameterized curves in the domain $(t,\mathbf{k})$ along which the first order PDE becomes a set of first-order ODEs.
For each one of those curves we have
\begin{eqnarray}
\mathcal{W}_\mathrm{r}\!\left[\mathbf{k},t\right] & = & \mathcal{W}_\mathrm{r}\!\left[ \mathbf{k}(\tau), t(\tau) \right] ,\\
\label{eq:cc} \frac{d}{d\tau} \mathcal{W}_\mathrm{r} & = & \frac{dt}{d\tau} \frac{\partial}{\partial t} \mathcal{W}_\mathrm{r} + \frac{d\mathbf{k}}{d\tau}^{\!\mathrm{T}} \boldsymbol{\nabla}_{\!\!\mathbf{k}} \mathcal {W}_\mathrm{r} \, ,
\end{eqnarray}
Next, we attempt to match the right-hand side of Eq.~\eqref{eq:cc} to the left-hand side of Eq.~\eqref{eq:phase}.
This results in a system of ODEs in the parameter $\tau$.
We will look for curves that synchronize with the initial time so that $t(0)=0$, $\mathbf{k} (0) = \mathbf{k}_0$.
The solution for the parameterization of the time coordinate is simple:
\begin{equation}
\frac{dt}{d\tau} = 1 \; \Rightarrow \; t(\tau) = \tau \, .
\end{equation}
On the other hand, finding the parameterization for the Fourier transform of the phase-space variables is a bit more involved. It is characterized by the linear ODE system
\begin{equation}
\frac{d}{d\tau} \mathbf{k}^{\!\mathrm{T}}\!(\tau) = + \mathbf{k}^{\!\mathrm{T}}\!(\tau) \, \boldsymbol{\mathcal{H}}(\tau) \label{eq:ccfps} .
\end{equation}
and its solutions can be written as
\begin{equation}
\mathbf{k}(\tau) = \boldsymbol{\Phi}_k (\tau) \, \mathbf{k}_0 \label{eq:k(tau)} ,
\end{equation}
where $\boldsymbol{\Phi}_k (\tau)$ is the matrix propagator associated with the transpose of Eq.~\eqref{eq:ccfps} and equals the identity matrix at $\tau=0$.
For local dissipation, $\boldsymbol{\mathcal{H}}$ is time independent and the propagator is simply given by $\boldsymbol{\Phi}_k^\mathrm{T} (\tau)= e^{+\tau \, \boldsymbol{\mathcal{H}}}$, which equals $\boldsymbol{\Phi}^{\!-1}(\tau)$.
Such a relation between the matrix propagator of the integro-differential Langevin equation~\eqref{eq:langevin_bold} and the local equation~\eqref{eq:ccfps} actually holds in general.
Indeed, taking into account Eq.~\eqref{eq:simplifiedH}, it follows that the propagator for the characteristic curves $\boldsymbol{\Phi}_k^\mathrm{T}(\tau)$ must satisfy the equation
\begin{eqnarray}
\frac{d}{d\tau} \boldsymbol{\Phi}_k^\mathrm{T}(\tau) &=& -\boldsymbol{\Phi}_k^\mathrm{T}(\tau) \, \dot{\boldsymbol{\Phi}}(t) \, \boldsymbol{\Phi}^{\!-1}(t) ,
\end{eqnarray}
which is equivalent to the relation
\begin{eqnarray}
\frac{d}{d\tau} \left( \boldsymbol{\Phi}_k^\mathrm{T}(\tau) \, \boldsymbol{\Phi}(\tau) \right) &=& 0 .
\end{eqnarray}
Together with $\boldsymbol{\Phi}_k^\mathrm{T}(0) \, \boldsymbol{\Phi}(0) = \mathbf{I}$, since both $\boldsymbol{\Phi}_k (\tau)$ and $\boldsymbol{\Phi}(\tau)$ equal the identity matrix at the initial time, this implies that $\boldsymbol{\Phi}_k^\mathrm{T} (\tau) = \boldsymbol{\Phi}^{\!-1}(\tau)$.

We now have the rules for transforming back and forth between the domain coordinates $\left( t, \mathbf{k} \right)$ and the characteristic curve coordinates $\left( \tau, \mathbf{k}_0 \right)$;
$\mathbf{k}_0$ uniquely specifies each characteristic curve parameterized by $\tau$.
Using these results, we can immediately apply the method of characteristic curves to solving Eq.~\eqref{eq:phase} as follows:
\begin{align}
\frac{d}{d\tau} \mathcal{W}_\mathrm{r}\!\left[ \mathbf{k}(\tau), t(\tau) \right] =&\;  -\mathbf{k}^{\!\mathrm{T}} \mathbf{D}(t) \, \mathbf{k} \, \mathcal{W}_\mathrm{r}\!\left[ \mathbf{k}(\tau), t(\tau) \right] ,\\
\frac{d}{d\tau} \mathcal{W}_\mathrm{r}\!\left[ \boldsymbol{\Phi}_k(\tau) \mathbf{k}_0, t(\tau) \right] =&\; - \mathbf{k}_0^{\mathrm{T}} \boldsymbol{\Phi}_k^{\mathrm{T}}(\tau) \, \mathbf{D}(\tau) \, \boldsymbol{\Phi}_k(\tau) \, \mathbf{k}_0 \nonumber \\
& \times \mathcal{W}_\mathrm{r}\!\left[ \boldsymbol{\Phi}_k(\tau) \mathbf{k}_0, t(\tau) \right] .
\end{align}
The last equation is a linear ODE whose solution can be easily found to be
\begin{align}
& \mathcal{W}_\mathrm{r}\!\left[ \boldsymbol{\Phi}_k(\tau) \mathbf{k}_0, \tau \right] = \\
& \mathcal{W}_\mathrm{r}\!\left[ \mathbf{k}_0, 0 \right] e^{-\int_0^\tau \! d\tau' \left(\mathbf{k}_0^{\mathrm{T}} \boldsymbol{\Phi}_k^{\mathrm{T}}(\tau') \, \mathbf{D}(\tau') \, \boldsymbol{\Phi}_k(\tau') \mathbf{k}_0 \right)} , \nonumber
\end{align}
where $\mathcal{W}_\mathrm{r}\!\left[ \mathbf{k}_0, 0 \right]$ is the initial characteristic function at $t=0$.
We can now express the solution back in terms of $\mathbf{k}$ and $\boldsymbol{\Phi}$ to get the final result
\begin{equation}
\mathcal{W}_\mathrm{r}\!\left[ \mathbf{k}, t \right] = \mathcal{W}_\mathrm{r}\!\left[ \boldsymbol{\Phi}^{\!\mathrm{T}}\!(t) \, \mathbf{k}, 0 \right] e^{- \frac{1}{2} \mathbf{k}^{\!\mathrm{T}}\! \boldsymbol{\sigma}_T(t) \, \mathbf{k}} \label{eq:solm},
\end{equation}
with thermal covariance defined
\begin{equation}
\boldsymbol{\sigma}_T(t) \equiv 2 \int_0^t \!\! d\tau \, \boldsymbol{\Phi}(t,\tau) \, \mathbf{D}(\tau) \, \boldsymbol{\Phi}^{\!\mathrm{T}}\!(t,\tau) \label{eq:tsig-nl},
\end{equation}
and note that $\boldsymbol{\Phi}(t,\tau)$ here does not have time-translational invariance for nonlocal dissipation, where $\boldsymbol{\Phi}(t,\tau) = \boldsymbol{\Phi}(t) \, \boldsymbol{\Phi}^{\!-1}(\tau) \neq \boldsymbol{\Phi}(t\!-\!\tau)$; see the discussion in \ref{sec:nonlocal_prop}.

\subsubsection{Equivalence with the Result from the Langevin Equation}
\label{sec:equivalence}

We have shown that the form of the solution from the master equation is equivalent to that derived from the Langevin equation in Sec.~\ref{sec:MSSL}.
What remains to be shown is that the thermal covariances are indeed equivalent.
To do this one can differentiate Eq.~\eqref{eq:tsig-nl} with respect to time and get the following result:
\begin{equation}
\dot{\boldsymbol{\sigma}}_T(t) = - \boldsymbol{\mathcal{H}}(t) \, \boldsymbol{\sigma}_T(t) - \boldsymbol{\sigma}_T(t) \, \boldsymbol{\mathcal{H}}^{\!\mathrm{T}}\!(t) + 2 \, \mathbf{D}(t) \label{eq:cov_diff}.
\end{equation}
This equation is also satisfied by the thermal covariance expression directly derived from the Langevin equation, as can be seen from Eq.~\eqref{eq:Ddynamic}.
Furthermore, the thermal covariances given by Eqs.~\eqref{eq:tsig-nl} and \eqref{eq:sigma1} both have vanishing initial conditions: $\boldsymbol{\sigma}_T(0)=0$.
Therefore, since they are both solutions of the same ordinary differential equation and have the same initial conditions, they must be equivalent.

\section{Evolution of States}
\label{sec:solutions}

\subsection{General Solutions}

Whether derived via the Langevin equation in Sec.~\ref{sec:MESL} or solving the master equation in Sec.~\ref{sec:MESM}, the evolution of the system state is most easily represented in terms of the characteristic function (the Fourier transform) of the reduced Wigner distribution:
\begin{equation}
\mathcal{W}_\mathrm{r}\!\left[ \mathbf{k}, t \right] = \mathcal{W}_\mathrm{r}\!\left[ \boldsymbol{\Phi}^{\!\mathrm{T}}\!(t) \, \mathbf{k}, 0 \right] e^{- \frac{1}{2} \mathbf{k}^{\!\mathrm{T}}\! \boldsymbol{\sigma}_T(t) \, \mathbf{k}} \label{eq:solg},
\end{equation}
with the thermal covariance $\boldsymbol{\sigma}_T(t)$ given by
\begin{equation}
\boldsymbol{\sigma}_T(t) = \int_0^t \!\! d\tau \! \int_0^t \!\! d\tau' \, \boldsymbol{\Phi}(t\!-\!\tau) \, \boldsymbol{\nu}(\tau,\tau') \, \boldsymbol{\Phi}^{\!\mathrm{T}}\!(t\!-\!\tau') , \label{eq:tsig}
\end{equation}
where $\boldsymbol{\Phi}(t)$ is the phase-space propagator for the Langevin equation defined in Eq.~\eqref{eq:Phi}.

The solution in Eq.~\eqref{eq:solg} consists of two factors.
The first one tends to unity in the long time limit and encodes the disappearance of the initial state (we will call it the \emph{death factor}).
The second factor describes the appearance of a Gaussian state that evolves in time and tends asymptotically to a state that corresponds to thermal equilibrium (we will refer to this as the \emph{birth factor}).
Assuming dissipation, all initial distributions evolve towards this final Gaussian state, with thermal covariance $\boldsymbol{\sigma}_T(t)$.
This state does not look like the thermal state of a free harmonic oscillator because of the coupling to the environment.
It more likely results from considering the thermal equilibrium state for the whole system (system plus environment) including the system-environment interaction, which gives rise to a non-trivial correlation between them, and tracing out the environment.

The \emph{death factor} contains the information on the initial conditions;
it describes the gradual disappearance of the initial distribution and it is always temperature independent.
The free evolution of the Wigner function corresponds to rotation in phase space (when properly rescaled) at constant angular velocity.
Dissipation will modify this rotation to inspiralling of the trajectories down to the origin, or decay to the origin without completing a full rotation in the case of overdamping.
More generally, for nonlocal dissipation the trajectories will correspond to those of a parametrically damped oscillator, which in some cases could be quite complicated.

The \emph{birth factor} describes the complicated birth and settlement of a state of thermal equilibrium.
This factor is always Gaussian with a covariance matrix given by Eq.~\eqref{eq:tsig}, which involves a convolution of the noise kernel with propagators that reflect the natural oscillatory decay of the system.
This covariance matrix vanishes at the initial time and tends at late times to an equilibrium covariance matrix which can be easily determined from the Lyapunov equation \eqref{eq:thermal_relation}.
The thermal covariance matrix is always positive definite.


\subsubsection{Trajectories of the Cumulants}
\label{sec:wigner-sol}

\begin{figure}[h]
\centering
\includegraphics[width=0.5\textwidth]{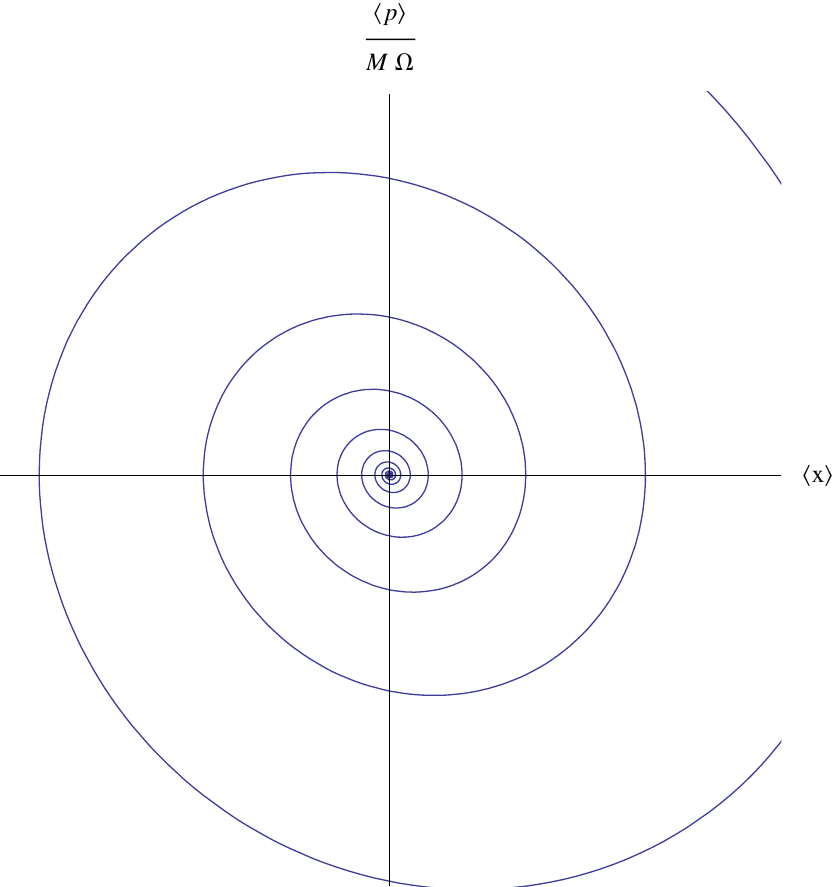}
\caption{\label{fig:spiral} The trajectory of the expectation values
$\langle x \rangle, \langle p \rangle$.}
\end{figure}

As we have already mentioned, the Fourier transfom of the reduced Wigner function corresponds to its characteristic function, from which the correlation functions for the phase-space variables can be easily derived using Eq.~\eqref{eq:correlations1}.
The general expressions for the cumulants can be obtained straightforwardly from the logarithm of the reduced Wigner function in Fourier space as follows:
\begin{equation}
\sum_{n=1}^{\infty} \frac{1}{n!} \kappa^{(n)}_{i_1 \ldots i_n}(t) \prod_{l=1}^{n} \imath k^{i_l} = \log{\mathcal{W}_\mathrm{r}(\mathbf{k},t)} ,
\end{equation}
where $k^{i_l}$ denotes the components of the vector $\mathbf{k}$ and we used the Einstein summation convention for pairs of repeated indices (\emph{i.e.}, it is implicitly understood that a sum $\sum_{i_l = 1}^{2}$ should be preformed over each pair of repeated indices $i_l$).
$\boldsymbol{\kappa}^{(n)}$ is the $n^{th}$ cumulant and acts as a tensor of order $n$ contracted with $n$ copies of $\mathbf{k}$.
Using the result for $\mathcal{W}_\mathrm{r}(t,\mathbf{k})$ from Eq.~\eqref{eq:solg} we have
\begin{align}
\sum_{n=1}^{\infty} \frac{1}{n!} \kappa^{(n)}_{i_1 \ldots i_n}(t) \prod_{l=1}^{n} \imath k^{i_l} =&\; \sum_{n=1}^{\infty} \frac{1}{n!} \kappa^{(n)}_{i_1 \ldots i_n}(0) \prod_{l=1}^{n} \imath \left( \boldsymbol{\Phi}^{\!\mathrm{T}}\!(t) \, \mathbf{k}\right)^{i_l} \nonumber \\
& - \frac{1}{2} \mathbf{k}^{\!\mathrm{T}}\! \boldsymbol{\sigma}_T(t) \, \mathbf{k} \label{eq:cumulants2} ,
\end{align}
where $\kappa^{(n)}_{j_1 \ldots j_n} (0)$ are the cumulants associated with the initial distribution.
Eq.~\eqref{eq:cumulants2} implies
\begin{equation}
\kappa^{(n)}_{i_1 \ldots i_n}(t) = \kappa^{(n)}_{j_1 \ldots j_n} (0) \prod_{l=1}^{n} \imath \left( \boldsymbol{\Phi}^{\!\mathrm{T}}\!(t) \right)^{j_l i_l} + \delta_{n2} \, \sigma_T^{i_1 i_2} (t) .
\end{equation}
We can see that the only cumulant with a non-vanishing asymptotic value, which is a consequence of the thermal fluctuations, is the covariance matrix (with $n=2$).
The closely related second momenta of the distribution are given by
\begin{equation}
\langle \mathbf{z} \mathbf{z}^{\!\mathrm{T}} \rangle(t) = \boldsymbol{\Phi}(t) \, \langle \mathbf{z} \mathbf{z}^{\!\mathrm{T}} \rangle_\mathbf{q_0} \, \boldsymbol{\Phi}^{\!\mathrm{T}}\!(t) + \boldsymbol{\sigma}_T(t) \label{eq:covariance2} ,
\end{equation}
where $\langle \cdots \rangle_\mathbf{q_0}$ denotes the expectation value with respect to the reduced Wigner function at the initial time,
\footnote{Note that the expectation value of any phase-space function with respect to the  reduced Wigner function is equivalent to a quantum expectation value with respect to the corresponding reduced density matrix where the arguments $x$ and $p$ of the phase-space function are promoted to operators and the Weyl ordering prescription is employed.
In particular, for the second-order cumulants this corresponds to considering symmetrized two-point quantum correlation functions.}
as defined in Eq.~\eqref{eq:initial_average}.
All other cumulants experience whatever oscillatory decay is inherent in the homogeneous solution of the Langevin equation.
In particular, the expectation value
\begin{equation}
\langle \mathbf{z} \rangle(t) = \boldsymbol{\Phi}(t) \, \langle \mathbf{z} \rangle_\mathbf{q_0} \, ,
\end{equation}
follows a trajectory like that plotted in Fig.~\ref{fig:spiral} for local dissipation, where one can see that the trajectory of the expectation values $\langle x \rangle, \langle p \rangle$ for any initial distribution inspiral into the origin.
This captures the behavior of Gaussians plotted by Unruh and Zurek \cite{UnruhZurek89}.

\subsubsection{Thermal Covariance}
\label{sec:thermal_contour}

As we have seen, the only additional quantity that needs to be calculated besides the propagator is the thermal covariance.
Here we discuss the full-time evolution of the thermal covariance, which can be most easily obtained from Eq.~\eqref{eq:tsig}.
Using the addition formula for the argument of the cosine function appearing in the definition of the noise kernel, one obtains the following expressions for the components of the thermal covariance, which only involve calculating a single time integral besides the integral over frequencies:
\begin{align}
\sigma_T^{xx}(t) =&\; \int_0^\infty \!\!\!d\omega \, I(\omega) \coth\!\left( \frac{\omega}{2 T} \right) \left[ G(t) * \cos{(\omega t)} \right]^2 \\
& +  \int_0^\infty \!\!\!d\omega \, I(\omega) \coth\!\left( \frac{\omega}{2 T} \right) \left[ G(t) * \sin{(\omega t)} \right]^2  , \nonumber \\
\sigma_T^{xp}(t) =&\; \frac{1}{2} M \dot{\sigma}_T^{xx}(t)  , \\
\sigma_T^{pp}(t) =&\; \int_0^\infty \!\!\!d\omega \, I(\omega) \coth\!\left( \frac{\omega}{2 T} \right) \left[ M \dot{G}(t) * \cos{(\omega t)} \right]^2 \nonumber \\
& + \int_0^\infty \!\!\!d\omega \, I(\omega) \coth\!\left( \frac{\omega}{2 T} \right) \left[ M \dot{G}(t) * \sin{(\omega t)} \right]^2 .
\end{align}
These results are expressed in terms of Laplace convolutions of the propagator with sinusoidal functions which become trivial in Laplace domain, although one must eventually transform back to compute the squares.
Moreover, integrating by parts in the Laplace convolutions and taking into account that $G(0)=0$, the momentum covariance can be expressed in the alternative form
\begin{align}
\sigma_T^{pp}(t) =&\; \int_0^\infty \!\!\!d\omega \, \omega^2 I(\omega) \coth\!\left( \frac{\omega}{2 T} \right) \left[ M G(t) * \cos{(\omega t)} \right]^2 \nonumber \\
&+ \int_0^\infty \!\!\!d\omega \, \omega^2 I(\omega) \coth\!\left( \frac{\omega}{2 T} \right) \left[ M G(t) * \sin{(\omega t)} \right]^2 \nonumber \\
& + M^2 \, \nu(0) \, G(t)^2 ,
\end{align}
which is completely analogous to that for the position covariance, but
with an effectively higher-order spectral density due to the additional factor of $\omega^2$, plus a simple cut-off sensitive transient term which decays with the characteristic relaxation rate.
It becomes then obvious that the momentum covariance will contain the dominant contribution to any potential ultraviolet sensitivity of the thermal covariance, whereas the position covariance will contain the dominant contribution to any possible infrared sensitivity.

In order to compute the evolution of the thermal covariance, especially when calculating it numerically, it is often convenient to use the following alternative expressions, which can be derived by differentiating with respect to time the $xx$ and $pp$ components of Eq.~\eqref{eq:tsig}:
\begin{eqnarray}
\dot{\sigma}_T^{xx}(t) &=& 2\, G(t) \left[ \nu(t)*G(t) \right]  , \label{eq:sigmadot1a}\\
\dot{\sigma}_T^{pp}(t) &=& 2 M^2 \dot{G}(t) \frac{d}{dt}\!\left[ \nu(t)*G(t) \right] , \\
\sigma_T^{xp}(t) &=& \frac{M}{2} \dot{\sigma}_T^{xx}(t)
= M\, G(t) \left[ \nu(t)*G(t) \right]  \label{eq:sigmadot1c},
\end{eqnarray}
where the convolution of the propagator with the noise kernel should be performed before the frequency integral of the noise kernel.
This will typically result in expressions more amenable to numerics since one can avoid increasingly oscillatory integrands.


For odd meromorphic spectral functions the frequency integral can be evaluated by contour integration (and the residue theorem) using the rational expansion of the hyperbolic cotangent
\begin{eqnarray}
\coth\!\left( \frac{\omega}{2T} \right) &=& \frac{2 T}{\omega} + \frac{2}{\pi} \sum_{k=1}^\infty \frac{ \frac{\omega}{2 \pi T} }{k^2 + \left( \frac{\omega}{2 \pi T} \right)^2 } . \label{eq:coth}
\end{eqnarray}
One should then be left with a sum of terms rational in the Laplace domain, which can be contracted into digamma or harmonic-number functions [respectively $\psi(z)$ or $\mathrm{H}(z)$], which are asymptotically logarithmic.
When transforming back to the time domain, the residues of the hyperbolic cotangent additionally give rise to products of rational functions of $k$ with $e^{-2\pi T t k}$.
These terms contain all effects which decay at temperature-dependent rates and can be expressed in terms of Lerch transcendent functions, $\varPhi\!\left( z, 1, e^{-2\pi T t} \right)$, which are useful for numerical calculations but not particularly insightful.

Fortunately, one can also derive a simple analytic expression for the late-time thermal covariance, as shown in \ref{sec:late_cov_der}:
\begin{align}
\boldsymbol{\sigma}_T(\infty) &= \int_0^\infty \!\!\! d\omega \, I(\omega) \coth\!\left( \frac{\omega}{2T} \right) |\hat{G}(\imath \omega)|^2 \left[ \begin{array}{cc} 1 & 0 \\ 0 & M^2\omega^2 \end{array} \right] , \label{eq:tclflt}
\end{align}
which reduces the calculation of late-time uncertainties to a single integral.
This relation confirms that for late times the momentum covariance has precisely $\omega^2$ more frequency sensitivity in its integrand.

\subsubsection{Linear Entropy}
\label{sec:entropy}

In this subsection we investigate the linear entropy~\cite{Wlodarz03},
which can be easily obtained from the Wigner distribution as follows:
\begin{equation}
S_\mathrm{L} = 1 - \mbox{Tr}(\rho^2_\mathrm{r}) = 1 - 2\pi \int \!\! d^2\mathbf{z} \, W_{\!\mathrm{r}}^2 (\mathbf{z},t) .
\end{equation}
In Fourier space it becomes
\begin{equation}
S_\mathrm{L} = 1 - \frac{1}{2\pi} \int \!\! d^2\mathbf{k} \, |\mathcal{W}_\mathrm{r}(\mathbf{k},t)|^2 ,
\end{equation}
and using the result in Eq.~\eqref{eq:solg} we finally get
\begin{equation}
S_\mathrm{L} = 1 - \frac{1}{2\pi} \int \!\! d^2\mathbf{k} \, \left|\mathcal{W}_\mathrm{r}\!\left( 0, \boldsymbol{\Phi}^{\!\mathrm{T}}\!(t) \, \mathbf{k} \right) \right|^2 e^{- \mathbf{k}^{\!\mathrm{T}}\! \boldsymbol{\sigma}_T(t) \, \mathbf{k}} \label{eq:lin-ent1} .
\end{equation}
At the initial time the linear entropy is that of the initial state, and at late times it tends to
\begin{equation}
S_\mathrm{L} = 1 - \frac{1}{2 \sqrt{\det\boldsymbol{\sigma}_T^\infty}} .
\end{equation}

Alternatively, one can express the linear entropy in terms of an integral of the Fourier-transformed reduced Wigner function at the initial time by introducing the change of variables $\mathbf{k}_0 = \boldsymbol{\Phi}^{\!\mathrm{T}}\!(t) \, \mathbf{k}$.
Eq.~\eqref{eq:lin-ent1} can then be written as
\begin{align}
S_\mathrm{L} =&\; 1 - \frac{1}{2\pi} \int \!\! d^2 \mathbf{k}_0 \, \frac{1}{\det\!{\left[ \boldsymbol{\Phi}(t) \right]}} \, |\mathcal{W}_\mathrm{r}\!\left( 0, \mathbf{k}_0 \right)|^2 \nonumber \\
& \times e^{- \mathbf{k}_0^{\mathrm{T}} \boldsymbol{\Phi}^{\!-1}(t) \, \boldsymbol{\sigma}_T(t) \, \boldsymbol{\Phi}^{-\mathrm{T}}(t) \, \mathbf{k}_0} \nonumber \\
=&\; 1 - \frac{1}{2 \sqrt{\det[\boldsymbol{\sigma}_T(t)]}} \int \!\! d^2\mathbf{k}_0 \, |\mathcal{W}_\mathrm{r}\!\left( 0, \mathbf{k}_0 \right)|^2 \nonumber \\
& \times N\!\!\left( \mathbf{0}, \frac{1}{2} \boldsymbol{\Phi}^{\!\mathrm{T}}\!(t) \, \boldsymbol{\sigma}_T^{-1}\!(t) \, \boldsymbol{\Phi}(t); \mathbf{k}_0 \right) ,
\end{align}
where $N(\boldsymbol{\mu},\boldsymbol{\sigma}; \mathbf{k}_0)$ is a normalized Gaussian distribution for the variable $\mathbf{k}_0$ with mean $\boldsymbol{\mu}$ and covariance $\boldsymbol{\sigma}$.
For small times this integral is similar to that for the initial state, whereas for long times the normalized Gaussian distribution becomes increasingly close to a delta function.

For a Gaussian initial state
\begin{equation}
\mathcal{W}_\mathrm{r} (0, \mathbf{k}_0) = \exp\left(- \mathbf{k}_0^{\mathrm{T}} \boldsymbol{\sigma}_0 \, \mathbf{k}_0 - \imath \, \mathbf{k}_0^\mathrm{T} \langle \mathbf{z} \rangle_0\right) ,
\end{equation}
the integral in Eq.~\eqref{eq:lin-ent1} can be explicitly computed:
\begin{align}
S_\mathrm{L} &= 1 - \frac{1}{2\pi} \int \!\! d^2\mathbf{k} \, e^{- \mathbf{k}^{\!\mathrm{T}}\! \left( \boldsymbol{\Phi}(t) \, \boldsymbol{\sigma}_0 \, \boldsymbol{\Phi}^{\!\mathrm{T}}\!(t) + \boldsymbol{\sigma}_T(t) \right) \mathbf{k}} \nonumber \\
&= 1 - \frac{1}{2 \sqrt{\det\!\left[ \boldsymbol{\Phi}(t) \, \boldsymbol{\sigma}_0 \, \boldsymbol{\Phi}^{\!\mathrm{T}}\!(t) + \boldsymbol{\sigma}_T(t) \right]}} .
\end{align}
For these Gaussian states, reasonable linear entropy is synonymous with reasonable uncertainty functions (\emph{i.e.}, the linear entropy will be positive if and only if the Heisenberg uncertainty principle is satisfied).
We will find that the late time uncertainty is well behaved.
The uncertainty at the initial and intermediate times should not violate the Heisenberg uncertainty principle either.

\subsubsection{Decoherence of a Quantum Superposition}
\label{sec:decoherence}

In this section we will illustrate how one can get a useful qualitative picture of the phenomenon of environment-induced decoherence from the the solutions of the master equation given by Eqs.~\eqref{eq:solg}-\eqref{eq:tsig}.
In order to do that we will consider a quantum superposition, $|\psi\rangle = \big( |\psi_+\rangle + |\psi_-\rangle \big)/\sqrt{K}$, of a pair of states $|\psi_\pm\rangle$ which correspond to a pair of Gaussian wavefunctions in position space separated by a distance $2 \delta_x$ and where $K$ is an appropriate normalization constant.
Specifically, we have
\begin{eqnarray}
\psi_\pm(x) &=& \psi_0(x \mp \delta_x) \label{eq:gaussian} , \\
\psi_0(x) &=& \sqrt{N(0 , \sigma_0^{xx} ; x)} .
\end{eqnarray}
where $N({\mu},{\sigma}^2; {x})$ is a normalized Gaussian distribution for the variable $x$ with mean ${\mu}$ and variance ${\sigma^2}$,
and $\psi_0(x)$ is a reference Gaussian state centered at the origin.

\begin{figure}[h]
\centering
\includegraphics[width=0.5\textwidth]{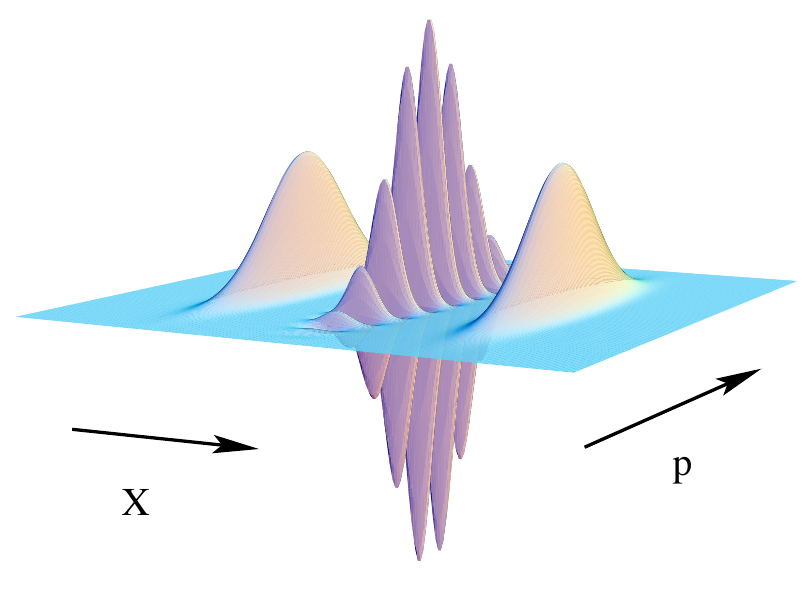}
\caption{\label{fig:wigner} Wigner function associated with a state $|\psi\rangle = \big( |\psi_1\rangle + |\psi_2\rangle \big)/\sqrt{K}$ which corresponds to the coherent quantum superposition of two Gaussian wavefunctions in position space shifted by a distance $\delta_x$.}
\end{figure}

Taking into account the definition of the Wigner function,
\begin{equation}
W\!(x,p)=\frac{1}{2\pi}\int_{-\infty }^{+\infty} \!\!\! dy \, e^{ipy}\rho (x\!-\!y/2,x\!+\!y/2) \label{eq:wigner_def2},
\end{equation}
and applying it to the density matrix $\rho(x,x') = \langle x | \psi \rangle \langle \psi | x' \rangle$ we get
\begin{align}
W\!(\mathbf{z}) &= \frac{1}{K} \Big[ W_{\!+}(\mathbf{z}) + W_{\!-}(\mathbf{z}) + 2 \cos(2 \delta_x p) \, W_0(\mathbf{z}) \Big] \label{eq:wigner1} ,
\end{align}
where $W_{\!+}$, $W_{\!-}$ and $W_0$ are respectively the Wigner functions of the states $|\psi_+\rangle$, $|\psi_-\rangle$ and $|\psi_0\rangle$.
This Wigner function, plotted in Fig.~\ref{fig:wigner}, exhibits oscillations of size $1/\delta_x$ along the $p$ direction.
These oscillations are closely connected to the coherence of the quantum superposition (and the existence of non-diagonal terms in the density matrix) and are absent in the Wigner function for the incoherent mixture $W\!(\mathbf{z}) = (1/2) [ W_{\!+}(\mathbf{z}) + W_{\!-}(\mathbf{z}) ]$.

In this context the decoherence effect due to the interaction with the environment corresponds to the washing-out of the oscillations in the reduced Wigner function as it evolves according to the master equation.
This can be seen rather simply from the result for the solutions of the master equation obtained in this section and given by Eqs.~\eqref{eq:solg}-\eqref{eq:tsig}.
Taking into account that the inverse Fourier transform of Eq.~\eqref{eq:solg} corresponds to a convolution of the homogeneously evolving initial state and a Gaussian function with the thermal covariance $\boldsymbol{\sigma}_T(t)$ as its covariance matrix, the Wigner function can then be expressed as
\begin{equation}
W_{\!\mathrm{r}}(t,\mathbf{z}) = \int d\mathbf{z}' \frac{ N\!(\mathbf{0}, \boldsymbol{\sigma}_T(t); \mathbf{z}\!-\!\mathbf{z}') }{\det\!{\left[ \boldsymbol{\Phi}(t) \right]}}
W_{\!\mathrm{r}}\!\left(0, \boldsymbol{\Phi}^{\!-1}(t)\, \mathbf{z}'\right)
\label{eq:smearing1} ,
\end{equation}
where the thermal Gaussian acts as a Gaussian smearing function which starts as a delta function at the initial time and broadens with the passage of time until it eventually reaches its asymptotic thermal-equilibrium value.
Therefore, several aspects will be at play. On the one hand, the initial state evolves as a phase-space distribution with trajectories corresponding to the homogeneous solutions of the Langevin equation~\eqref{eq:langevin_bold} and with the same qualitative behavior depicted in Fig.~\ref{fig:spiral} for the trajectories of $\langle x \rangle$ and $ \langle p \rangle$.
On the other hand, by diagonalizing $\boldsymbol{\sigma}_T(t)$ at each instant of time one gets the principal directions and the widths $(\sigma_1,\sigma_2)$ of the Gaussian smearing function, which will average out any details of those sizes along the corresponding directions.
When $\boldsymbol{\sigma}_T(t)$ along the direction of the interference oscillations of the Wigner function becomes comparable to their wavelength, they get washed out and the Wigner function becomes equivalent to that of the completely incoherent mixture. The time it takes for this to happen is known as the decoherence time $t_\mathrm{dec}$.

Knowledge of the qualitative behavior of $\boldsymbol{\sigma}_T(t)$, combined with the fact that the phase-space distribution $\det\!{\left[ \boldsymbol{\Phi}(t) \right]}^{-1} \, W_{\!\mathrm{r}}\!\left(0, \boldsymbol{\Phi}^{\!-1}(t)\, \mathbf{z}'\right)$ is rotating with the characteristic oscillation frequency and shrinking with the characteristic relaxation time is all that one needs to understand how different initial states decohere as time goes by.
In particular, if the decoherence time-scale, given by $t_\mathrm{dec}$, is much shorter than the characteristic oscillation period and the relaxation time (but sufficiently longer than $1/\Lambda$), one can approximate the phase-space distribution by the initial reduced Wigner function (after any possible initial kick).
For instance, for an Ohmic environment in the high-temperature regime one can, under those circumstances, approximately take $\sigma_T^{pp} (t) \sim D^\infty_{\!pp} \, t$ with $D^\infty_{\!pp} \sim 2 M \gamma_0 T$ and from the condition $\sqrt{\sigma_{\!pp} (t)} \sim 1/\delta_x$ obtain an estimated decoherence time $t_\mathrm{dec} \sim 1/(2 M \gamma_0 T \delta_x^2)$, in agreement with the standard result for this situation \cite{Zurek91,Paz93}.
On the other hand, if $M$, $\gamma_0$ or $\delta_x$ are very small $t_\mathrm{dec}$ can become comparable or larger than the dynamical timescales $1/\Omega$ or $1/\gamma$, and the previous estimate can no longer be applied because one needs to take into account the evolution of $\boldsymbol{\sigma}_T(t)$, which is then less simple (it will roughly oscillate with frequency $\Omega$ around a central value which increases with a characteristic timescale $1/\gamma$ until it approaches the asymptotic thermal value), as well as the rotation and shrinking of the initial Wigner function under the homogeneous evolution.
Note also that if we had considered an initial superposition of Gaussian states peaked at the same location but with different momenta, which corresponds to a Wigner function along the position rather than momentum direction, the decoherence time would typically be much longer, since $\sigma_T^{xx} (t)$ vanishes at the initial time and grows with a characteristic timescale of order $1/\Omega$.
In that case, the rotation of the Wigner function becomes important since the oscillations can then be averaged out due to the larger values of $\sigma_T^{pp}(t)$.

The zero-temperature regime for an Ohmic environment is also qualitatively different.
There is a substantial contribution to $\sigma_T^{pp}(t)$ from a jolt of the diffusion coefficient $D_{\!pp}$ for times of order $1/\Lambda$.
However, this is actually regarded as an unphysical consequence of having considered a completely uncorrelated initial state for the system plus environment, and this kind of highly cut-off sensitive features at early times of order $1/\Lambda$ should disappear if one considers a finite (cut-off independent) preparation time for the initial state of the system coupled to the environment \cite{Anglin97}.
For further discussion on this point as well as a possible way of avoiding these spurious effects and generating a properly correlated initial state by using a finite switch-on time for the system-environment interaction see \ref{sec:initial_coupling}.
For sufficiently weak coupling, $M$, or $\delta_x$, $t_\mathrm{dec}$ can become comparable or larger than the relaxation time more easily than at high temperatures since the components $\boldsymbol{\sigma}_T(t)$ are much smaller in this case.
For example, the asymptotic thermal value of $\sigma_{\!pp}$ is of order $M \Omega$ (for weak coupling), much smaller than the high-temperature results, which is of order $M T$.
In such situations, the main effect of considering a sufficiently long time is through the shrinking of $\det\!{\left[ \boldsymbol{\Phi}(t) \right]}^{-1} \, W_{\!\mathrm{r}}\!\left(0, \boldsymbol{\Phi}^{\!-1}(t)\, \mathbf{z}'\right)$ and the size of its oscillations.

We have focused in this subsection on describing the qualitative features of the environment-induced decoherence of an initial coherent superposition that can be easily inferred from our general result for the evolution of the reduced Wigner function. A much more quantitative study is possible by using the exact analytical results for the diffusion coefficients and, especially, $\boldsymbol{\sigma}_T(t)$, which will be presented in Secs.~\ref{sec:ohmic} and~\ref{sec:non-ohmic}. We expect agreement with the numerical results obtained in  Ref.~\cite{Paz93}, although significant deviations may appear when the nonlocal effects of dissipation are important (such as in the sub-ohmic case) since previously obtained master equations are not valid in those regimes.

\subsection{Late-Time Dynamics}
\label{sec:late_approx}

We now focus our attention on the dynamics generated by the stationary limit of the master equation, assuming that one exists.
For an Ohmic spectrum with a large cut-off the pseudo-Hamiltonian $\boldsymbol{\mathcal{H}}$ will reach its asymptotic value within the cut-off timescale, whereas the diffusion $\mathbf{D}$ within the typical system timescales (although certain terms contributing to the diffusion coefficients will decay at a temperature-dependent rate whenever this is faster); see Sec.~\ref{sec:ohmic} for a detailed analysis of all these questions.
In the weak-coupling regime this leaves the majority of the system evolution within this late-time regime wherein the master equation is effectively stationary.
However, the existence of such a regime is not guaranteed in general.
For instance, in the sub-ohmic case the evolution can be persistently nonlocal and the effectively local late-time regime discussed here need not exist, as will be shown in Sec.~\ref{sec:sub-ohmic}.

\subsubsection{Late-Time Propagator}
\label{sec:latePhi}

If the late-time stationary limit of the master equation exists, the late-time pseudo-Hamiltonian operator will take the form
\begin{eqnarray}
\boldsymbol{\mathcal{H}} &=& \left[ \begin{array}{cc} 0 & -\frac{1}{M} \\ M \Omega^2_\mathrm{R} & 2\Gamma \end{array} \right] ,
\end{eqnarray}
and can be effectively represented as arising from the propagator
\begin{eqnarray}
\hat{G}_\mathrm{R}(s) &=& \frac{\frac{1}{M}}{s^2 + 2 \Gamma s + \Omega_\mathrm{R}^2} , \\
G_\mathrm{R}(t) &=& \frac{1}{M \tilde{\Omega}_\mathrm{R}} \sin\!\left( \tilde{\Omega}_\mathrm{R} t \right) e^{- \Gamma t} , \label{eq:green_local}
\end{eqnarray}
with $\tilde{\Omega}_\mathrm{R} = \sqrt{ \Omega^2_\mathrm{R} - \Gamma^2 }$.
This effective propagator $G_\mathrm{R}(t)$ is not equivalent to the late time limit of the true propagator $G(t)$, but they should share the same asymptotic dynamics.
Specifically if one can take the asymptotic expansion
\begin{eqnarray}
G(t) &=& G_{\!\infty}(t) + \delta G(t) ,
\end{eqnarray}
where $G_{\!\infty}(t)$ contains the asymptotic limiting behavior and $\delta G(t)$ contains the early time corrections, which decay faster at late times, then $G_{\!\infty}(t)$ should directly yield $\tilde{\Omega}_\mathrm{R}$ and $\Gamma$ in its arguments, although a phase and amplitude difference between $G_{\!\infty}(t)$ and $G_\mathrm{R}(t)$ may exist.
This can be rigorously justified if $\hat{\gamma}(s)$ and, thus, $\hat{G}(s)$ are rational, which implies that the time dependence of $G(t)$ corresponds to damped oscillations with various timescales.
On the other hand, the sub-ohmic spectral distribution that will be studied in Sec.~\ref{sec:sub-ohmic} provides a pertinent counter-example [in that case $G(t)$ decays as a negative power-law rather than exponentially] which shows that this situations does not necessarily exist when the spectral density function is not meromorphic.

If we indeed have a rational spectral density, then from the nonlocal propagator one only needs to solve the characteristic equation
\begin{eqnarray}
f^2 + 2 \hat{\gamma}(f) \, f + \Omega^2 = 0 , \label{eq:char_rate}
\end{eqnarray}
to obtain all the rates $f$ associated with the propagator (this is the same equation whose roots need to be found when decomposing the propagator in Laplace domain into simple fractions). From Eq.~\eqref{eq:char_rate} and the positivity of the damping kernel, it follows that the real part of $f$ will always be negative definite. Those with the smallest real part in absolute value give the late-time coefficients: the real part corresponds to $-\Gamma$ and the imaginary one to $\Omega_\mathrm{R}$. A specific example can be found in Sec.~\ref{sec:ohmic}, where the Ohmic case with a finite cut-off is studied in detail.
On the other hand, if one treats the system-environment interaction perturbatively, one can show that the late-time weak-coupling coefficients take the following form:
\begin{eqnarray}
f_\pm &=& -\Gamma \pm \imath \Omega_\mathrm{R} , \\
\Gamma &=& \mathrm{Re}[ \hat{\gamma}(\imath \Omega) ] + \mathcal{O}(\gamma^2), \\
\Omega_\mathrm{R} &=& \Omega - \mathrm{Im}[ \hat{\gamma}(\imath \Omega) ] + \mathcal{O}(\gamma^2) ,
\end{eqnarray}
which is in agreement with the results for the weak-coupling master equation obtained in Ref.~\cite{FRHQOS}.
Any additional timescales would then be perturbations of the cut-off or other timescales intrinsic to the spectral function.

It should be noted that in general the late-time propagator discussed here cannot be employed to calculate the diffusion coefficients or the thermal covariance, not even at late times. This is because both quantities evaluated at an arbitrary time $t$ get non-negligible contributions involving the propagator at early times, as can be seen for instance from Eqs.~\eqref{eq:diff_simple} and \eqref{eq:tsig}. Nevertheless, one can still employ the late-time propagator to obtain the late-time evolution of the thermal covariance (and the diffusion coefficients) provided that one already has an accurate result for its constant asymptotic value [obtained for example with Eq.~\eqref{eq:tclflt}], as will be illustrated next. In addition, one can also use the propagator $G_\mathrm{R}(t)$ given by Eq.~\eqref{eq:green_local}, which corresponds to the limit of local dissipation, to calculate the thermal covariance and diffusion coefficients for an Ohmic environment with a sufficiently large cut-off, since in that case the contribution from the extra early-time term of the propagator can be neglected when calculating these quantities for times later than $\Lambda^{\!-1}$, as will be shown in Sec.~\ref{sec:ohmic}.

\subsubsection{Late-Time Diffusion and Covariance}
\label{sec:late_cov}

Given late-time master equation coefficients which have all taken their asymptotic values, one can show that the evolution of the covariance in that regime is given by
\begin{align}
\boldsymbol{\sigma}(t) &= \boldsymbol{\sigma}_T^\infty + \boldsymbol{\Phi}(t\!-\!t_\mathrm{i}) \left[ \boldsymbol{\sigma}(t_\mathrm{i}) - \boldsymbol{\sigma}_T^\infty \right] \boldsymbol{\Phi}^{\!\mathrm{T}}\!(t\!-\!t_\mathrm{i})
\label{eq:late_covariance},
\end{align}
which is a solution of Eq.~\eqref{eq:cov_diff} as long as one assumes $\boldsymbol{\mathcal{H}}(t)$ and $\mathbf{D}(t)$ to be time-independent after some time $t_\mathrm{i}$ in the late-time regime.
Note that we have assumed that the master equation coefficients reached their asymptotic values much faster than the relaxation time (as illustrated in \ref{sec:mod-time_app} with the example of the ohmic distribution, this may be the case for finite temperature, but not necessarily so for zero temperature).

The asymptotic value of the late-time thermal covariance $\boldsymbol{\sigma}_T^\infty$ has been reduced to a single integral in \ref{sec:late_cov_der}.
From this single integral formulation, it is actually easier to obtain first $\boldsymbol{\sigma}_T^\infty$, and then obtain the late-time diffusion coefficients using the Lyapunov equation~\eqref{eq:thermal_relation}.
However, it is interesting to note the inverse relation
\begin{eqnarray}
\boldsymbol{\sigma}_T^\infty &=& \left[ \begin{array}{cc} \frac{1}{M \Omega_\mathrm{R}^2} \left( \frac{1}{2 \Gamma}D_{\!pp}^\infty - D_{\!xp}^\infty \right) & 0 \\ 0 & \frac{M}{2 \Gamma} D_{\!pp}^\infty \end{array} \right] \label{eq:ltsig} ,
\end{eqnarray}
for the following reason.
As we have pointed out in Sec.~\ref{sec:thermal_contour}, only the momentum covariance can contain the highest frequency sensitivities.
From the Lyapunov solution we can see that the regular diffusion coefficient would also contain such high-frequency sensitivities as it alone determines the late-time momentum covariance.
Therefore, the anomalous diffusion coefficients must act as an ``anti-diffusion'' coefficient in keeping the position covariance free of such sensitivities.
On the other hand, only the position covariance can contain the lowest frequency sensitivities and these must, therefore, be entirely contained in the anomalous diffusion coefficient if they exist.

In summary, any specific features of the initial distribution decay away and at late times the state tends generically to a Gaussian with a covariance matrix given by Eq.~\eqref{eq:tclflt}. As follows from Eq.~\eqref{eq:covariance2}, the late-time position and momentum uncertainties are, therefore, entirely given by the asymptotic values of the thermal covariance:
\begin{eqnarray}
(\Delta x)^2 &=& (\boldsymbol{\sigma}^\infty_T)_{xx} \label{deltaX}, \\
(\Delta p)^2 &=& (\boldsymbol{\sigma}^\infty_T)_{pp} \label{deltaP}.
\end{eqnarray}

\section{Ohmic Case with Finite Cut-off}
\label{sec:ohmic}

\subsection{The Nonlocal Propagator}
\label{sec:nonlocalG}

The arguably simplest example of ohmic dissipation with finite cut-off that one can construct corresponds to the following damping kernel:
\begin{equation}
\hat{\gamma}(s) = \frac{\gamma_0}{1+\frac{s}{\Lambda}} .
\end{equation}
This damping kernel is constant at frequencies much smaller than the cut-off,
but vanishes in the high frequency limit.
The corresponding spectral density also exhibits a rational cut-off function, which decays quadratically for large frequencies:
\begin{eqnarray}
I(\omega) &=& \frac{2}{\pi} M \gamma_0 \, \omega \left[ 1 + \left( \frac{\omega}{\Lambda} \right)^2 \right]^{-1}  \label{eq:spectral} .
\end{eqnarray}
Calculating the Green function amounts to factoring a cubic polynomial.
Specifically, one needs to factor $(s^2 + \Omega^2)(s + \Lambda) + 2 \gamma_0 \Lambda s$ in the denominator of the Green function $\hat{G}(s)$.
For the underdamped system the effect of a large finite cut-off is to shift the system relaxation and oscillation timescales slightly:
\begin{eqnarray}
\gamma_\star &=& \gamma_0 \left[ 1 + 2\frac{\gamma_0}{\Lambda} + \mathcal{O}\!\left( \frac{1}{\Lambda^2} \right) \right] , \\
\Omega_\star^2 &=& \frac{\Lambda}{\Lambda-2\gamma_\star} \Omega^2 .
\end{eqnarray}
and to add an additional relaxation timescale comparable to the cut-off:
\begin{eqnarray}
\Lambda_\star &=& \Lambda - 2\gamma_\star .
\end{eqnarray}
If we parametrize everything in terms of these phenomenological frequencies, the Green function for the fully nonlocal damping kernel can always be expressed as
\begin{eqnarray}
\hat{G}(s) &=& \frac{1}{M} \frac{ s + \Lambda }{\left( s + \Lambda_\star \right) \left( s^2 + 2 \gamma_\star s + \Omega_\star^2 \right)} , \label{eq:G-nonlocal}
\end{eqnarray}
without the need to explicitly factor a cubic polynomial, while the original parameters are given by
\begin{eqnarray}
\gamma_0 &=& \frac{\Lambda_\star^2 + 2\gamma_\star\Lambda_\star + \Omega_\star^2}{(\Lambda_\star + 2\gamma_\star)^2} \gamma_\star \label{eq:rate1a}, \\
\Omega^2 &=& \frac{\Lambda_\star}{\Lambda_\star + 2\gamma_\star} \Omega_\star^2 , \\
\Lambda &=& \Lambda_\star + 2 \gamma_\star \label{eq:rate1c}.
\end{eqnarray}
then we never have to actually factor the cubic polynomial.

After using partial fraction decomposition in Eq.~\eqref{eq:G-nonlocal}, one can easily transform back to the time domain and obtain the exact propagator for the nonlocal case:
\begin{align}
G(t) &= \frac{\Lambda_\star^2 + \Omega_\star^2}{(\Lambda_\star \!-\! \gamma_\star)^2 \!+\! \tilde{\Omega}_\star^2} \left[ G_\mathrm{R}(t) \!-\! \frac{2 \gamma_\star}{\Lambda_\star^2 \!+\! \Omega_\star^2} \left( \!\dot{G}_\mathrm{R}(t) \!-\! \frac{e^{-\Lambda_\star t}}{M} \right) \right] , \label{eq:G(t)-nonlocal}
\end{align}
where $G_\mathrm{R}(t)$ is the late-time local propagator introduced in Eq.~\eqref{eq:green_local}. Note that as long as $\Lambda_\star > \gamma_\star$
the term proportional to $e^{-\Lambda_\star t}$ can be neglected at sufficiently late times, when the terms involving $G_\mathrm{R}(t)$ dominate. This corresponds to the late-time regime discussed in Sec.~\ref{sec:latePhi} [the term proportional to $\dot{G}_\mathrm{R}(t)$ simply causes a phase shift] and the late-time master equation coefficients are, therefore,
\begin{eqnarray}
\Gamma = \gamma_\star , & & \Omega_\mathrm{R} = \Omega_\star .\label{eq:lateMECR}
\end{eqnarray}
In the high cut-off limit one recovers the usual coefficients $\gamma_0$ and $\Omega$. Furthermore, in that limit one can approximate $G(t)$ by $G_\mathrm{R}(t)$ since the extra terms are suppressed by inverse powers of $\Lambda^2$.
For $G(t)$ this is true even at arbitrarily early times of order $\Lambda^{\!-1}$: although the exponential factor is not suppressed, the prefactor $1/\Lambda_\star^2$ is sufficient to suppress its contribution to $G(t)$.
This is not true, however, for $\ddot{G}(t)$ (or higher-order derivatives), which also appears in $\boldsymbol{\Phi}(t)$. From Eqs.~\eqref{eq:sigma1} and \eqref{eq:Phi(t)} we can see that the component involving $\ddot{G}(t)$ does not contribute to the thermal covariance, but whether it contributes to its time derivative $\dot{\boldsymbol{\sigma}}_T(t)$ as well as to the diffusion coefficients, which are related to $\dot{\boldsymbol{\sigma}}_T(t)$ through Eq.~\eqref{eq:Ddynamic}, is a bit more subtle.
In order to analyze this point it is convenient to consider Eq.~\eqref{eq:diff_simple}. On the one hand, the time derivative acting on $\boldsymbol{\Phi}(t\!-\!\tau)$ in the second term on the right-hand side of that equation will give rise to $\ddot{G}(t\!-\!\tau)$ and an unsuppressed contribution from $e^{-\Lambda_\star (t\!-\!\tau)}$.
[Analogously to what was explained above for Eq.~\eqref{eq:sigma1}, there is no contribution from the components of the transition matrix involving $\ddot{G}(t)$, and it can only arise when time derivatives act on other components.]
On the other hand, the additional time integral in that term when considering such a contribution will generate an extra $1/\Lambda_\star$ factor as compared to the first term on the right-hand side of Eq.~\eqref{eq:diff_simple}.
Thus, the final conclusion is that we can use the approximate local propagator $G_\mathrm{R}(t)$ to calculate the diffusion coefficients at arbitrary times in the large cut-off limit.
Comparison of the results evaluated using the exact expressions and plotted in Sec.~\ref{sec:jolts} and the approximate results for the large cut-off limit 
also support this conclusion.

We close this subsection with a brief discussion of the possible dissipative regimes when considering finite values of the cut-off in our spectral function, since the presence of this new scale can give rise to a richer set of possibilities.
\begin{figure}[h]
\centering
\includegraphics[width=0.5\textwidth]{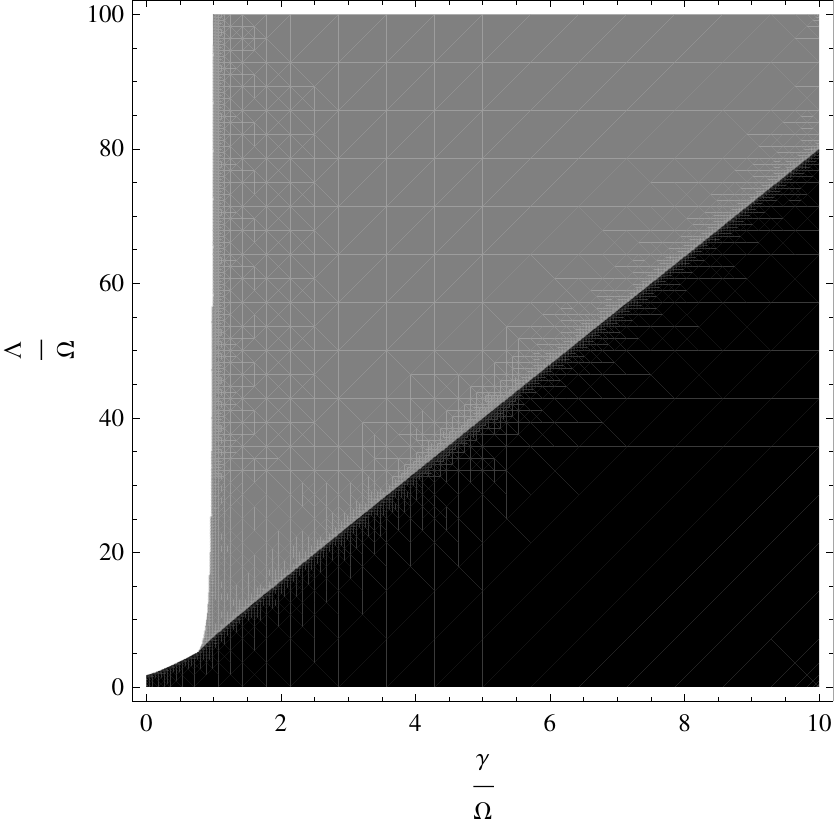}
\caption{Dissipative phases for Ohmic damping with finite rational cut-off. From left to right they are \emph{underdamped} in white, \emph{overdamped} in grey, and \emph{strong coupling} in black.}
\label{fig:ohmic-phase}
\end{figure}
For our rational cut-off function we have three different dissipative regimes corresponding to the three shaded regions in Fig.~\ref{fig:ohmic-phase}.
The boundary between different regions corresponds to the values of the parameters for which a pair of roots of the denominator of $\hat{G}(s)$ degenerate and change character, i.e. they change from a complex conjugate pair to two real ones.
Atop the diagram where $\Lambda \gg \Omega$, lies the regime of local dissipation,
whereas along the bottom of the diagram where $\Lambda \ll \Omega$, lies an effectively sub-ohmic regime as $\Lambda$ becomes an IR cut-off.
The white shaded vertical stripe to the left lies completely in the weak coupling regime and constitutes the \emph{underdamped} regime.
This regime is as described previously with slowly decaying oscillations and a cut-off-dependent decay rate.
The grey shaded middle region denotes the \emph{overdamped} regime.
This regime is also analogous to that of the simple and overdamped harmonic oscillator but with an additional cut-off-dependent decay rate.
The black shaded region to the right denotes a new nonlocal \emph{strong-coupling} regime that emerges for a sufficiently strong coupling (such that $\gamma_0$ is large compared to the cut-off).
Specifically, as derived from Eqs.~\eqref{eq:rate1a}-\eqref{eq:rate1c}, the relevant scales for this regime in the limit of very strong coupling are
\begin{eqnarray}
\Lambda_\star &=& \frac{\Omega^2}{2\gamma_0} - \frac{\Omega^4}{4\Lambda\gamma_0^2} + \mathcal{O}\!\left( \frac{1}{\gamma_0^3} \right) , \\
\gamma_\star &=& \frac{\Lambda}{2} - \frac{\Omega^2}{4\gamma_0} + \mathcal{O}\!\left( \frac{1}{\gamma_0^2} \right) , \\
\Omega_\star &=& 2 \Lambda \gamma_0 + \Omega^2 + \mathcal{O}\!\left( \frac{1}{\gamma_0} \right) .
\end{eqnarray}
Hence, we can see that one has moderately damped, rapid oscillations plus an additional slow decay rate.

\subsection{Initial Jolts}
\label{sec:jolts}

Early studies by Unruh and Zurek \cite{UnruhZurek89} as well as HPZ \cite{HPZ92} already revealed that at low temperatures the normal diffusion coefficient $D_{\!pp}(t)$ of an ohmic environment exhibited a strong cut-off sensitivity for very early times of order $1/\Lambda$. As shown in the next section and \ref{sec:mod-time_app}, in the large cut-off limit where the use of the local propagator is a good approximation one can obtain relatively simple analytic results. They confirm that for zero temperature the normal diffusion coefficient, which vanishes at the initial time, exhibits an initial jolt with an amplitude of order $\Lambda$ peaked around a time of order $1/\Lambda$ and then decays roughly like $1/t$ (for times much earlier than $1/\Omega$ and $1/\gamma_0$).

\begin{figure}[h]
\centering
\includegraphics[width=0.5\textwidth]{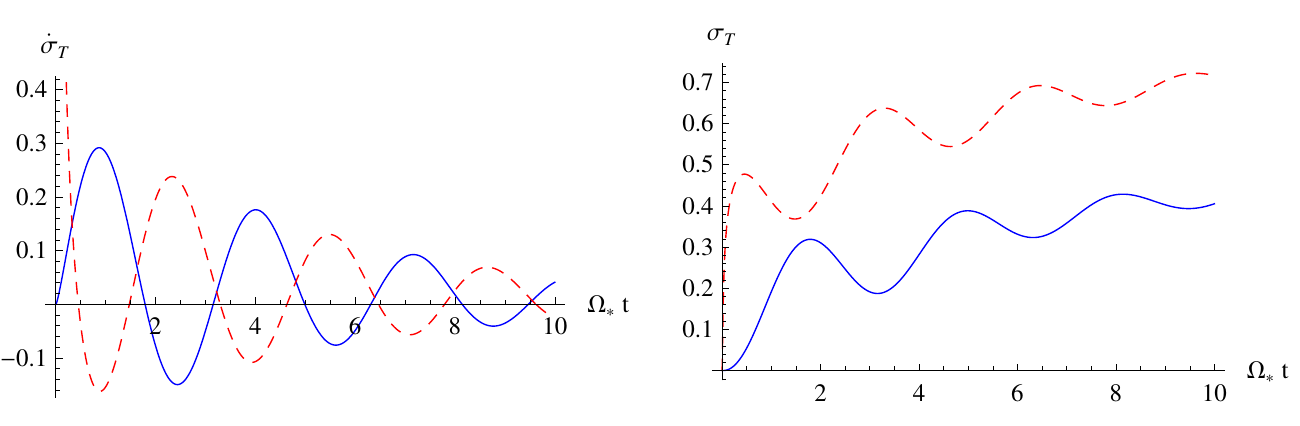}
\caption{\label{fig:jolt}Exact thermal covariance dynamics for \textcolor{blue}{$\cdot$ normalized position uncertainty $M \Omega_\star \sigma_{xx}^T(t)$} and \textcolor{red}{$\cdots$ normalized momentum uncertainty $\frac{\sigma_{pp}^T(t)}{M \Omega_\star}$} in the highly non-Markovian regime with $T=\gamma_\star = \frac{\Omega_\star}{10}$, $\Lambda_\star = 100\, \Omega_\star$.}
\end{figure}

\begin{figure}[h]
\centering
\includegraphics[width=0.5\textwidth]{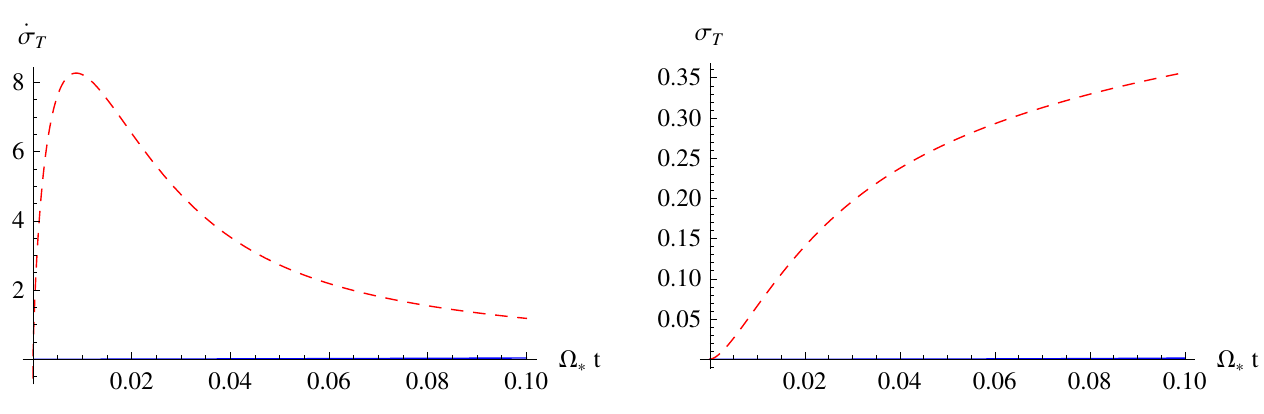}
\caption{\label{fig:jolt0}Same plot as in the previous figure, but with a much larger time resolution, which reveals the presence of the initial jolt in \textcolor{red}{$\dot{\sigma}_{pp}(t)$} peaked around $t\sim1/\Lambda_*$, while \textcolor{blue}{$\sigma_{xx}(t)$} and \textcolor{blue}{$\dot{\sigma}_{xx}(t)$} remain essentially zero at those timescales.}
\end{figure}

Alternatively, one can obtain the exact analytic results for finite cut-off by computing $\dot{\boldsymbol{\sigma}}_T(t)$ using Eqs.~\eqref{eq:sigmadot1a}-\eqref{eq:sigmadot1c}, as explained in Sec.~\ref{sec:thermal_contour}. The resulting expressions are rather lengthy and not particularly insightful, and will not be reported here, but they have been employed to plot some examples of exact results for the diagonal components of $\dot{\boldsymbol{\sigma}}_T(t)$ and $\boldsymbol{\sigma}_T(t)$ in Figs.~\ref{fig:jolt} and \ref{fig:jolt0}. From the different components of the thermal covariance and its time derivative one can obtain the diffusion coefficients using Eq.~\eqref{eq:cov_diff}, and in particular one can see from Fig.~\ref{fig:jolt0} the presence of the jolt mentioned above.

It is important to emphasize that such kind of behavior, as well as an associated rapid growth of $\sigma_{pp}(t)$ and a slower growth of $\sigma_{xx}(t)$ (which eventually decays exponentially within the relaxation time-scale $1/\Gamma$) until they both reach values which depend logarithmically on $\Lambda$ for large values of $\Lambda$, is a consequence of having started with a completely uncorrelated initial stated. A possible way of generating a properly correlated initial state is by smoothly switching on the system-environment interaction within a time-scale much longer than $1/\Lambda$, but longer than the other relevant time-scales of the system. This is discussed in some detail in \ref{sec:initial_coupling}. It also contains a number of technical details concerning the effects of the switch-on function appearing in the dissipation kernel, which can be be mainly reabsorbed in redefinition of the initial sate. The key point, however, is the role played by the switch-on function appearing in the noise kernel, which eliminates the strong cut-off sensitivities and jolts mentioned above when calculating correlation functions (the covariance matrix) and its derivatives.

We conclude this subsection briefly mentioning some generally applicable bounds on the growth of the different thermal covariance components.
First, we note from Eq.~\eqref{eq:tsig} that the thermal covariance is positive definite as the noise kernel is a positive definite function.
We also note that the thermal covariance begins with $\boldsymbol{\sigma}_T(0) = 0$ and $\dot{\boldsymbol{\sigma}}_T(0) = 0$.
Given that this matrix is positive definite, the off-diagonal entries must be smaller than the average (arithmetic or geometric) diagonal entries.
But the off-diagonal $\sigma_T^{xp}(t)$ is proportional to $\dot{\sigma}_T^{xx}(t)$ and we have, therefore, the constraint
\begin{eqnarray}
\left| \dot{\sigma}_T^{xx}(t) \right| &\leq& \frac{2}{M} \sqrt{ \sigma_T^{xx}(t) \sigma_T^{pp}(t) } ,
\end{eqnarray}
which is also generally less than the late-time uncertainty as both $\sigma_T^{xx}(t)$ and $\sigma_T^{pp}(t)$ begin increasing and then proceed to undergo damped oscillations, wherein each cycle there is a net increase in uncertainty.
This constrains the growth of position uncertainty.
If the uncertainty function takes reasonable values, then the position uncertainty must have reasonable growth.

An analogous constraint can be placed upon the growth in momentum uncertainty by considering the positive definite matrix $\dot{\boldsymbol{\Phi}} \cdot \boldsymbol{\nu} \cdot \dot{\boldsymbol{\Phi}}^{\!\mathrm{T}}$ which yields
\begin{eqnarray}
\left| \dot{\sigma}_T^{pp}(t) \right| &\leq& 2 M \sqrt{ \sigma_T^{pp}(t) \left( \ddot{G} \cdot \nu \cdot \ddot{G} \right)\!(t) } .
\end{eqnarray}
So while the growth in position uncertainty is well constrained, growth in momentum is much less constrained.
Corresponding to this, we show in Sec.~\ref{sec:late_cov} that the late-time momentum uncertainty has much more sensitivity to the high frequency modes of the bath.
In terms of ohmic coupling, the initial linear jolts, $\dot{\sigma}_{pp}^T \sim \Lambda$, and late-time logarithmic cut-off sensitivity only occurs in the momentum uncertainty.
The position uncertainty is relatively well behaved in both respects, having only initial logarithmic jolts and no late-time cut-off sensitivity at all.
The (linear) momentum jolting occurs only for a short period of time, $\Delta t \sim \Lambda^{\!-1}$.
The result is a rapid momentum dispersion near the initial time, but bounded logarithmically.

\subsection{Full-Time Diffusion Coefficients for Large Cut-off}
\label{sec:mod-time}

Full-time solutions for finite cut-off are completely possible given our analytic spectrum, the exact nonlocal propagator in Sec.~\ref{sec:nonlocalG}, and the contour integrals detailed in Sec.~\ref{sec:thermal_contour}.
Such resulting solutions were used to plot the early time evolution in Fig.~\ref{fig:jolt}, but they are a bit cumbersome for publishing.
Therefore, for pedagogical reasons we will restrict ourselves to the high cut-off regime in this subsection since substantial additional simplifications can be employed in that case. For nonlocal dissipation it is in general much easier to calculate first the thermal covariance than the diffusion coefficients, but the situation will be different here. The key point that will be exploited in this subsection is that for large cut-off the propagator in the ohmic case can be approximated by the local one, $G_\mathrm{R}(t)$, as discussed in Sec.~\ref{sec:nonlocalG}.
The advantage of using the local propagator $G_\mathrm{R}(t)$ is that only the term involving a single time integral contributes to the expression for the diffusion coefficients in Eq.~\eqref{eq:diff_simple}. On the other hand, if one is only interested in the late-time asymptotic values of the diffusion coefficients, one can obtain simple analytic results without the need to restrict oneself to large values of the cut-off by using the results that will be presented in the next subsection.

The details of the derivation and the complete results for the diffusion coefficients at arbitrary times
are provided in \ref{sec:mod-time_app}. Here we simply highlight the main results and discuss some of their implications. Both diffusion coefficients can be written in the following compact form:
\begin{align}
D_{\!xp}(t) =&\; D_{\!xp}(\infty) \label{eq:Dxp2a} \\
& - M \gamma_0 \left\{ \dot{G}_\mathrm{R}(t) + G_\mathrm{R}(t) \left( \!2 \gamma_0 \!-\! \frac{d}{dt} \right) \right\} \mbox{DF}(t) , \nonumber \\
D_{\!pp}(t) =&\; D_{\!pp}(\infty) \label{eq:Dpp2a} \\
& - M \gamma_0 \left\{ \dot{G}_\mathrm{R}(t) \left( \!\gamma_0 \!+\! \frac{d}{dt} \right) + G_\mathrm{R}(t) \, \Omega^2 \right\} \mbox{DF}(t) , \nonumber
\end{align}
where $D_{\!xp}(\infty)$ and $D_{\!pp}(\infty)$ are immediately obtained by multiplying Eqs.~\eqref{eq:lateDxp}-\eqref{eq:lateDpp} by $s$ and taking the limit $s\to0$.
The general expression for DF$(t)$ is given by Eq.~\eqref{eq:DFLerch}, but a simple result for the zero temperature case is provided in Eq.~\eqref{eq:DF_T0}.
Essentially, DF$(t)$ decays in a manner slightly more complicated than that of exponential integrals with system, coupling, and temperature timescales but such that temperature is the most dominant.

It is important to note that the coefficients $D_{\!xp}(t)$ and $D_{\!pp}(t)$ both exhibit logarithmic divergences in the limit $\Lambda \rightarrow \infty$.
This has been pointed out for $D_{\!xp}(t)$ in Ref.~\cite{Lombardo05}, where the coefficients of the master equation were calculated perturbatively to second order in the system-environment coupling constants (linear order in $\gamma_0$).
The fact that there is also a logarithmic divergence in $D_{\!pp}(t)$ was not seen in that reference because it is quartic in the system-environment coupling constants (quadratic in $\gamma_0$).
Moreover, strictly speaking such kinds of perturbative calculations cannot be employed to study the long time behavior since they are only valid for $t \ll \gamma_0^{-1}$ and they miss for instance the exponential decay of the second and third terms on the right-hand side of Eqs.~(\ref{eq:Dxp2})-(\ref{eq:Dpp2}).

We close this subsection with some remarks about the late-time diffusion coefficients in the weak coupling regime. Expanding Eqs.~\eqref{eq:lateDxp}-\eqref{eq:lateDpp} perturbatively in $\gamma_0$ we get
\label{sec-caldeira}
\begin{align}
D_{\!xp}(\infty) &= \frac{2}{\pi} \gamma_0 \, \mbox{Re}\!\left[ \mbox{H}\!\left( \frac{\Lambda}{2 \pi T} \right) - \mbox{H}\!\left( \frac{\imath \Omega}{2 \pi T} \right) \right] + \mathcal{O}(\gamma_0^2) , \label{eq:weakDxp}\\
D_{\!pp}(\infty) &= \gamma_0 \, \Omega \coth\!{\left( \frac{\Omega}{2T} \right)} + \mathcal{O}(\gamma_0^2).
\end{align}
In comparison to the weak coupling master equation of Caldeira
\emph{et al.}\ \cite{Caldeira89}, the normal diffusion coefficient is
the same to lowest order in the coupling, but the anomalous diffusion
coefficient is \emph{completely absent} there.
\begin{figure}[h]
\centering
\includegraphics[width=0.5\textwidth]{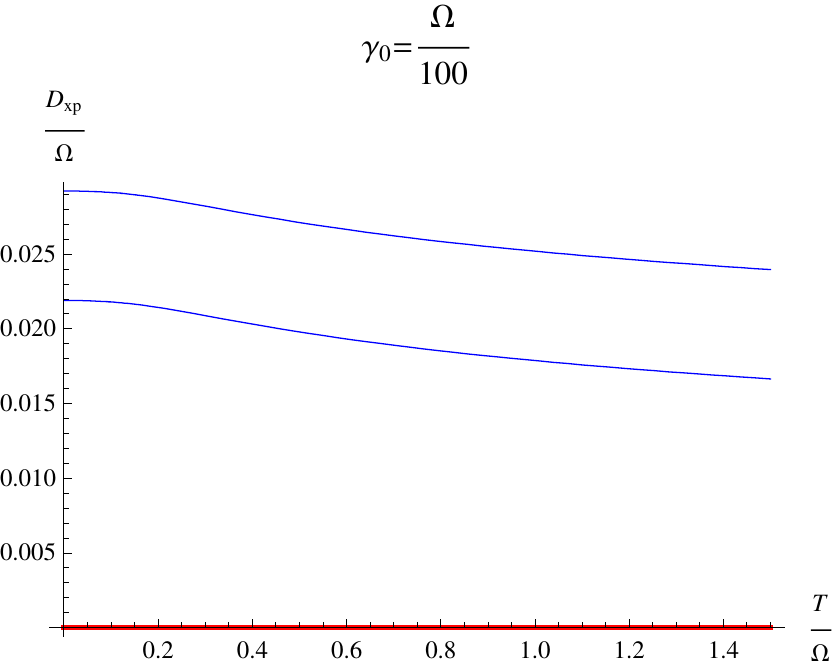}
\caption{Late time $D_{\!xp}$ for \textcolor{red}{$\bullet$ high temperature or equivalently Caldeira}, \textcolor{blue}{$\cdot$ HPZ at $\Lambda=10^3 \Omega$ and $\Lambda=10^4 \Omega$}.}
\label{fig:CCR2}
\end{figure}
The largest contribution (in the weak coupling regime) to the
anomalous diffusion coefficient comes from the cut-off and it does
not vanish at finite temperature (see Fig.~\ref{fig:CCR2}). This
logarithmic sensitivity does not enter into the normal diffusion
coefficient until second order, but in the anomalous diffusion
coefficient it is only proportional to one power of the coupling
constant, which is the order to which the master equation of Caldeira
\emph{et al.}\ \cite{Caldeira89} should be valid.
In this weak-coupling perturbative expansion, both diffusion coefficients are of order $\gamma_0$ plus higher-order corrections, but they give contributions of different orders to the late-time thermal covariance $\boldsymbol{\sigma}_T^\infty$, Sec.~\ref{sec:late_cov}.
Whereas $D_{\!pp}^\infty$ gives contributions of order $1$ because it appears multiplied by a factor $1/\gamma_0$, $D_{\!xp}^\infty$ gives contributions of order $\gamma_0$.
That is why the correct thermalization in the weak-coupling limit was obtained in Ref.~\cite{Caldeira89} despite having completely neglected the anomalous diffusion coefficient.
The origin of the mixed orders in $\gamma_0$ appearing on the right-hand side of Eq.~\eqref{eq:ltsig} can be ultimately traced to the fact that $\boldsymbol{\mathcal{H}}$ contains terms both of order $1$ and $\gamma_0$, whose implication for $\boldsymbol{\sigma}_T^\infty$ can be straightforwardly seen from the Lyapunov equation \eqref{eq:thermal_relation}.

\subsection{Late-Time Covariance for Finite Cut-off}
\label{sec:coefficients_late} \label{sec:uncertainty}

In Sec.~\ref{sec:nonlocalG} the late-time dissipation and renormalized frequency coefficients were directly inferred from the nonlocal propagator to be $\gamma_\star$ and $\Omega_\star$, the result of factoring a cubic polynomial in the nonlocal Green function.
These coefficients are entirely non-perturbative in both coupling and cut-off and completely determine the late-time propagator.
The remaining part of the solution pertains to the emergence of the thermal covariance, whose late-time dynamics can be described as in Sec.~\ref{sec:late_cov}, given the late-time propagator and the late-time thermal covariance.
The late-time thermal covariance can also be related to the late-time diffusion coefficients through the Lyapunov equation, Eq.~\eqref{eq:thermal_relation}, but the thermal covariance is an easier quantity to compute.
If interested in the diffusion coefficients, one can then obtain them straightforwardly using Eq.~\eqref{eq:thermal_relation}.

For our spectral density the simplified integrals derived in \ref{sec:late_cov_der} are contour integrals and can be evaluated via the residue theorem after using the rational expansion of the hyperbolic cotangent, Eq.~\eqref{eq:coth}.
The result for the late-time, but non-perturbative thermal covariance obtained in this way is
\begin{align}
\sigma^{xx}_T &= \frac{T}{M \Omega^2} + \frac{1}{2 M \tilde{\Omega}_\star} \mathrm{Im}\!\left[ \mathcal{R} \right] , \label{eq:latecov_xx} \\
\sigma^{pp}_T &= M T + \frac{M \tilde{\Omega}_\star}{2} \mathrm{Im}\!\!\left[ \left(1\!-\!\imath \frac{\gamma_\star}{\tilde{\Omega}_\star} \right)^{\!\!\!2} \mathcal{R} \right] \label{eq:latecov_pp}, \\
\mathcal{R} &\equiv \frac{2}{\pi} \frac{\Lambda_\star\!+\!\gamma_\star\!-\!\imath \tilde{\Omega}_\star}{\Lambda_\star\!-\!\gamma_\star\!-\!\imath \tilde{\Omega}_\star} \left\{ \mathrm{H}\!\left( \!\frac{\gamma_\star\!+\!\imath \tilde{\Omega}_\star}{2 \pi T}\! \right) \!-\! \mathrm{H}\!\left( \!\frac{\Lambda_\star}{2 \pi T}\! \right) \right\} ,
\end{align}
where we assumed, as before, that $\tilde{\Omega}_\star = \sqrt{\Omega_\star^2-\gamma_\star^2}$ is real and H$[z]$ denotes the harmonic number function defined in Sec.~\ref{sec:harmonic}.
If one expands those expressions, and the expressions below, under the assumption that $\tilde{\Omega}_\star$ is real, e.g. using $\mathrm{Im}[z]=(z-\bar{z})/(2\imath)$, then one will have the more general expressions which will apply even in the overdamped regime.

At \textbf{high temperature} all of the harmonic number functions vanish, leaving only the first terms in Eqs.~\eqref{eq:latecov_xx}-\eqref{eq:latecov_pp}, which are proportional to temperature:
\begin{eqnarray}
\sigma^{xx}_T &=& \frac{T}{M \Omega^2} + \mathcal{O}(T^0) , \\
\sigma^{pp}_T &=& M T + \mathcal{O}(T^0) .
\end{eqnarray}
This corresponds to the high-temperature result of classical statistical mechanics.
It is interesting that this can happen for a finite cut-off and, therefore, outside the strict Markovian limit.

At \textbf{zero temperature} the first terms in Eqs.~\eqref{eq:latecov_xx}-\eqref{eq:latecov_pp} vanish and all of the harmonic number functions can be equivalently evaluated as logarithms,
so that $\mathcal{R}$ simplified as follows:
\begin{align}
\mathrm{H}\!\left( \!\frac{\gamma_\star\!+\!\imath \tilde{\Omega}_\star}{2 \pi T}\! \right) \!-\! \mathrm{H}\!\left( \!\frac{\Lambda_\star}{2 \pi T}\! \right) =&\; \imath \cos^{\!-1}\!\!\left( \frac{\gamma_\star}{\Omega_\star} \right) \!-\! \log\!\left( \frac{\Lambda_\star}{\Omega_\star} \right) \nonumber \\
& + \mathcal{O}(T) .
\end{align}
This generalizes the results of Unruh and Zurek \cite{UnruhZurek89}, who explored the zero temperature regime in the limit of local dissipation.

\begin{figure}[h]
\centering
\includegraphics[width=0.5\textwidth]{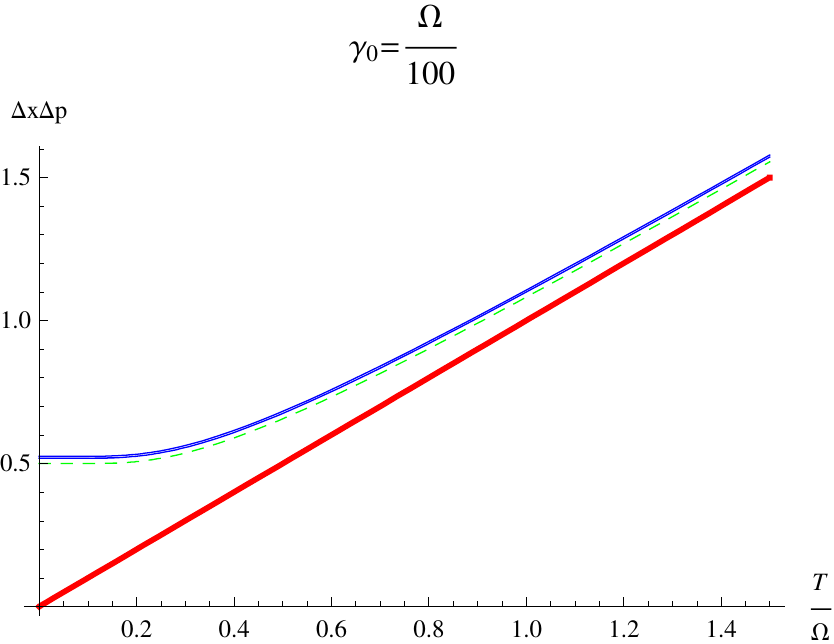}
\caption{\label{fig:dxdp}Late time $\Delta x \Delta p$ for \textcolor{red}{$\bullet$ high temperature, classical statistical mechanics}, \textcolor{green}{$\cdots$ weak coupling approximation $\frac{1}{2}\coth{\frac{\Omega}{2T}}$} \textcolor{blue}{$\cdot$ HPZ at $\Lambda=10^3 \Omega$ and $\Lambda=10^4 \Omega$}.}
\end{figure}

\begin{figure}[h]
\centering
\includegraphics[width=0.5\textwidth]{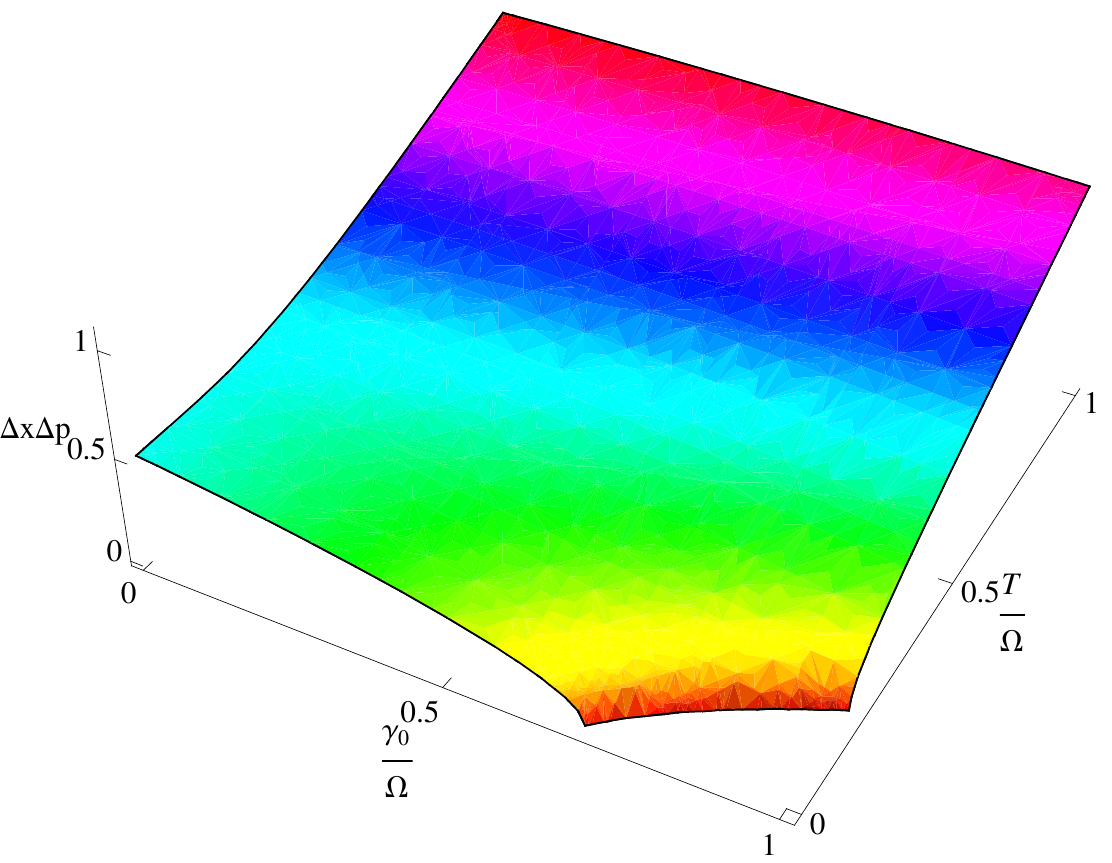}
\caption{\label{fig:dxdp3d-ren}Late time $\Delta x \Delta p$ for the unphysical, subtracted theory.}
\end{figure}

\begin{figure}[h]
\centering
\includegraphics[width=0.5\textwidth]{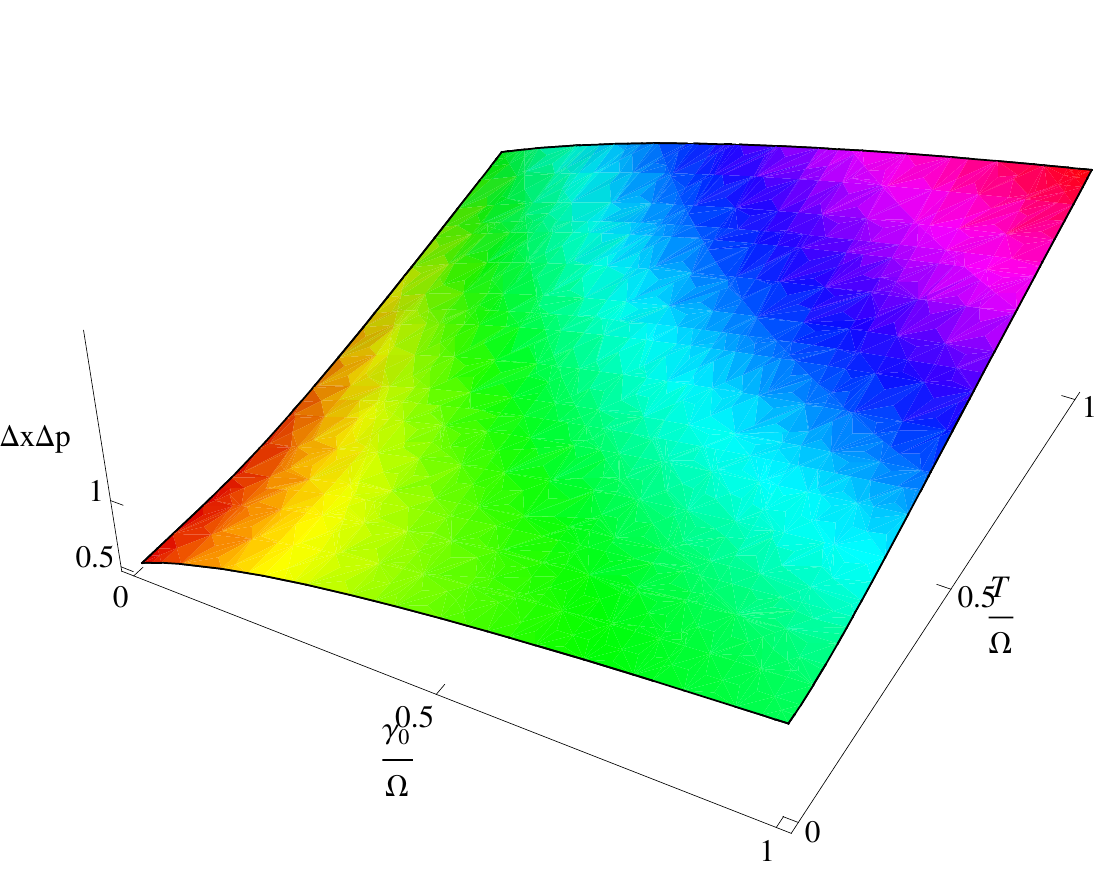}
\caption{\label{fig:dxdp3d-unren}Late time $\Delta x \Delta p$ for the $\Lambda = 10^3 \Omega$ theory.}
\end{figure}

Finally, in the \textbf{weak coupling} limit these expressions correctly reproduce the free thermal state:
\begin{eqnarray}
\sigma^{xx}_T &=& \frac{1}{2 M \Omega}\coth\!\left( \frac{\Omega}{2T} \right) + \mathcal{O}(\gamma_0) , \label{eq:free_xx}\\
\sigma^{pp}_T &=& \frac{M \Omega}{2}\coth\!\left( \frac{\Omega}{2T} \right) + \mathcal{O}(\gamma_0) \label{eq:free_pp}.
\end{eqnarray}
One can also see that at weak coupling the uncertainty function agrees with the weak coupling approximation for moderate values of the cut-off scale, as shown in Fig.~\ref{fig:dxdp}.
Had one naively tried to have finite diffusion in the limit $\Lambda \rightarrow \infty$,
subtracting by hand the $\log(\Lambda / \Omega)$ term,
one would find a violation of the Heisenberg uncertainty principle at low temperature and strong coupling (see Fig.~\ref{fig:dxdp3d-ren}), which renders the theory unphysical.
Of course this does not happen with the unsubtracted theory, as seen in Fig.~\ref{fig:dxdp3d-unren}.
It is, thus, clear that the logarithmic dependence on the ultraviolet cut-off that appears in the diffusion is a physically important parameter and not something that can be subtracted away.

While the logarithmic sensitivity appears in both diffusion coefficients, it is suppressed in the position uncertainty by inverse powers of the cut-off.
For the momentum uncertainty, the logarithmic sensitivity appears already to first order in $\gamma_0$ (which is itself quadratic in the system-environment coupling constant) and is otherwise unsuppressed.
This behavior had already been noticed for Gaussian wave-packets in the Ohmic environment \cite{Hakim85,UnruhZurek89},
and as we have discussed in Sec.~\ref{sec:late_cov}, the position uncertainty will be free of the highest cut-off sensitivities for \emph{any} spectral density.

\label{sec:strong-couple}
Finally, given that our results are \emph{nonperturbative}, it is also interesting to point out what happens in the highly nonlocal \textbf{strong coupling} regime mentioned Sec.~\ref{sec:nonlocalG}.
The late-time thermal covariance for this case essentially corresponds to taking the large $\Omega_\star$ limit limit of Eqs.~\eqref{eq:latecov_xx}-\eqref{eq:latecov_pp}:
\begin{eqnarray}
\sigma^{xx}_T &=& \frac{T}{M\Omega^2} + \frac{1}{2 M \Omega_\star}
+ \mathcal{O}\!\left( \frac{1}{\Omega_\star^2} \right) , \\
\sigma^{pp}_T &=& \frac{M \Omega_\star}{2} + \mathcal{O}(\Omega_\star^0) .
\end{eqnarray}
For this model of strong coupling to the environment, and yet finite cut-off, the Brownian particle will become strongly localized in position at late time and sufficiently low temperatures.
And although the particle is localized in position, the uncertainty principle is not violated but at most minimized in the zero temperature limit.

\section{Sub-ohmic and Supra-ohmic Cases}
\label{sec:non-ohmic}

\subsection{Sub-ohmic with no Cut-off}
\label{sec:sub-ohmic}

As an example where the nonlocal effects of dissipation are important, we will consider one of the most common and well-behaved sub-ohmic spectral densities, $I(\omega) \propto \sqrt{\omega}$, which requires neither a UV nor an IR cut-off in the final results (although one still needs to renormalize the frequency introducing a logarithmically divergent bare counterterm).
Our formulas will take a simpler form if we express our spectral density in terms of  a quadratic coupling constant $\gamma_\star$ as follows:
\begin{eqnarray}
I(\omega) &=& \frac{2}{\pi} M \gamma_\star \sqrt{\omega_\star \, \omega} , \\
\omega_\star^2 & \equiv & \Omega^2 + \gamma_\star^2 .
\end{eqnarray}
It is then a straightforward calculation to find the propagator
\begin{eqnarray}
\hat{G}(s) &=& \frac{\frac{1}{M}}{ s^2 + 2 \Gamma \sqrt{2 \omega_\star \, s} + \Omega^2 } ,
\end{eqnarray}
which is amenable to partial fraction decomposition in $\sqrt{s}$ since $s$ is strictly positive.
As we have defined our nonlinear coupling strength in anticipation of this polynomial,
the roots of the quartic denominator $r_k: k \in \{1,2,3,4\}$ can be shown to be the conjugate pairs
\begin{eqnarray}
r_{1,2} &=& \frac{1}{\sqrt{2}} \left( +\sqrt{\omega_\star} \pm \imath \sqrt{\omega_\star + 2 \gamma_\star} \right) , \\
r_{3,4} &=& \frac{1}{\sqrt{2}} \left( -\sqrt{\omega_\star} \pm \imath \sqrt{\omega_\star - 2 \gamma_\star} \right) .
\end{eqnarray}
After partial fraction decomposition, we may cast our propagator in the form
\begin{eqnarray}
\hat{G}(s) &=& \sum_{k=1}^4 \frac{A_k}{M} \frac{1}{ \sqrt{s} - r_k } , \\
A_j &=& \prod_{\substack{k=1\\k \neq j}}^4 \frac{1}{r_j - r_k} ,
\end{eqnarray}
with inverse Laplace transform
\begin{eqnarray}
G(t) &=& \sum_{k=1}^4 \frac{A_k}{M} r_k \, e^{r_k^2 t} \, \mbox{erfc}\!\left(-r_k \sqrt{t} \right) \label{eq:subohmicG},
\end{eqnarray}
where $\mathrm{erfc}(z)$ is the cumulative error function of the normal distribution, defined in \ref{sec:erfc}.
There are additional terms from the individual Laplace transforms, like $t^{-1/2}$, but they vanish in the sum.
Using Eq.~\eqref{eq:erfc_asymp} for the asymptotic expansion of $\mathrm{erfc}(z)$ in order to expand the Green function in Eq.~\eqref{eq:subohmicG} at late times, we obtain terms of the form
\begin{align}
z\, e^{z^2} \, \mbox{erfc}(z) =&\; \frac{1}{\sqrt{\pi}} \sum_{k=0} (-1)^k \frac{(2k)!}{k!} \frac{1}{(2z)^{2k}} \nonumber \\
& + \left\{ \begin{array}{cc}
0 & \mbox{Re}[z] \geq 0 \\
2 z \, e^{z^2} & \mbox{Re}[z] \leq 0
\end{array} \right. ,
\end{align}
which we can use to expand the Green function in Eq.~\eqref{eq:subohmicG}.
\begin{figure}[h]
\centering
\includegraphics[width=0.5\textwidth]{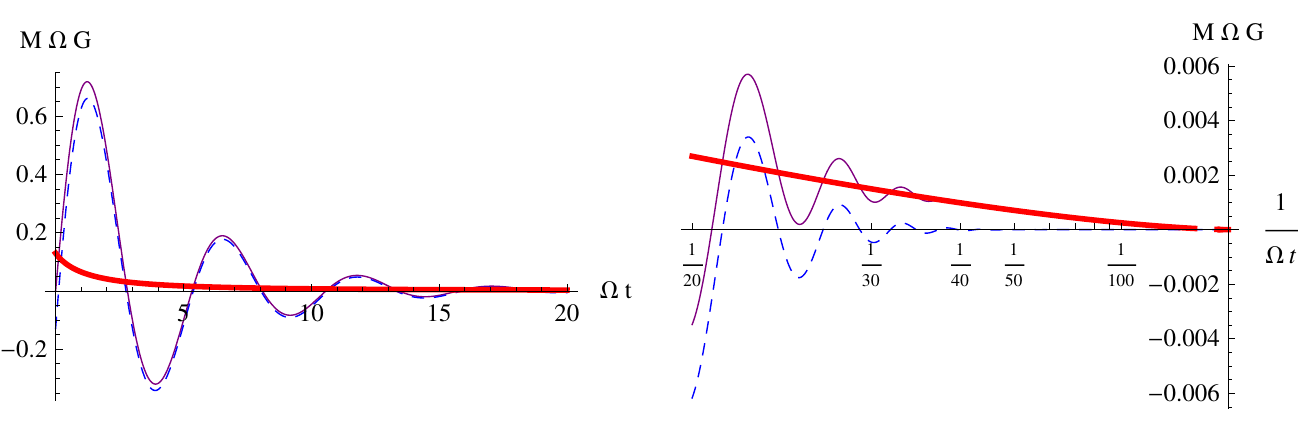}
\caption{\label{fig:sub-ohmicG}Asymptotic expansion of sub-ohmic \textcolor{magenta}{$\cdot$ propagator $G(t)$} into \textcolor{blue}{$\cdots$ the local contribution} and \textcolor{red}{$\bullet$ the nonlocal contribution} for $\gamma_\star = \frac{\Omega}{4}$. The local contribution is initially more significant, but the nonlocal contribution dominates eventually.}
\end{figure}
After grouping all the contributions together, we will find exponential terms with characteristic frequencies $f = -\Gamma \pm \imath \sqrt{\omega_\star^2+2\gamma_\star \omega_\star}\,$,
which are actually the solutions to the characteristic rate equation \eqref{eq:char_rate} with smallest \emph{negative} real part.
These are the only terms that one would have considered if the local propagator $G_\mathrm{R}(t)$ within the late-time approximation of Sec.~\ref{sec:late_approx} had been employed.
In addition, and more importantly are the power-law decay terms which admit no local representation.

This sub-ohmic model provides a perfect example showing when effectively local treatments, such as that in Sec.~\ref{sec:late_approx}, will fail completely.
At first the local contribution will dominate and the master equation coefficients will appear to trend towards $\Gamma \approx \gamma_\star$ and $\Omega_\mathrm{R} \approx \omega_\star + \gamma_\star$.
However, the nonlocal contribution (the power-law terms) will eventually dominate the more swiftly decaying local contribution (the exponential terms) and a correct treatment of the nonlocal dynamics will be required.
In fact, as the nonlocal contribution becomes comparable to the local contribution, the master equation coefficients will become periodically divergent [this is related to the fact that $\det \boldsymbol{\Phi}(t)$ vanishes and changes sign at those times.].
The underlying homogeneous evolution is well behaved and strictly dissipative (the damping kernel is positive definite), but the localizing perspective of the master equation becomes divergently unnatural.
Any errors, numeric or analytic, can be catastrophic in the master equation perspective. In this respect, the subtleties missed in previous derivations of the master equation, as pointed out in Sec.~\ref{sec:MED} and which are relevant whenever nonlocal effects are important, will likely give rise to substantial discrepancies in this case.

The full-time evolution is rather complicated, but the late-time limit is very manageable.
For example, from Eq.~\eqref{eq:tclflt} we can express the late-time thermal position uncertainty as
\begin{equation}
\sigma_T^{xx}(\infty) = 2 \int_0^\infty \!\! I(\omega) \coth\!{\left( \frac{\omega}{2 \pi T} \right)} |\hat{G}(\imath \omega)|^2 \, 2 \sqrt{\omega} \, d\sqrt{\omega} ,
\end{equation}
where we have used the relation $d\omega = 2 \sqrt{\omega} \, d\sqrt{\omega}$.
The integrand is amenable to partial fraction decomposition, after a rational expansion of the hyperbolic cotangent with Eq.~\eqref{eq:coth}, and can therefore be integrated without resorting to numerics.
Additionally, and in contrast to the ohmic case, the integrand is even in $\sqrt{\omega}$ for all temperatures, including zero, and contour integration techniques are more generally applicable.

\begin{figure}[h]
\centering
\includegraphics[width=0.5\textwidth]{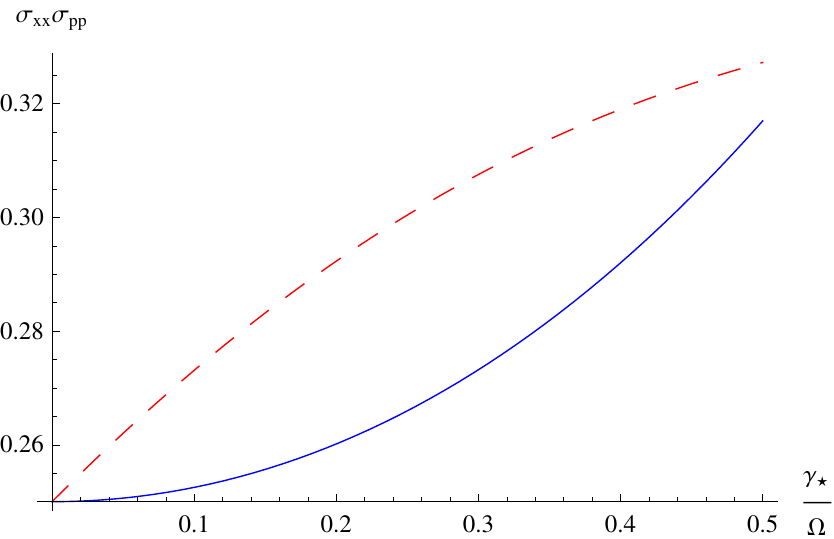}
\caption{\label{fig:sub-ohmicS}Late-time sub-ohmic uncertainty function at zero temperature with the \textcolor{blue}{$\cdot$ exact nonlocal solution} and \textcolor{red}{$\cdots$ fictitious effectively local solution}. In the limit of vanishing dissipation, one has the minimal uncertainty ground state (zero temperature thermal state) in each case.}
\end{figure}

Strictly speaking we cannot compare exact sub-ohmic solutions to those obtained with an incorrect master equation since the master equation will yield nonsense,
but we can compare the exact nonlocal dynamics to those obtained by extracting the local dynamics and assuming it to be the dominant behavior.
Obviously the effectively local approximation is incorrect, but it should be good to zeroth order in the coupling and one might naively expect that it might also behave reasonably for finite coupling strength.
However, in Fig.~\ref{fig:sub-ohmicS} we compare the late-time uncertainty functions and show there to be sharp disagreement to the first two orders in the coupling constant squared (the slope and the curvature of the curves on the plot).

\subsection{Supra-ohmic with Finite Cut-off}
\label{sec:supra-ohmic}

The conventional wisdom has been to consider supra-ohmic spectral densities of the form
\begin{eqnarray}
I_n(\omega) &=& \frac{2}{\pi} M \, \gamma_n\, \omega \left(\frac{\omega}{\Lambda}\right)^n \, \chi\!\left({\small\frac{\omega}{\Lambda}}\right) ,
\end{eqnarray}
where $\chi: [0,\infty) \to [1,0)$ denotes the cut-off regulator.
Without a cut-off regulator, all supra-ohmic couplings have greater than logarithmic high frequency divergence in the diffusion and thermal covariance integrals (see \ref{sec:late_cov_der} for exact integrals in the infinite time limit).
Even when regulated, the mere potential for divergence therefore corresponds to cut-off sensitivity from the high frequency portion of noise integrals, which is balanced by the extra inverse powers of the cut-off in the pre-factor of the above spectral density.

Here we will restrict our investigation to the following spectral density
\begin{eqnarray}
I(\omega) &=& \frac{2}{\pi} M \, \gamma_2\, \omega \frac{\left(\frac{\omega}{\Lambda}\right)^2}{\left(1+\left(\frac{\omega}{\Lambda}\right)^2\right)^2} ,
\end{eqnarray}
because this example is exactly solvable.
The corresponding damping kernel in Laplace space is
\begin{eqnarray}
\hat{\gamma}(s) &=& \frac{\gamma_2}{2} \frac{\frac{s}{\Lambda}}{\left(1+\frac{s}{\Lambda}\right)^2} .
\end{eqnarray}
One might be inclined to view this damping kernel as a tiny mass renormalization plus even less significant higher order terms, but the effect quite different from that, as we will see.
After factoring the fourth-order polynomial, the fully nonlocal propagator can be decomposed by partial fractions into two sets of timescales.
Expanding perturbatively in $\gamma_2$, the first set of timescales correspond to the system frequency with weak damping
\begin{eqnarray}
\gamma_\star &=& \gamma_2 \frac{\left(\frac{\Omega}{\Lambda}\right)^2}{\left(1+\left(\frac{\Omega}{\Lambda}\right)^2\right)^2} + \mathcal{O}(\gamma_2^2) , \\
\Omega_\star &=& \Omega \left( 1 - \frac{\gamma_2}{\Lambda} \frac{1-\left(\frac{\Omega}{\Lambda}\right)^2}{\left(1+\left(\frac{\Omega}{\Lambda}\right)^2\right)^2} + \mathcal{O}(\gamma_2^2) \right) ,
\end{eqnarray}
while the second set of timescales correspond to quickly decaying nonlocal contributions associated with the cut-off scale:
\begin{eqnarray}
\gamma_\Lambda &=& \Lambda - \gamma_\star , \\
\Omega_\Lambda &=& \frac{\Omega}{\Omega_\star} \Lambda .
\end{eqnarray}
The situation is analogous to ohmic case with a finite cut-off except that the nonlocal part of the propagator is also oscillating at the rate $\tilde{\Omega}_\Lambda \approx \sqrt{\gamma_2 \Lambda}$,
for weak coupling and high cut-off.

This form of spectral density was constructed only with well-behaved high frequency contributions in mind.
\begin{figure}[h]
\centering
\includegraphics[width=0.5\textwidth]{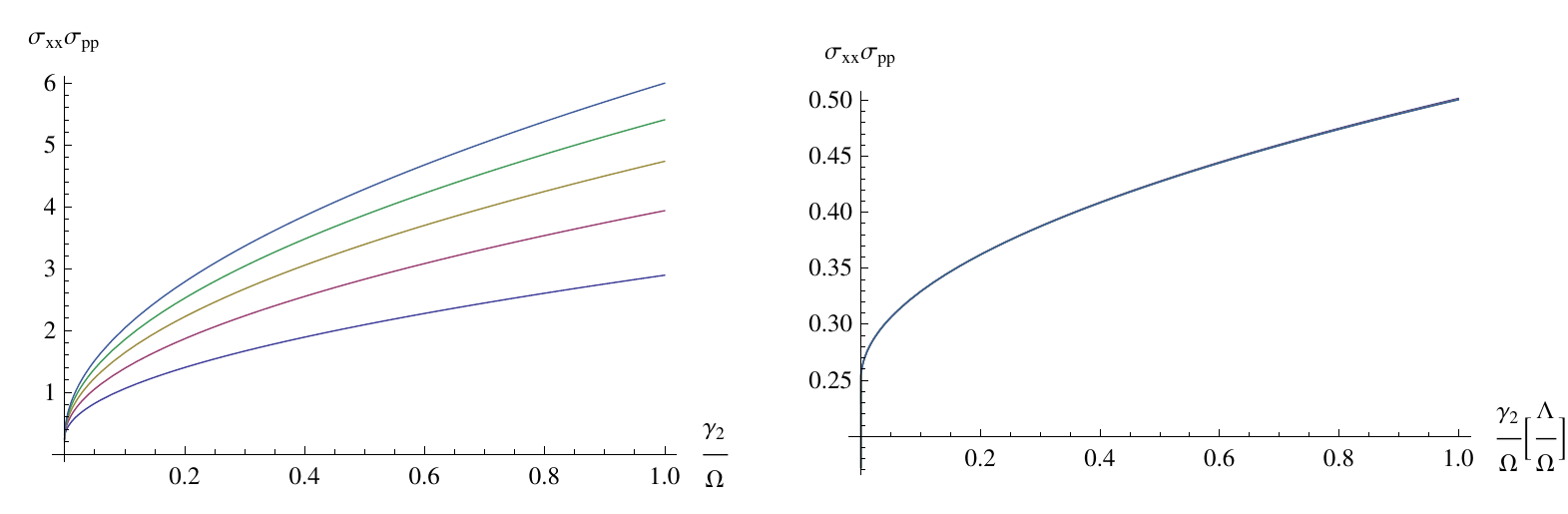}
\caption{Late-time supra-ohmic uncertainty function at zero temperature for cut-offs between $100 \Omega$ and $500 \Omega$. The left plot is with a conventional coupling scale, while the right plot has decreased the coupling strength by an extra power of the cut-off.}
\label{fig:supra-ohmicS}
\end{figure}
Nevertheless, as shown in Fig.~\ref{fig:supra-ohmicS}, we find the conventional form of spectral density to be inadequate.
There is clearly some cut-off sensitivity in the thermal covariance which is remedied by introducing an additional power of cut-off suppression.
E.g. the conventional form of spectral density is not well behaved, but the substitution
\begin{eqnarray}
\gamma_2 &\to& \frac{\Omega}{\Lambda} \gamma_2 ,
\end{eqnarray}
is well behaved.

An explanation only emerges after a more thorough examination of the contour integrals.
The high-frequency regime, $\omega \gg \Lambda$, is already rendered well behaved by the conventional cut-off-dependent prefactor.
The near-resonance regime, $\omega \approx \Omega$, which produces the weak coupling limit, also appears to be well behaved.
There is only one remaining suspect and it proves to be the culprit.
The previously unaccounted for cut-off sensitivity arises here from the nonlocal timescales of the propagator, i.e. the $\omega \approx \Lambda$ regime.
This is quite surprising as unlike sub-ohmic coupling, supra-ohmic coupling does yield a well-behaved local representation for its late-time dynamics.
But residues of the contour integral which correspond to the nonlocal timescales reveal the correct dominant behavior $\sigma_{pp} \approx \frac{1}{2} M \tilde{\Omega}_\Lambda = \frac{1}{2} M \sqrt{\gamma_2 \Lambda}$,
for weak coupling and high cut-off.
Therefore the conventional, linear coupling $\gamma_2$ must be suppressed by an additional factor of the cut-off, else the momentum covariance will be plagued by a $\sqrt{\Lambda}$ sensitivity.

\section{Generalizations of the Theory}
\label{sec:generalization}
\subsection{Influence of a Classical Force}
\label{sec:force}

In this section we consider the case of a classical force $F(t)$ acting on the quantum oscillator.
This is done by adding a time-dependent potential $-F(t)x$ to the system Lagrangian:
\begin{equation}
\mathsf{L}_\mathrm{s} = \frac{1}{2}M \left( \dot{x}^2 - \Omega^2 x^2 \right) + F(t) x ,
\end{equation}
which gives rise to the following additional source on the right-hand side of Eq.~\eqref{eq:langevin_bold}:
\begin{equation}
\mathbf{F}(t) = \left[ \begin{array}{c} 0 \\ F(t) \end{array} \right] .
\end{equation}
Following our master equation derivation in Sec.~\ref{sec:MED}, it is easy to see that such a deterministic source in the Langevin equation simply adds a driving term to the master equation, which becomes
\begin{align}
& \frac{\partial}{\partial t} W_{\!\mathrm{r}}\!\left(\mathbf{z},t\right) = \label{eq:forced_master} \\
& \left\{ \boldsymbol{\nabla}_{\!\!\mathbf{z}}^{\mathrm{T}} \, \boldsymbol{\mathcal{H}}(t) \, \mathbf{z} - \boldsymbol{\nabla}_{\!\!\mathbf{z}}^\mathrm{T} \, \mathbf{F}_{\!\mathrm{eff}}(t) + \boldsymbol{\nabla}_{\!\!\mathbf{z}}^{\mathrm{T}} \, \mathbf{D}(t) \, \boldsymbol{\nabla}_{\!\!\mathbf{z}} \right\} W_{\!\mathrm{r}}\!\left(\mathbf{z},t\right) , \nonumber
\end{align}
where the effective force $\mathbf{F}_{\!\mathrm{eff}}(t)$ is given by
\begin{equation}
\mathbf{F}_{\!\mathrm{eff}}(t) \equiv \mathbf{F}(t) + \int_0^t \!\! d\tau \left\{ \left[ \frac{d}{dt} + \boldsymbol{\mathcal{H}}(t) \right] \boldsymbol{\Phi}(t\!-\!\tau) \right\} \mathbf{F}(\tau) \label{eq:eff_force}.
\end{equation}
Note that the last term in Eq.~\eqref{eq:eff_force}  is a consequence of having nonlocal dissipation and, as we saw in Sec.~\ref{sec:MED}, it vanishes for local dissipation.

Similarly, the method of Sec.~\ref{sec:MESL}, based on the solutions of the Langevin equation, can be straightforwardly generalized to this case and one obtains the following result for the time evolution of the reduced Wigner function:
\begin{equation}
\mathcal{W}_\mathrm{r}\!\left[t, \mathbf{k} \right] = \mathcal{W}_\mathrm{r}\!\left[ 0, \boldsymbol{\Phi}^{\!\mathrm{T}}\!(t) \, \mathbf{k} \right] e^{- \frac{1}{2} \mathbf{k}^{\!\mathrm{T}}\! \boldsymbol{\sigma}_T(t) \, \mathbf{k}} \, e^{-\imath \mathbf{k}^{\!\mathrm{T}} \left\langle \mathbf{z} \right\rangle_{\!F}(t)}
\label{Fsolution1} ,
\end{equation}
with a driven mean $\left\langle \mathbf{z} \right\rangle_{\!F}\!(t)$ given by
\begin{eqnarray}
\left\langle \mathbf{z} \right\rangle_{\!F}\!(t) &=& (\boldsymbol{\Phi} * \mathbf{F})(t)
\label{eq:qF1}.
\end{eqnarray}
On the other hand, one can alternatively use the method of characteristic curves to solve the master equation, as done in Sec.~\ref{sec:charcurv}.
Fourier-transforming Eq.~\eqref{eq:forced_master}, one gets an equation analogous to Eq.~\eqref{eq:phase} but with an extra term $-\imath \mathbf{k}^{\!\mathrm{T}} \mathbf{F}_{\!\mathrm{eff}}(t)$ on the right-hand side. Following the same procedure as in Sec.~\ref{sec:charcurv}, one finally obtains the same result as in Eq.~\eqref{Fsolution1} but with
\begin{eqnarray}
\left\langle \mathbf{z} \right\rangle_{\!F}\!(t) &=& \int_0^t \!\! d\tau \, \boldsymbol{\Phi}(t,\tau) \, \mathbf{F}_{\!\mathrm{eff}}(\tau)
\label{eq:qF2}.
\end{eqnarray}
Eqs.~\eqref{eq:qF1} and \eqref{eq:qF2} can be shown to be equivalent as follows. First, one rewrites Eq.~\eqref{eq:eff_force} as
\begin{equation}
\mathbf{F}_{\!\mathrm{eff}}(\tau)
=  \left[ \frac{d}{d\tau} + \boldsymbol{\mathcal{H}}(\tau) \right] \int_0^\tau \!\! d\tau \,
\boldsymbol{\Phi}(\tau\!-\!\tau') \mathbf{F}(\tau')
\label{eq:eff_force2}.
\end{equation}
Next, one substitutes Eq.~\eqref{eq:eff_force2} into Eq.~\eqref{eq:qF2}  and performs an integration by parts of the derivative term. Finally, one takes into account that $(d/d\tau) \left( \boldsymbol{\Phi}^{\!-1}(\tau) \right) \linebreak[3] = \boldsymbol{\Phi}^{\!-1}(\tau) \, \boldsymbol{\mathcal{H}}(\tau)$, which follows from Eq.~\eqref{eq:simplifiedH}, and the result in Eq.~\eqref{eq:qF1} is recovered.
Hence, we see that although $\boldsymbol{\Phi}(t,\tau)$ and $\boldsymbol{\Phi}(t\!-\!\tau)$ are different for nonlocal dissipation, this is exactly compensated by the contribution from the second term on the right-hand side of Eq.~\eqref{eq:eff_force}, which does not vanish in that case.

Note that just as all the temperature dependence appears entirely in the second cumulant, or covariance, the external force only affects the first cumulant, or mean.
Eq.~\eqref{Fsolution1} shows that the mean, $\langle \mathbf{z}\rangle (t)$, is shifted by $\langle \mathbf{z} \rangle_{\!F}(t)$, which characterizes the response to the driving force.
In fact, using Eq.~\eqref{eq:Phi} one can immediately see that it corresponds to shifting $\langle x \rangle$ and $\langle p \rangle$ respectively by $(G * F) (t)$ and $(M \dot{G} * F) (t)$, as one would expect.

\subsection{$N$-Oscillator Master Equation}

Our compact matrix notation allows a number of generalizations in a fairly straightforward fashion.
As an illustration we present the generalization of our results for the master equation and its solutions to the case of multiple system oscillators $\{ x_\alpha \}$ (which includes the case of a higher dimensional oscillator) with arbitrarily bilinear coupling to themselves and to the bath oscillators $\{y_j \}$.
We consider the system Lagrangian for $N$ oscillators and a generic bilinear term for the system-bath interaction:
\begin{align}
\mathsf{L}_\mathrm{s} &= \frac{1}{2} \left( \dot{\mathbf{x}}^\mathrm{T} \mathbf{M} \, \dot{\mathbf{x}} - \mathbf{x}^\mathrm{T} \mathbf{M}\boldsymbol{\Omega}^2 \, \mathbf{x} \right) , \label{eq:Nsystem}  \\
&= \frac{1}{2} \left( \dot{x}^\alpha M_{\alpha\beta} \dot{x}^\beta - x^\alpha M\Omega_{\alpha\beta}^2 x^\beta \right) , \nonumber\\
\mathsf{L}_\mathrm{int} &= \mathbf{y}^\mathrm{T} \mathbf{c} \, \mathbf{x}  =  y^i c_{i \beta} x^\beta  , \label{eq:Nbath}
\end{align}
where we used Einstein's summation convention for repeated indices and the matrix $\mathbf{c}$ connects system positions (denoted by Greek indices) to bath positions (denoted by Latin indices).
The matrices $\mathbf{M}$ and $\mathbf{M}\boldsymbol{\Omega}^2$ are symmetric and positive definite.
The eigenvalues of $\boldsymbol{\Omega}$ correspond to the normal-mode frequencies, as can be inferred from our Langevin equation~\eqref{eq:langevin-N}.

The effects of the environment for the generalized situation described by Eqs.~\eqref{eq:Nsystem}-\eqref{eq:Nbath} can be entirely encoded in a simple generalization of the spectral density as well as the noise and damping kernels:
\begin{align}
I_{\alpha\beta}(\omega) &= \sum_k \delta\!\left( \omega \!-\! \omega_k \right) \frac{c_{k \alpha} c_{k \beta}}{2 m_k \omega_k} , \label{eq:spectral_matrix} \\
\boldsymbol{\nu}(t,\tau) &= \int_0^\infty \!\!\! d\omega \, \mathbf{I}(\omega) \coth\!\left( \frac{\omega}{2T} \right) \cos\!\left[\omega(t\!-\!\tau)\right] , \\
\mathbf{M} \boldsymbol{\gamma}(t,\tau) &= \int_0^\infty \!\!\! d\omega \frac{\mathbf{I}(\omega)}{\omega} \cos\!\left[\omega(t\!-\!\tau)\right] .
\end{align}
In fact, one can directly specify the system-environment coupling by giving the spectral density matrix $\mathbf{I}(\omega)$, which must be symmetric and positive semi-definite, as implied by Eq.~\eqref{eq:spectral_matrix}.
After taking the Laplace transform, the Langevin equation in position space is then given by
\begin{align}
\mathbf{M} \left( s^2 + 2 s\, \hat{\boldsymbol{\gamma}}(s) + \boldsymbol{\Omega}^2 \right) \hat{\mathbf{x}}(s) &= \mathbf{M} \left( s \, \mathbf{x}_0 + \dot{\mathbf{x}}_0 \right) + \hat{\boldsymbol{\xi}}(s) \label{eq:langevin-N},
\end{align}
which can be solved via matrix inversion to find $\hat{\mathbf{G}}(s)$, with which one can construct $\boldsymbol{\Phi}(t)$ and $\boldsymbol{\sigma}_T(t)$.
However, closed-form evaluation of $\mathbf{G}(t)$ can be rather involved: even for local dissipation the two-oscillator problem requires factoring a fourth-order polynomial.
In general, the $N$-oscillator problem will require factoring a polynomial of order $2N$ for local dissipation and of order $2N+1$ or more for nonlocal dissipation.
We leave more thorough discussion to future work, where we will derive block-matrix equations for the positon and momentum parts in the phase-space representation analogous to those herein.

\section{Discussion}
\label{sec:discussion}

Quantum Brownian motion of an oscillator coupled to a thermal
reservoir of quantum oscillators has been  the canonical model for
the study of open quantum systems where one can use it to investigate
all the environmental effects on an open quantum system it interacts
with, even of macroscopic scale, such as quantum dissipation,
diffusion, decoherence and entanglement. It also provides important
information on quantum measurement, such as noise, fluctuations,
correlations, uncertainty relation and standard quantum limit in
mesoscopic systems. Many experiments have been carried out for
testing these processes. An exact master equation  was reported some
years ago \cite{HPZ92} governing the reduced density matrix of an open
quantum system coupled to a general environment of arbitrary spectral
density and temperature. Subsequently there have also been claims of
exact solutions \cite{FordOconnell01}. We have found many previous derivations
to be correct for local dissipation, but containing errors or
omissions for nonlocal dissipation; in their place we have presented
the most complete and correct derivation of the QBM master equation
to date. In this paper we report on solutions to this equation for a
fairly general set of physical conditions and a generalization of the
QBM master equation to a system with an arbitrary number of
oscillators. Most of the previous results required one to solve
integro-differential equations numerically, whereas we have reduced
everything to quadrature, which can be further simplified in many
cases using contour integration techniques.
We expect these results to be useful in realistic settings for the analysis of many problems which can be described by this model.

More specifically, we have found a compact expression for the general
solution of this master equation,  showing that at late times it
tends to a Gaussian state entirely characterized by its asymptotic
covariance matrix. For odd meromorphic spectral densities, and many
others, the result for this late-time covariance matrix can be
evaluated as a simple contour integral. As an example we provide
explicit exact nonperturbative results for an ohmic environment with
a finite cut-off which are valid for an arbitrarily strong coupling.
At sufficiently low temperatures and strong coupling this equilibrium
state becomes highly squeezed and the system becomes extremely
localized in position space, a phenomenon with potentially
interesting applications in the realm of mesoscopic systems.

The general solution of the master equation involves the matrix
propagator of a linear integro-differential equation. We have been
able to solve these equations exactly for several ohmic, sub-ohmic
and supra-ohmic environments with a finite cut-off and studied the
evolution of the system for finite times. This is achieved using
Laplace transforms and eventually transforming back to time domain.
From such exact (and simple) solutions for the propagator one gains
highly valuable information. For instance, one can justify that using
the local propagator is a valid approximation for the ohmic
environment in the large cut-off limit. This approximation leads to
great simplifications and we are then able to provide relatively
simple analytic expressions for the diffusion coefficients of the
master equation at all times. Similarly, our exact solutions for the
propagator in specific examples of sub-ohmic and super-ohmic
environments reveal a dominant contribution from nonlocal dissipation
effects. In the first case it is a consequence of long-time
correlations, due to the low-frequency modes of the environment, that
become important at late times. In contrast, the source of
nonlocality in the supra-ohmic case is the UV regulator function, and
it gives rise to a marked cut-off sensitivity of the momentum
covariance which had not been noticed so far. On the other hand, it
should be pointed out that although the  results for the exact
propagator of the integro-differential equation are rather simple,
some of the general expressions for the solutions of the master
equation are rather lengthy and have not been reported here. They
have, nevertheless, been employed to evaluate and plot the exact time
evolution of the thermal covariance for an ohmic environment with a
finite cut-off in Sec.~\ref{sec:jolts}.

It is important to discuss  the cut-off sensitivity of the late-time
covariance and diffusion coefficients for an ohmic environment in the
weak coupling regime. While $\sigma_{\!xx}^\infty$ is finite in the
infinite cut-off limit, $\sigma_{\!pp}^\infty$ depends
logarithmically on $\Lambda$ for large $\Lambda$ already at order
$\gamma_0$ [Eq.~\eqref{eq:latecov_pp}]. This means that it is
absolutely necessary to consider a finite cut-off. The kind of
divergences that appear otherwise cannot be dealt with by
renormalizing the frequency or other bare parameters of the theory.
In fact, as shown in Sec.~\ref{sec:uncertainty}, subtracting the
divergent term would lead to inconsistencies (violation of
Heisenberg's uncertainty principle). Furthermore, from the late-time
thermal covariance one can immediately obtain the late-time diffusion
coefficients as well (see the discussion at the end of
Sec.~\ref{sec:mod-time}). One finds then that both the normal and
anomalous diffusion coefficients are logarithmically sensitive to
large cut-offs. However, while this dependence appears in $D_{\!xp}$
[Eq.~\eqref{eq:weakDxp}] at order $\gamma_0$, in $D_{\!pp}$ it only
appears at order $\gamma_0^2$, and it had been missed in previous
analytic studies which treated $\gamma_0$ perturbatively to lowest
order.

We would also like to stress the following point. When studying an
ohmic environment with a finite but large cut-off, it can be a good
approximation to consider local dissipation (infinite cut-off limit
for the damping kernel) while keeping the cut-off finite in the noise
kernel. This has already been discussed above and justifies
calculations like those of Ref.~\cite{UnruhZurek89} up to corrections
suppressed by inverse powers of the cut-off. However, the opposite is
not true: it is essential to keep a finite cut-off in the noise
kernel to avoid the divergences discussed in the previous paragraph.
This is precisely the origin of the divergences and pathological
behavior found in Ref.~\cite{FordOconnell01}, where a finite cut-off was
employed in the damping kernel but not in the noise kernel. Instead
one should use the same spectral function everywhere, which means
having a finite cut-off in both kernels, and everything would be well
defined then.
Note that these divergences would appear in the momentum covariance even at asymptotically late times, as discussed in the previous paragraph.
There is a different kind of sensitivity to large values of the cut-off that is due to having started with a uncorrelated state for the system and the environment.
This gives rise to a jolt in the normal diffusion coefficient at early times of order $1/\Lambda$ with an amplitude proportional to $\Lambda$, as well as a logarithmic dependence on the cut-off of $\sigma_{\!xx}$ (and $\sigma_{\!pp}$) that decays exponentially with the relaxation time-scale $1/\Gamma$.
They would not be present if one had started with an appropriately correlated initial sate, and then prepared the system in a finite time (not suddenly). Alternatively, this can be implemented by switching on the system-environment interaction smoothly in a finite time much larger than $1/\Lambda$, but shorter than the other dynamical scales of the system.
\footnote{A detailed discussion is provided in \ref{sec:initial_coupling}.
There are plenty of technical details concerning the initial kick and the effect of the switch-on function on the damping kernel,
but the real key point is the effect of the switch-on function in the noise kernel,
when evaluating either the diffusion coefficients or the correlation functions (the covariance matrix).}

As a further generalization of the QBM master equation we have included the influence of external forces.
This modifies the dynamics by driving the mean position and momentum just as with a classical driven system (even for nonlocal dissipation).
In this model we found that the force has no effect upon the width of the wave-packet or any cumulant other than the mean.
These results may be useful for the study of low-temperature measurements of driven oscillators, which are relevant for experiments with nanomechanical resonators \cite{Naik06,Lahaye04}.
They also play a crucial role in future schemes for the detection of gravitational waves with high-intensity laser interferometers, where the radiation pressure effects on the cavity mirrors are important \cite{Kimble02,Buonanno01}.

Finally, we have extended the model of one quantum oscillator
bilinearly coupled to a thermal reservoir of oscillators to a model
of multiple oscillators bilinearly coupled to themselves and the bath
in an arbitrary fashion. With this generalization, the potential for
application  \cite{Paz08,Paz09} becomes almost endless and we leave
further study to future research~\cite{NQBM}.

\section*{Acknowledgement}
We would like to thank Juan Pablo Paz for interesting conversations
which encouraged us to look into the sub-ohmic case. C.~F.\ and
B.~L.~H.\ are  supported in part by grants from the NSF-ITR program
(PHY-0426696) and the DARPA-QuEST program (DARPAHR0011-09-1-0008).
A.~R.\  was partly supported by LDRD funds from Los Alamos National
Laboratory and by a DOE grant.

\appendix

\section{Special Functions}

\subsection{Harmonic Number}
\label{sec:harmonic}

The \emph{harmonic number} $\mathrm{H}(n)$ is a function similar to a logarithm, whose definition and main properties are
\begin{eqnarray}
\mbox{H}(n) & = & \sum_{k=1}^n \frac{1}{k}\, , \qquad n \in \mathbb{Z}^+ \\
\mbox{H}(0) & = & 0 , \\
\gamma_\mathrm{E} & = & \lim_{n \rightarrow \infty} [\mbox{H}(n)-\log{(n)})],
\end{eqnarray}
where $\gamma_\mathrm{E}$ is known as the Euler-Mascheroni constant.
Its generalization to the complex plane exhibits similar properties and is given by
\begin{equation}
\mbox{H}(z) = \gamma_\mathrm{E} + \psi(z\!+\!1) ,  \qquad z \in \mathbb{C},
\end{equation}
where $\psi(z)$ is the \emph{digamma function}, defined as
\begin{equation}
\psi(z) = \frac{\Gamma'(z)}{\Gamma(z)}.
\end{equation}
It satisfies the recurrence relation
\begin{equation}
\psi(z+1) = \psi(z) + \frac{1}{z} ,
\end{equation}
and its Taylor expansion around $1$ as well as its asymptotic expansion for $|z| \to \infty$ are given respectively by
\begin{align}
\psi(z\!+\!1)  =&\;  -\gamma_\mathrm{E} + \sum_{k=1}^\infty \zeta(k\!+\!1) (-z)^k \nonumber \\
& \mathrm{~for~} |z|<1 , \\
\psi(z)  \sim&\;  \ln z -\frac{1}{2 z} - \frac{1}{12 z^2} + \cdots \nonumber \\
& \mathrm{~if~} \left| \arg{(z)} \right| < \pi \label{psi2} ,
\end{align}
where $\zeta(n)$ is the Riemann zeta function.

\subsection{Exponential Integral}
\label{sec:exp_int}

The \emph{exponential integral} is a special function which is defined for $\left|\arg(z)\right| < \pi$ as
\begin{equation}
\mathrm{E}_1(z) = \int_z^\infty \frac{e^{-z'}}{z'} dz' ,
\end{equation}
and has a branch cut along $\left|\arg(z)\right| = \pi$.
Its series expansion is
\begin{equation}
\mathrm{E}_1(z) =   -\gamma_\mathrm{E} - \ln z
- \sum_{n=1}^\infty \frac{(-1)^n}{n \, n!} z^n,
\end{equation}
and its asymptotic expansion for $|z| \to \infty$ is given by
\begin{equation}
\mathrm{E}_1(z) = \frac{e^{-z}}{z} \left( 1 - \frac{1}{z} + \frac{2}{z^2}
+ \cdots \right) .
\end{equation}



\subsection{Error Function}
\label{sec:erfc}

The \emph{error function} is defined as
\begin{equation}
\mathrm{erf}(z) = \frac{2}{\sqrt{\pi}} \int_0^z \!\! e^{-w^2} dw ,
\end{equation}
where the path integration is subject to the restriction
$\lim_{|w|\to\infty} |\mathrm{arg}(w)| < \pi/4$. In addition, the \emph{complementary error function} is defined as
\begin{equation}
\mathrm{erfc}(z) = \frac{2}{\sqrt{\pi}} \int_z^\infty \!\! e^{-w^2} dw = 1 - \mathrm{erf}(z) .
\end{equation}
The series expansion is
\begin{equation}
\mathrm{erf}(z) = \frac{2}{\sqrt{\pi}}\sum_{n=1}^\infty \frac{(-1)^n}{(2n\!+\!1)n!} z^{2n+1} ,
\end{equation}
and the asymptotic expansion for $|z| \to \infty$ (and $|\mathrm{arg}(z)| < 3\pi/4$) is given by
\begin{equation}
\mathrm{erfc}(z) = \frac{e^{-z^2}}{\sqrt{\pi}\, z} \left( 1 - \frac{1}{2z^2}
+ \frac{3}{4 z^4} + \cdots \right) \label{eq:erfc_asymp} ,
\end{equation}
which along with the fact that erf$(z)$ is odd, is sufficient to create an accompanying asymptotic expansion for $|\mathrm{arg}(z)| > 3\pi/4$

\section{Some properties of Laplace Transforms}
\label{sec:laplace}

Given a real function $f(t)$, defined for all real numbers $t\geq0$, its Laplace transform is defined as
\begin{equation}
\hat{f}(s) = \mathcal{L}\big\{f(t)\big\}(s) = \int_{0^-}^\infty \!\! e^{-s t} f(t) dt ,
\end{equation}
where the one-sided limit from the left for the lower limit of integration is chosen so that the transform of the Dirac delta function is one, i.e. $\mathcal{L}\{\delta(t)\}=1$.
The main properties used in the paper are the following. First, the Laplace transform of a derivative is given by
\begin{equation}
\mathcal{L}\!\left\{\dot{f}(t)\right\}\!(s) = s \hat{f}(s) - f(0) .
\end{equation}
And from this one can easily infer that the Laplace transform of an integral:
\begin{equation}
\mathcal{L}\!\left\{\int_0^t \!\! d\tau \, f(\tau) \right\}\!(s) = \frac{1}{s} \hat{f}(s) .
\end{equation}
Second, multiplying $f(t)$ by an exponential corresponds to a translation of the Laplace transform:
\begin{equation}
\mathcal{L}\!\left\{e^{at} f(t)\right\}\!(s) = \hat{f}(s\!-\!a) .
\end{equation}
Third, if the inverse Laplace transform of $\hat{f}(s)$ is $f(t)\, \theta(t)$, multiplying $\hat{f}(s)$ by an exponential corresponds to a translation of the inverse Laplace transform:
\begin{equation}
\mathcal{L}^{-1} \!\left\{e^{as} \hat{f}(s) \right\}\!(s) = f(t\!+\!a)\, \theta(t\!+\!a) .
\end{equation}
Fourth, the Laplace transform of a \emph{Laplace convolution} is given by the product of the Laplace transforms:
\begin{equation}
\mathcal{L}\big\{(f*g)(t)\big\}(s) = \hat{f}(s)\, \hat{g}(s) ,
\end{equation}
where
\begin{equation}
(f*g)(t) = \int_0^t dt' f(t\!-\!t')g(t') .
\end{equation}
Fifth, the \emph{initial value theorem} relates the initial value of a function $f(t)$ and the infinite limit of its Laplace transform as follows:
\begin{equation}
f(0^+) = \lim_{s \to \infty} s \hat{f}(s) \label{eq:initial_value}.
\end{equation}
Sixth, the \emph{final value theorem} relates the infinite limit of a function $f(t)$ and the initial value of its Laplace transform as follows:
\begin{equation}
f(\infty) = \lim_{s \to 0} s \hat{f}(s)  \label{eq:final_value},
\end{equation}
provided that all the poles of $\hat{f}(s)$ are on the $\mathrm{Re}(s) < 0$ half of the $s$ complex plane.\\
Seventh, the inverse Laplace transform of $\hat{f}(s)$ can be calculated using \emph{Bromwich's integral}, which involves an analytic continuation of $\hat{f}(s)$ in the complex plane:
\begin{equation}
f(t) = \mathcal{L}^{-1} \big\{\hat{f}(s)\big\}(s) = \frac{1}{2\pi \imath}
\int_{\!\alpha - \imath\infty}^{\alpha + \imath\infty} \!\! e^{s t} \hat{f}(s) \, ds ,
\end{equation}
where $\alpha$ is a real number chosen so that the integration path lies within the region of convergence of $\hat{f}(s)$, i.e., $\alpha > \mathrm{Re}(s_j)$ for every singularity $s_j$ of $\hat{f}(s)$.

Bromwich's integral illustrates the close relationship between the Laplace transform and the Fourier transform through analytic continuation. However, even if all the singularities of $\hat{f}(s)$ lie on the $\mathrm{Re}(s) < 0$ half of the complex plane, the relation is not direct because the Laplace transform involves an integral with domain $[0,\infty)$ rather than $(-\infty,\infty)$. The precise relationship can be understood as follows. Consider a real function $f(t)$ defined for all real values of $t$ and whose Fourier transform is $\tilde{f}(\omega).$ It is useful to define the following additional Fourier transforms:
\begin{equation}
\tilde{f}_\pm(\omega) = \int_{\!-\infty}^{+\infty} \!\!\! dt\, e^{-\imath \omega t} f(t) \,\theta(\pm t) \,,
\label{eq:fourier_pm}
\end{equation}
such that $\tilde{f}(\omega) = \tilde{f}_+(\omega) + \tilde{f}_-(\omega)$ and which satisfy the property $\tilde{f}_\pm(-\omega) = \big(\tilde{f}_\pm(\omega)\big)^*$ since $f(t)$ is real. Assuming that the Laplace transform $\hat{f}(s)$ has no singularities for $\mathrm{Re}(s) > 0$, it can be related by analytic continuation to $\tilde{f}_+(\omega)$:
\begin{equation}
\tilde{f}_+(\omega) = \lim_{\epsilon \to 0} \hat{f}(\epsilon \!+\! \imath\omega) \label{eq:laplace_fourier1}.
\end{equation}
If $f(t)$ is an even function, one has $\tilde{f}_-(\omega) = \tilde{f}_+(-\omega)$, and using Eq.~\eqref{eq:laplace_fourier1} one can then write
\begin{equation}
\tilde{f}(\omega) = \tilde{f}_+(\omega) +  \tilde{f}_+(-\omega) = \lim_{\epsilon \to 0} \Big[ \hat{f}(\epsilon \!+\! \imath\omega) + \hat{f}(\epsilon \!-\! \imath\omega) \Big]
\label{eq:laplace_fourier2}.
\end{equation}
Similarly, if $f(t)$ is an odd function, one has $\tilde{f}_-(\omega) = - \tilde{f}_+(-\omega)$, which implies
\begin{equation}
\tilde{f}(\omega) = \tilde{f}_+(\omega) -  \tilde{f}_+(-\omega) = \lim_{\epsilon \to 0} \Big[ \hat{f}(\epsilon \!+\! \imath\omega) - \hat{f}(\epsilon \!-\! \imath\omega) \Big]
\label{eq:laplace_fourier3}.
\end{equation}

\section{System-Environment Interaction and Renormalization}
\label{sec:divergences}

\subsection{Frequency Renormalization and the Damping Kernel}

In this subsection we discuss the relationship between the Lagrangians in Eqs.~\eqref{eq:lagrangian1} and \eqref{eq:switch}. The question of frequency renormalization plays a central role in this discussion and it will be analyzed by rewriting the equation of motion in terms of the damping kernel, given by Eq.~\eqref{damping2}, rather than the dissipation kernel. Here we will take $\theta_\mathrm{s}(t)=1$ and leave the examination of effects due to a non-vanishing switch-on time for the next subsection.

If one starts with the Lagrangian in Eq.~\eqref{eq:lagrangian1}, the homogeneous part of the Langevin integro-differential equation analogous to Eq.~\eqref{eq:integro-diff1} is then 
\begin{equation}
(L \cdot x)(t) = M \ddot{x}(t) + M\Omega^2_\mathrm{bare}\, x(t) + 2\!\int_0^t \! d\tau \, \mu(t\!-\!\tau) \, x(\tau) . \label{eq:integro-diff4}
\end{equation}
Using the expression for the dissipation kernel in terms of the damping kernel, $\mu(t\!-\!\tau) = M (\partial/\partial t) \gamma(t\!-\!\tau)$, and integrating by parts, one obtains
\begin{align}
(L \cdot x)(t) =&\; M \ddot{x}(t) + 2M\!\!\int_0^t \!\! d\tau \, \gamma(t\!-\!\tau) \, \dot{x}(\tau) \label{eq:integro-diff5} \\
&+ M \big(\Omega^2_\mathrm{bare} \!-\! \delta\Omega^2\big)\, x(t) + 2M \gamma(t) \, x(0)  \nonumber ,
\end{align}
where the time-independent frequency renormalization $\delta\Omega^2$ is given by
\begin{equation}
\delta\Omega^2 = 2\, \gamma(0) = \frac{2}{M} \int_0^\infty \!\! d\omega \, 
\frac{I(\omega)}{\omega}
\label{eq:freq_ren} .
\end{equation}
By choosing a bare frequency with an appropriate counterterm in order to cancel the frequency renormalization, $\Omega^2_\mathrm{bare} = \Omega^2 + \delta\Omega^2$, one would be finally left with Eq.~\eqref{eq:integro-diff2} and an effective frequency for the system oscillator $\Omega$.
This is the same equation that one obtains if one starts with the Lagrangian in Eq.~\eqref{eq:switch} and takes $\theta_\mathrm{s}(t)=1$, which can be easily understood as follows. Recalling the definition of the spectral function $I(\omega)$ in terms of the coupling constants $c_n$, it is immediate to see that the square of the last term on the right-hand side of Eq.~\eqref{eq:switch} gives $-(1/2)M \delta\Omega^2 x^2$. Therefore, the Lagrangians in Eqs.~\eqref{eq:lagrangian1} and \eqref{eq:switch} are equivalent provided that one makes the choice mentioned above for $\Omega^2_\mathrm{bare}$.

\subsection{Initial-Time Divergences, Coupling Switch-on and Initial-State Distortion}
\label{sec:initial_coupling}

\subsubsection{Initial-Time Divergences and Coupling Switch-on}
The derivation of the HPZ master equations relies upon the key assumption that the system and environment are initially uncorrelated.
For an ohmic environment, this gives rise to an initial ``jolt'' in the normal diffusion coefficient of the master equation with a characteristic time-scale of order $\Lambda^{\!-1}$ and an amplitude proportional to $\Lambda$.
Similarly,
the frequency $\Omega^2_\mathrm{R}(t)$ in the master equation starts with a large value of the order of $\Lambda$ and decreases to moderate values in a time of order $\Lambda^{\!-1}$.

The physical origin of the jolts in the coefficients of the master
equation as well as other initial time divergences, such as the
divergent contributions to correlation functions of system observables
that are due to divergent boundary terms at the initial time (see
Appendix~D in Ref.~\cite{HRV04}), can be understood as follows. In
general when a system couples to an environment with an infinite
number of modes, well-behaved states exhibit correlations with
arbitrarily high-frequency modes. In contrast, states that are
uncorrelated for sufficiently high frequencies (such as completely
factorizable states) are pathological. For instance, in the limit of infinite cut-off they have infinite energy (even with an origin of energies such that the ground state of the whole interacting system has vanishing energy) and their Hilbert space is unitarily
inequivalent to the space of physical states, spanned by the basis of
energy eigenvectors of the whole system Hamiltonian including the
system-environment interaction. (Of course for a finite UV cut-off
there are no divergences or unitary inequivalence, but the potentially divergent terms are very
sensitive to changes in the value of the cut-off.) Physically
acceptable initial states that correspond to the thermal equilibrium
state for the whole system can be obtained using Euclidean path
integrals \cite{Grabert88}. However, the instantaneous preparation
functions employed in Ref.~\cite{Grabert88} to produce other states in
addition to the thermal equilibrium state still give rise to initial
divergences, as explained in Ref.~\cite{Romero97}. In order to obtain
finite results, one needs to prepare the new initial state within a
non-vanishing time \cite{Anglin97}, which corresponds to a physically
more realistic situation. The alternative approach that we follow here
is to switch on the system-environment interaction smoothly within a time $t_\mathrm{s}$ much
longer than $\Lambda^{\!-1}$ but shorter than any other relevant
time-scale of the problem. In this way the factorized initial state,
which is perfectly acceptable in the uncoupled case, becomes
adequately correlated with the arbitrarily high-frequency modes in a
regular fashion.

When adding the short time switch-on function to the
system-environment coupling to turn on the interaction gradually,
as in Eq.~\eqref{eq:switch}, the initial
jolt is no longer present in the results for the diffusion
coefficients, which behave smoothly during the switch-on time.
Furthermore, for times much longer than $t_\mathrm{s}$ the contribution to Eq.~\eqref{eq:diff_simple} from the switch-on period is negligible and one can simply use that equation without including any switch-on function. This point is implicitly exploited throughout the paper: unless explicitly stated, our calculations of the diffusion coefficients do not take into account the switch-on functions and the results for those coefficients should only be regarded as valid for times sufficiently larger than $t_\mathrm{s}$, while their values during that period should be smoothly interpolated so that they vanish at the initial time.

Either the quick transition from the bare frequency to the renormalized one (in the absence of a smooth switch-on function) or switching on the interaction in a finite time $t_\mathrm{s}$ can have a non-negligible effect on the homogenous solutions of the Langevin equation even for times much larger than $\Lambda^{\!-1}$ or $t_\mathrm{s}$.
Fortunately, as we will show in the remaining subsections, the effect can be entirely accounted for by a finite shift of the initial momentum and the corresponding transformation of the initial state.

\subsubsection{Initial Kick (finite cut-off, vanishing switch-on time)}
\label{sec:kick_NS}

We start by considering the case in which there is \textbf{no switch-on} time and analyze the effect of the \emph{slip} term, which corresponds to the last term on the right-hand side of Eq.~\eqref{eq:integro-diff2} for the Langevin operator. It can be interpreted as a transient driving force
\begin{eqnarray}
F_{\!\gamma}(t) &=& 2 \gamma(t) x_0 ,
\end{eqnarray}
whose contribution to the solution is simply adding a term $G(t)*F_{\!\gamma}(t)$.
The infinite cut-off limit can be analyzed using distributions in the time domain or working with Laplace transforms. As derived in Eqs.~\eqref{eq:laplace0}-\eqref{eq:green1}, the solution of Eq.~\eqref{eq:integro-diff2} (including the slip term) is given in Laplace space by
\begin{eqnarray}
\hat{x}(s) &=& M \left( s x_0 + \dot{x}_0 \right)\hat{G}(s) + \hat{G}(s) \hat{\xi}(s) ,
\end{eqnarray}
whereas one can easily infer that the solution without the slip term would be
\begin{equation}
\hat{y}(s) = M \left( s y_0 + \dot{y}_0 + 2 \hat{\gamma}(s) \, y_0 \right)\hat{G}(s) + \hat{G}(s) \hat{\xi}(s) . \label{eq:NoSlip}
\end{equation}
In the limit of local dissipation (large cut-off limit) $\hat{\gamma}(s)=\gamma_0$ and one can see that the effect of the slip term is an initial \emph{kick} to the homogeneous solutions, whose values before and after the kick can be related via
\begin{eqnarray}
y_0 &=& x_0 , \label{eq:DistortX} \\
\dot{y}_0 &=& \dot{x}_0 - 2 \gamma_0 x_0 . \label{eq:DistortV}
\end{eqnarray}
This induces a distortion of the reduced Wigner function associated with the transformation $\dot{x}_0 \to \dot{x}_0 - 2\gamma_0 x_0$ which occurs within the cut-off timescale.
The effect of such an initial kick can be entirely absorbed in a redefinition of the initial state, as will be discussed in Sec.~\ref{sec:state_distortion}.

\subsubsection{Initial Kick (large cut-off, non-vanishing switch-on time)}
\label{sec:kick_NS2}

Next, we consider the case with a non-vanishing \textbf{switch-on} time $t_\mathrm{s}$ and smooth switch-on function such that $\theta_\mathrm{s}(0)=0$ and $\theta_\mathrm{s}(t)=1$ for $t\geq t_\mathrm{s}$.
Integrating the dissipation kernel by parts in Eq.~\eqref{eq:integro-diff1}, the homogeneous part of the Langevin equation becomes
\begin{align}
& \ddot{x}(t) + 2\! \int_0^t \!\! d\tau \, \gamma_\mathrm{P}(t,\tau) \, \dot{x}(\tau) + \Omega^2 x(t) = \nonumber \\
& - 2\,\theta_\mathrm{s}(t)\! \int_0^t \!\! d\tau \, \gamma(t\!-\!\tau) \, \dot{\theta}_\mathrm{s}(\tau) \, x(\tau) \label{eq:eom4}, 
\end{align}
in terms of the positive-definite kernel
\begin{equation}
\gamma_\mathrm{P}(t,\tau) \equiv \gamma(t\!-\!\tau) \, \theta_\mathrm{s}(t) \theta_\mathrm{s}(\tau) ,
\end{equation}
where we have not made any approximations concerning the timescales of the dissipation kernel and switch-on function yet,
but have expressed our result in terms of the damping kernel defined in Eq.~\eqref{damping2} and taken into account Eq.~\eqref{eq:freq_ren}.
Either in the limit of local dissipation or vanishing switch-on time, the term on the right-hand side of Eq.~\eqref{eq:eom4} gives rise to a slip term analogous to that found in the previous subsection. This is because for $x(t)$ evolving slowly compared to $\Lambda^{\!-1}$ or $t_\mathrm{s}$, one has a a convolution of the distributions $\gamma(t)$ and $\dot{\theta}_\mathrm{s}(t)$, which is also a distribution localized near the initial time. In particular, it is immediate to see that the results of Sec.~\ref{sec:kick_NS} are recovered in the limit of vanishing $t_\mathrm{s}$ [see the remark below about $\dot{\theta}_\mathrm{s}(t)$ in that limit].

If we take the high cut-off limit $\gamma(t\!-\!\tau) = \gamma_0\, \delta(t\!-\!\tau)$, which should be a good approximation for $\Lambda \gg t_\mathrm{s}^{-1}$, the right-hand side of Eq.~\eqref{eq:eom4} takes a simple form and we are left with
\begin{align}
\ddot{x}(t) + 2 \gamma_0 \, \theta_\mathrm{s}^2(t) \, \dot{x}(t) + \Omega^2 x(t) &= - \gamma_0 \, \delta_\mathrm{s}(t) \, x(t) \label{eq:eom5} \\
\delta_\mathrm{s}(t) & \equiv  \frac{d}{dt} \theta_\mathrm{s}^2(t),
\end{align}
where $\delta_\mathrm{s}(t)$ is a representation of the delta function in the limit of vanishing switch-on time; however, its support is entirely contained in the $t\geq0$ interval, so that $\int_0^\infty \delta_\mathrm{s}(t) dt = 1$.
[We also took the local dissipation limit on the left-hand side of Eq.~\eqref{eq:eom5} for simplicity, but we needn't have done so: that is a slowly varying term which does not play an important role here.]
For a very rapid switch-on function we have $\delta_\mathrm{s}(t) \, x(t) \, \approx \, \delta_\mathrm{s}(t) \, x(0)$ and this term produces an initial kick, $\dot{x}_0 \to \dot{x}_0 - \gamma_0 x_0$, analogous to that described in the previous subsection.
This kick of the homogeneous solutions will produce a distortion of reduced Wigner function which occurs within the switch-on timescale.
For times much larger than $t_\mathrm{s}$, this effect can also be entirely absorbed into a redefinition of the initial state, as described in the next subsection.

\subsubsection{Initial-State Distortion}
\label{sec:state_distortion}

In Sec.~\ref{sec:kick_NS} we calculated that in a particular limit of $\Omega \ll \Lambda \ll t_\mathrm{s}^{-1}$ one obtains a kick to the initial state of $\dot{x}_0 \to \dot{x}_0 - 2\gamma_0 x_0$ which occurs within the slower cut-off timescale.
Whereas in Sec.~\ref{sec:kick_NS2} we calculated that in a particular limit of $\Omega \ll t_\mathrm{s}^{-1} \ll \Lambda$ one obtains a kick to the initial state of $\dot{x}_0 \to \dot{x}_0 - \gamma_0 x_0$ which occurs within the slower switch-on timescale.
From the exact relation in Eq.~\eqref{eq:eom4}, if one tries to enforce both high cut-off and short switch-on time then there will be a kick $\dot{x}_0 \to \dot{x}_0 - c \gamma_0 x_0$ which occurs in the slower of the cut-off and switch-on timescales.
And if the stationary damping kernel $\gamma(t)$ and switch-on function's derivative $\dot{\theta}_\mathrm{s}(t)$ are suitably well-behaved distributions,
then this kick is bounded so that $0 \leq c \leq 2$.

From these results one might be tempted to consider modifying the Lagrangian by introducing a suitable time-dependent frequency renormalization counterterm $\delta\Omega_\mathrm{kick}^2(t) = - c \gamma_0 \delta(t)$.
However, even though an appropriate choice of time-dependent counterterm could compensate and effectively remove the effect of the initial kick in either case,
a truly finite cut-off is still necessary to have a finite thermal covariance,
and the switch-on function for the system-environment interaction is still essential to avoid the highly cut-off sensitive initial jolt in the normal diffusion coefficient
and other irregularities associated with an uncorrelated initial state
[the key point in these cases is the dependence on the switch-on function of the noise kernel shown in Eq.~\eqref{eq:noise}].

Moreover, the effect of any such kick can easily be accounted for by simply distinguishing between the ``bare'' initial state before the kick and the ``renormalized'' state immediately after the kick.
Following the approach in Ref.~\cite{CRV03} one can easily see that this initial kick translates into a distortion of the Wigner distribution from the bare initial state to a shifted one
\begin{equation}
W_{\!\mathrm{bare}}(x,p) \to W_{\!\mathrm{ren}}(x,p) = W_{\!\mathrm{bare}}(x,p \!-\! c M \gamma_0 x) \label{eq:phase_trans1} .
\end{equation}
This phase-space transformation has a Jacobian matrix $\mathbf{K}$ with determinant equal to one:
\begin{equation}
\mathbf{K} = \left[ \begin{array}{cc} 1 & 0 \\ - c M \gamma_0 & 1 \end{array} \right] \qquad  \det \mathbf{K} = 1 . \label{eq:jacobian}
\end{equation}
Therefore, it is simple to calculate renormalized expectation values
in terms of bare expectation values and vice versa:
\begin{eqnarray}
\langle A(x,p)\rangle_{\!\!\!{\mathrm{ren} \atop \mathrm{or}} \atop \mathrm{bare}} &=& \iint \!\! dx dp \, A(x,p) \, W_{\!\!\!\!{\mathrm{ren} \atop \mathrm{or}} \atop \mathrm{bare}} \!\!(x,p) , \\
\langle A(x,p) \rangle_\mathrm{ren} &=& \langle A(x,p \!+\! c M \gamma_0 x) \rangle_\mathrm{bare} \, .
\end{eqnarray}
We can immediately see that the normalization, linear entropy (see Sec.~\ref {sec:entropy}) and state overlap are all unchanged by the kick.
We can also check explicitly that the Heisenberg uncertainty relation is preserved as follows.
First, we start with the covariance matrix for $x$ and $p$ corresponding to the Wigner distribution
\begin{equation}
\boldsymbol{\sigma} = \left[ \begin{array}{cc} \sigma_{xx} & \sigma_{xp} \\ \sigma_{px} & \sigma_{pp} \end{array} \right] , \label{eq:covariance}
\end{equation}
with $\sigma_{xx} = \langle xx \rangle_\mathrm{ren}$, $\sigma_{xp} = \sigma_{px} = \langle xp \rangle_\mathrm{ren}$ and $\sigma_{pp} = \sigma_{pp} = \langle pp \rangle_\mathrm{ren}$,
and which transforms in the following way under linear phase-space transformations:
\begin{equation}
\boldsymbol{\sigma} \to \mathbf{K} \, \boldsymbol{\sigma} \, \mathbf{K}^\mathrm{T} \, .
\end{equation}
Hence, from Eq.~(\ref{eq:jacobian}) we have
\begin{equation}
\det{\boldsymbol{\sigma}}_\mathrm{bare} = \det{\boldsymbol{\sigma}}_\mathrm{ren} \, .
\end{equation}
Finally, one takes into account that
\begin{equation}
(\det{\boldsymbol{\sigma}}) \geq \frac{\hbar^2}{4} \label{eq:uncertainty1},
\end{equation}
corresponds to the formulation in terms of the Wigner function of the generalized Heisenberg uncertainty relation due to Schr\"odinger \cite{Robertson34,Trifonov02}:
\begin{equation}
\left\langle \Delta x^2 \right\rangle \left\langle \Delta p^2 \right\rangle - \left\langle \frac{1}{2} \{ \Delta x ,\Delta p \} \right\rangle^2 \geq \frac{\hbar^2}{4} \label{eq:uncertainty2},
\end{equation}
where $\{\hat{A},\hat{B}\} \equiv \hat{A}\hat{B} + \hat{B}\hat{A}$.

Furthermore, by switching to the density matrix formalism, we can see that pure states are mapped to pure states and positivity is preserved.
It is a straightforward calculation to show that
\begin{align}
\rho_\mathrm{bare}(x,y) &\to \rho_\mathrm{ren}(x,y) , \\
\rho_\mathrm{ren}(x,y) &= e^{+\imath \frac{c M \gamma_0}{2}x^2} \, \rho_\mathrm{bare} (x,y) \, e^{-\imath\frac{c M \gamma_0}{2}y^2} .
\end{align}
Therefore, if we start in a pure state, which acts as a projection operator
\begin{eqnarray}
\rho_\mathrm{bare}^2 &=& \rho_\mathrm{bare} \, ,
\end{eqnarray}
then it is fairly easy to see that this will hold for the distorted state.
Additionally, given the positivity condition
\begin{eqnarray}
\langle\psi| \rho_\mathrm{bare} |\psi\rangle &\geq& 0 ,
\end{eqnarray}
for all vectors $|\psi\rangle$, then it is easy to see that the distorted state will also fulfill this condition by simply considering the vectors $e^{\imath c M \gamma_0 x^2/2} \psi(x)$ in position representation.

In summary, the new Wigner function that results from the transformation defined by Eq.~(\ref{eq:phase_trans1}) always corresponds to a physical density matrix since the transformation preserves the normalization and the real-valuedness of the Wigner function (implying the normalization and hermiticity of the density matrix) as well as the positivity of the associated density matrix.
Therefore, if one is interested in analyzing the evolution of a certain state of the system better correlated with the environment, one can simply take such a state as $W_{\!\mathrm{ren}}(x,p)$ and study its evolution for $t \gg \max[t_\mathrm{s},\Lambda^{\!-1}]$ by considering the Langevin equation without the term that gives rise to the initial kick.
However, given any $W_{\!\mathrm{ren}}(x,p)$ it is always possible to follow in detail the evolution during the switch-on time by inverting Eq.~(\ref{eq:phase_trans1}) to obtain the corresponding initial Wigner function before the interaction was switched on and using the full Langevin equation with the contribution from the right-hand side of Eq.~(\ref{eq:eom4}) included.
In general this approach can be regarded simply as a formal procedure to generate a better correlated initial state, but the explicit construction involving unitary evolution for the whole system at all times guarantees that the result is well defined (i.e.\ the exact solutions of the master equation obtained in this way preserve the positivity of the density matrix).%
\footnote{Using this approach the system-environment correlations at high frequencies will be the same as those of other properly correlated states (such as the global equilibrium states considered in Ref.~\cite{Grabert88} or states prepared from those in a finite time).
However, in general the correlations for low frequencies will differ and the states of the whole system plus environment will not be equivalent even if their reduced Wigner functions are the same. In particular this implies that even if the reduced Wigner functions of the two states coincide at some given time,
they will in general evolve differently (until thermal equilibrium for the whole system is reached).}


\section{Peculiarities of Propagators and Green Functions Associated with
Integro-differential Equations}
\label{sec:nonlocal_prop}

In this Appendix we discuss a subtle mathematical point which, to the best of our knowledge, has been missed in the existing literature on master equations of QBM models. It has to do with properties of Green functions which are satisfied for ordinary differential equations but not for integro-differential equations. Thus, it becomes particularly relevant whenever the nonlocal aspects of the dissipation kernel cannot be neglected.

Consider an integro-differential equation of the form
\begin{equation}
\dot{\mathbf{z}}(t) + \int_0^t \!\! d\tau \, \boldsymbol{\mathsf{H}}(t\!-\!\tau) \, \mathbf{z}(\tau) = \boldsymbol{\xi}(t) \label{eq:langevin3},
\end{equation}
with the kernel $\boldsymbol{\mathsf{H}}(t\!-\!\tau)$ given by Eq.~\eqref{eq:langevin_kernel}. Its solutions can be written as
\begin{equation}
\mathbf{z}(t) = \boldsymbol{\Phi}(t) \, \mathbf{z}_0
+ (\boldsymbol{\Phi} * \boldsymbol{\xi})(t) , \label{eq:solution3}
\end{equation}
where $\mathbf{z}_0$ specifies the initial conditions and the matrix propagator $\boldsymbol{\Phi}(t)$ is given by Eq.~\eqref{eq:Phi(t)}.
As far as the homogeneous solutions are concerned, the values of a solution at two different times $\tau$ and $t$ are related by the \emph{transition matrix} $\boldsymbol{\Phi}(t) \, \boldsymbol{\Phi}^{\!-1}(\tau)$.
On the other hand, for some given initial conditions the inhomogeneous solutions are obtained by integrating the source with the \emph{retarded matrix propagator} $\boldsymbol{\Phi}_{\!\mathrm{ret}}(t\!-\!\tau) = \boldsymbol{\Phi}(t\!-\!\tau) \, \theta(t\!-\!\tau)$, as shown in Eq.~\eqref{eq:solution3}.

In the case of a linear differential equation (i.e.\ for a local damping kernel), the retarded matrix propagator and the transition matrix are related in a simple way: $\boldsymbol{\Phi}_{\!\mathrm{ret}}(t\!-\!\tau) = \boldsymbol{\Phi}(t) \, \boldsymbol{\Phi}^{\!-1}(\tau) \, \theta(t\!-\!\tau)$.
This can be seen by realizing that $\boldsymbol{\Phi}(t) \, \theta(t\!-\!\tau)$ satisfies the differential equation except for a delta function that results from differentiating the theta function, and that the two expressions are equal to the identity matrix at $t=\tau$.
In contrast, for an integro-differential equation (a nonlocal damping kernel) $\boldsymbol{\Phi}(t) \, \boldsymbol{\Phi}^{\!-1}(\tau) \, \theta(t\!-\!\tau)$ no longer corresponds to the retarded matrix propagator because $\boldsymbol{\Phi}(t) \, \theta(t\!-\!\tau)$ does not satisfy the integro-differential equation, which can be seen (for $t>0$) as follows:
\begin{align}
\dot{\boldsymbol{\Phi}}(t) = -\!\!\int_0^t \!\! d\tau' \, \boldsymbol{\mathsf{H}}(t\!-\!\tau') \, \boldsymbol{\Phi}(\tau')
&\neq -\!\!\int_\tau^t \!\! d\tau' \, \boldsymbol{\mathsf{H}}(t\!-\!\tau') \, \boldsymbol{\Phi}(\tau')  \label{eq:inequality} ,
\end{align}
where the right-hand side equates to
\begin{align}
-\!\!\int_\tau^t \!\! d\tau' \, \boldsymbol{\mathsf{H}}(t\!-\!\tau') \, \boldsymbol{\Phi}(\tau') &= -\!\!\int_0^t \!\! d\tau' \, \boldsymbol{\mathsf{H}}(t\!-\!\tau') \, \boldsymbol{\Phi}(\tau') \, \theta(\tau'\!\!-\!\tau).
\end{align}
The discrepancy is due to a term of the form $\int_0^\tau \! d\tau' \, \boldsymbol{\mathsf{H}}(t\!-\!\tau') \, \boldsymbol{\Phi}(\tau')$ (with $t>\tau$), which vanishes in the case of nonlocal damping kernel and hence a nonlocal kernel $\boldsymbol{\mathsf{H}}(t\!-\!\tau')$, but does not vanish in the nonlocal case.
On the other hand, $\boldsymbol{\Phi}_{\!\mathrm{ret}}(t\!-\!\tau)$ does satisfy the integro-differential equation with a delta source, as it should. This point, which can be alternatively seen in Laplace space fairly easily, follows from the fact that $\boldsymbol{\Phi}(t)$ is a solution of  the integro-differential equation by construction, together with the translational invariance of this kind of solutions [i.e.\ if $\boldsymbol{\Phi}(t)$ is a solution, $\boldsymbol{\Phi}(t\!-\!\tau)$ is also a solution%
\footnote{Note that if one uses a convention according to which $\boldsymbol{\Phi}(t)=0$ for $t<0$, then the notation $\boldsymbol{\Phi}(t) \, \theta(t)$ is redundant.}].
Such a translational invariance follows quite straightforwardly from the causal and translationally-invariant nature of the kernel $\boldsymbol{\mathsf{H}}(t\!-\!\tau')$ as well as the matrix propagator's support only for non-negative values of its argument:
\begin{align}
\dot{\boldsymbol{\Phi}}_\mathrm{ret}(t\!-\!\tau) &= -\!\!\int_0^{t-\tau}\!\!\! d\tau' \, \boldsymbol{\mathsf{H}}(t\!-\!\tau\!-\!\tau') \, \boldsymbol{\Phi}(\tau') + \mathbf{I}\, \delta(t\!-\!\tau) \nonumber \\
&= -\!\!\int_{\tau}^t \!\! d\tau'' \, \boldsymbol{\mathsf{H}}(t\!-\!\tau'') \, \boldsymbol{\Phi}(\tau''\!\!-\!\tau) + \mathbf{I}\, \delta(t\!-\!\tau) \nonumber \\
&= -\!\!\int_0^t \!\! d\tau'' \, \boldsymbol{\mathsf{H}}(t\!-\!\tau'') \, \boldsymbol{\Phi}(\tau''\!\!-\!\tau) + \mathbf{I}\, \delta(t\!-\!\tau),
\end{align}
where $\mathbf{I}$ is the identity matrix and we used the fact that $\boldsymbol{\Phi}(\tau')=0$ for $\tau'<0$ in the last equality.

From the previous discussion it immediately follows (taking $t>\tau$) that, contrary to the local case, the matrix propagator does not factorize in the nonlocal case, i.e.
\begin{equation}
\boldsymbol{\Phi}(t\!-\!\tau) \neq  \boldsymbol{\Phi}(t) \, \boldsymbol{\Phi}^{\!-1}(\tau)
\label{eq:factorization} .
\end{equation}
This lack of factorizability also implies that the Green function or, equivalently, the matrix propagator $\boldsymbol{\Phi}_\mathrm{f}(t,\tau)$ for the integro-differential equation when the boundary conditions are specified at some final time [and given by Eq.~\eqref{eq:final_propag}] is no longer an \emph{advanced} propagator, i.e.\ it is no longer true that $\boldsymbol{\Phi}_\mathrm{f}(t,\tau)=0$ for $t>\tau$. This can be proved by contradiction. If one considers $\tau>t>\tau'$ in Eq.~\eqref{eq:final_propag} and assumes that $\boldsymbol{\Phi}_\mathrm{f}(\tau,\tau')=0$, one is left with
\begin{equation}
0 = -\boldsymbol{\Phi}(\tau,t) \, \boldsymbol{\Phi}(t\!-\!\tau') + \boldsymbol{\Phi}(\tau\!-\!\tau')
\label{eq:proof1}.
\end{equation}
Taking the limit $\tau' \to t^-$ of Eq.~\eqref{eq:final_propag} and taking into account that $\lim_{u\to 0^+} \boldsymbol{\Phi} (u) = \mathbf{I}$, one finally obtains $\boldsymbol{\Phi}(\tau\!-\!t) = \boldsymbol{\Phi}(\tau) \, \boldsymbol{\Phi}^{\!-1}(t)$, which is in contradiction with Eq.~\eqref{eq:factorization}.
Therefore, the assumption $\boldsymbol{\Phi}_\mathrm{f}(\tau,\tau')=0$ for $\tau > \tau'$ cannot be true in the nonlocal case.

These facts or closely related ones have been missed in the existing literature on master equations for QBM models. As a consequence, the existing results for the coefficients of the master equation are mathematically incorrect unless strictly local dissipation is considered, and can give rise to significant discrepancies whenever nonlocal effects are important.
We close this appendix by briefly describing how this affects the different existing approaches to deriving the exact master equation for QBM models. One class of derivations \cite{HalliwellYu96,CRV03,CRV01} involve an intermediate step where the solution of an integro-differential equation like Eq.~\eqref{eq:langevin3} with specified boundary conditions (position and velocity) at a final time is needed. The previous discussion directly applies to this class of derivations and the main consequences are that the Green functions appearing there are not advanced and the explicit expressions which were provided, based on the assumption that those Green functions were advanced, are incorrect. Nevertheless, the results in those references can be easily corrected by removing the qualification of ``advanced'' propagator and discarding the explicit expressions for that Green function. The results would then become equivalent to the general result that we have obtained in Sec.~\ref{sec:MED}, although one would need to find a way to construct the Green function explicitly. We provide such an explicit construction of the corresponding matrix propagator $\boldsymbol{\Phi}_\mathrm{f}(\tau,\tau')$ in Eq.~\eqref{eq:final_propag}, where it is expressed in terms of known quantities, namely, $\boldsymbol{\Phi}(t)$ as given by Eq.~\eqref{eq:Phi(t)}.
Note, by the way, that if one had truly advanced propagators, one could show that the terms involving triple time integrals in the results for the diffusion coefficients [such as Eqs.~(B.17)-(B.18) in Ref.~\cite{CRV03}] actually vanish. In fact, these terms correspond to the last term on the right-hand side of our Eq.~\eqref{eq:diffusion0}, which only vanishes for local dissipation, as can be seen from Eqs.~\eqref{eq:diff_simple}, \eqref{eq:simplifiedH} and the discussion above.

A second class of derivations, including HPZ's original derivation for arbitrary temperature and spectral function, relies on the use of Green functions for the same integro-differential equation, but associated with mixed boundary conditions which correspond to specifying the position at the initial and final times.
Explicit expressions are provided for those Green functions $G(t,s)$ in terms of homogeneous solutions $u_1(\tau)$ and $u_2(\tau)$ which vanish at the final and initial times respectively. Unfortunately, although those expressions are standard results for ordinary differential equations, they are not valid for nonlocal integro-differential equations. This is because they involve the sum of two terms, each one of them being a certain solution of the  homogeneous integro-differential equation times $\theta(t\!-\!s)$ and $\theta(s\!-\!t)$ respectively [see Eq.~(2.34) in Ref.~\cite{HPZ92}].
However, for similar reasons to those given above and illustrated by Eq.~\eqref{eq:inequality}, when multiplied by the theta functions those solutions cease to satisfy the integro-differential equation.

Finally, a third class of derivations \cite{FordOconnell01} are based on showing that the solutions of the Langevin equation can be equivalently understood as solutions of a local ordinary differential equation rather than an integro-differential one.
This is true for the homogenous solutions of the Langevin equation and corresponds to the equivalence (after inverting and transposing) between the matrix propagator $\boldsymbol{\Phi}(t)$ associated with the Langevin equation and the matrix propagator $\boldsymbol{\Phi}_k(t)$ associated with the ordinary differential equation \eqref{eq:ccfps}, which we found in Sec.~\ref{sec:charcurv}.
However, such an equivalence is not true for inhomogenous solutions of the nonlocal Langevin equation.
One way of seeing this is by realizing that since Eq.~\eqref{eq:ccfps} is an ordinary differential equation, its retarded matrix propagator does factorize.
But if the inhomogeneous solutions of the local equation constructed with that propagator were also solutions of the inhomogeneous Langevin equation, it would imply that the retarded propagator associated with the latter also factorizes, which is not true for nonlocal dissipation, as we showed above.
\footnote{To use this argument directly one should consider the equation satisfied by $[\boldsymbol{\Phi}_k^\mathrm{T}(t)]^{-1}$ rather than Eq.~(\ref{eq:ccfps}), which is satisfied by $\boldsymbol{\Phi}_k(t)$.
That equation can be easily obtained by transposing and taking the matrix inverse of Eq.~(\ref{eq:ccfps}) applied to $\boldsymbol{\Phi}_k(t)$, and it is still a local linear differential equation.}
In particular, the derivation of Eq.~(2.18) in Ref.~\cite{FordOconnell01} is valid if one takes a vanishing inhomogeneous source $F(t)$.
Nevertheless, when deriving  Eq.~(2.18) for a non-vanishing source, the authors implicitly assumed that if the homogenous solutions of the Langevin equations satisfy a local differential equation, the inhomogeneous solutions of the Langevin equation should also satisfy the inhomogeneous version of the same local equation.
As we have explained, it turns out that this is only true for local dissipation.
Not surprisingly, making use of  Eq.~(2.18) the authors derive a master equation with diffusion coefficients lacking the terms with triple time integrals mentioned above, which in reality should only vanish for strictly local dissipation.

\section{Derivation of the Late-Time Thermal Covariance}
\label{sec:late_cov_der}

Here we present the derivation of the general single-integral representation of the late-time thermal covariance.
For the sake of brevity we will work out the explicit case of the late-time position uncertainty.
The late-time momentum uncertainty is analogous and the cross-correlation vanishes at late times, as implied by $\sigma_T^{xp}(t) = (M/2) \, \dot{\sigma}_T^{xx}(t)$ if $\sigma_T^{xx}(t)$ tends to a constant asymptotic value.

We start with the full-time, exact expression
\begin{align}
\sigma_T^{xx}(t) =&\; \int_0^\infty \!\!\! d\omega \, I(\omega) \coth\!\left( \frac{\omega}{2T} \right) \\
& \times \int_0^t \!\! d\tau_1 \! \int_0^t \!\! d\tau_2 \, G(\tau_1) \cos[\omega(\tau_1\!-\!\tau_2)] \, G(\tau_2) , \nonumber
\end{align}
where we have made the simple change of variables $\tau'_i = t-\tau_i$ for $i=1,2$.
Introducing the additional change of variables $\bar{\tau} = \tau_1 + \tau_2$, the result can be rewritten as
\begin{align}
& \sigma_T^{xx}(t) = \int_0^\infty \!\!\! d\omega \, I(\omega) \coth\!\left( \frac{\omega}{2T} \right) \\
& \times \int_0^t \!\!d\tau_2 \! \int_{\tau_2}^{\tau_2+t} \!\!\!d\bar{\tau} \, G(\bar{\tau}\!-\!\tau_2) \cos[\omega(\bar{\tau}\!-\!2\,\tau_2)] \, G(\tau_2) . \nonumber
\end{align}
The double time integration can then be split into two parts:
\begin{equation}
\int_0^t \!d\tau_2 \! \int_{\tau_2}^{\tau_2+t} \!\!\!d\bar{\tau} = \int_0^t \!d\tau_2 \! \int_{\tau_2}^{t} \!d\bar{\tau} + \int_0^t \!d\tau_2 \! \int_{t}^{t+\tau_2} \!\!\!d\bar{\tau} .
\end{equation}
At sufficiently late times the contribution form the second integration domain can be neglected and we can approximate the whole integral as follows:
\begin{equation}
\int_0^t \!d\tau_2 \! \int_{\tau_2}^{\tau_2+t} \!\!\!d\bar{\tau} \,\approx\, \int_0^t \!d\tau_2 \! \int_{\tau_2}^{t} \!d\bar{\tau} \,=\, \int_0^t \!d\bar{\tau} \! \int_0^{\bar{\tau}}\! d\tau_2 , \\
\end{equation}
The next step is to express the cosine in complex form with exponential functions.
Once that is done, it is not difficult to manipulate the result into the form of a Laplace convolution:
\begin{align}
\sigma_T^{xx}(t) \approx&\; \int_0^\infty \!\!\! d\omega \, I(\omega) \coth\!\left( \frac{\omega}{2T} \right) \\
&\times \int_0^t \!\!d\tau \, \mbox{Re}\!\left\{ \left[ e^{-\imath \omega \tau} G(\tau) \right] * \left[ e^{+\imath \omega \tau} G(\tau) \right] \right\} , \nonumber
\end{align}
where we renamed $\bar{\tau}$ as $\tau$.
Using the property of frequency shifting in the Laplace domain, i.e.\ $\mathcal{L} \{ e^{\lambda t} f(t) \} = \hat{f}(s\!-\!\lambda)$, we obtain
\begin{align}
\hat{\sigma}_T^{xx}(s) &\approx \int_0^\infty \!\!\! d\omega \, I(\omega) \coth\!\left( \frac{\omega}{2T} \right) \frac{1}{s} \, \hat{G}(s\!+\!\imath \omega) \, \hat{G}(s\!-\!\imath \omega) \label{eq:latexx1}.
\end{align}
Application of the final value theorem, as given by Eq.~\eqref{eq:final_value}, then immediately reveals the \emph{exact} late-time covariance
\begin{align}
\sigma_T^{xx}(\infty) &= \int_0^\infty \!\!\! d\omega \, I(\omega) \coth\!\left( \frac{\omega}{2T} \right) \hat{G}(+\imath \omega) \, \hat{G}(-\imath \omega)
\label{eq:latexx2}.
\end{align}
Proceeding in a completely analogous way, one can obtain the result for the momentum covariance and the cross correlation. For the cross correlation, the time derivative of one of the propagators gives an extra factor $(s+\imath\omega)$ in the expression in Laplace space. When taking the real part, as in Eq.~\eqref{eq:latexx1}, one is left only with $s$, which cancels out the factor $1/s$ in Eq.~\eqref{eq:latexx2}. Application of the final value theorem, as given by Eq.~\eqref{eq:final_value}, gives then a vanishing result for the asymptotic value of the cross-correlation: $\sigma_T^{xp}(\infty)=0$. As for the momentum covariance, the two time derivatives, one for each propagator, give an extra factor $(s^2+\omega^2)$ in the expression in Laplace space. When taking the real part and applying the final value theorem, one is left with
\begin{align}
\sigma_T^{pp}(\infty) &= M^2 \! \int_0^\infty \!\!\! d\omega \, I(\omega) \coth\!\left( \frac{\omega}{2T} \right) \omega^2 \hat{G}(+\imath \omega) \, \hat{G}(-\imath \omega)
\label{eq:latepp}.
\end{align}
Taking into account Eqs.~\eqref{eq:latexx2}-\eqref{eq:latepp} and the vanishing value of the asymptotic cross correlation, the asymptotic value of the thermal covariance matrix can be written as
\begin{align}
\boldsymbol{\sigma}_T(\infty) &= \mathrm{Sy} \int_0^\infty \!\!\! d\omega \, I(\omega) \coth\!\left( \frac{\omega}{2T} \right) \hat{\boldsymbol{\Phi}}(+\imath \omega) \left[ \begin{smallmatrix} 0 & 0 \\ 0 & 1 \end{smallmatrix} \right] \hat{\boldsymbol{\Phi}}^{\!\mathrm{T}}\!(-\imath \omega) ,
\end{align}
where $\mathrm{Sy}$ denotes matrix symmetrization.

\section{Moderate-Time Diffusion for Ohmic Case with Large Cut-off}
\label{sec:mod-time_app}

In this appendix we calculate the diffusion coefficients for the ohmic case using the local propagator $G_\mathrm{R}(t)$ instead of the exact one, which is a valid approximation in the high cut-off regime,
as discussed in Sec.~\ref{sec:nonlocalG}.
The big advantage of using $G_\mathrm{R}(t)$ is that only the first term on the right-hand side of Eq.~\eqref{eq:diff_simple}, which involves a single time integral, will give a non-vanishing contribution.
Furthermore, the Laplace transforms of the corresponding equations for the diffusion coefficients exhibit a rather simple form if one takes the following steps. First, one writes the cosine of the noise kernel in exponential form; next, manipulates the time integral until one has a Laplace convolution;
and then uses frequency shifting in the Laplace domain, i.e. $e^{\lambda t} f(t) \to \hat{f}(s\!-\!\lambda)$.
After some algebraic manipulations one finally gets
\begin{align}
\hat{D}_{\!xp}(s) &= - \frac{1}{s} \int_0^\infty \!\!\! d\omega \, I(\omega) \coth\!{\left( \frac{\omega}{2 T} \right)} \mbox{Re}\!\left[ \hat{G}_\mathrm{R} (s \!+\! \imath \omega) \right] , \label{eq:D-contour}\\
\hat{D}_{\!pp}(s) &= + \frac{1}{s} \int_0^\infty \!\!\! d\omega \, I(\omega) \coth\!{\left( \frac{\omega}{2 T} \right)} \mbox{Re}\!\left[ \hat{\dot{G}}_\mathrm{R} (s \!+\! \imath \omega) \right] . \label{eq:D-contour2}
\end{align}

Our late-time Green function \eqref{eq:green_local}  is rational in the Laplace domain [with late-time coefficients given by Eq.~\eqref{eq:lateMECR}].
Moreover, the spectral density $I(\omega)$ in Eq.~\eqref{eq:spectral} is meromorphic with a finite number of poles.
Together with the rational expansion of the hyperbolic cotangent in Eq.~\eqref{eq:coth}, this implies that the frequency integrals over $\omega$ in the above diffusion coefficients become sums over $k$ of trivial contour integrals in the Laplace domain.
Still in the Laplace domain, these sums can be identified as harmonic number functions (or, equivalently, digamma functions):
\footnote{\label{overdamping}Many of the expressions derived throughout this paper assume underdamping, i.e.\ $\gamma_0 < \Omega$ with $\tilde{\Omega} = \sqrt{\Omega^2 - \gamma_0^2}$.
They can be used for the overdamping regime by making the following analytical continuation: $\tilde{\Omega} \to \imath \tilde{\gamma}$ with $\tilde{\gamma} =\sqrt{\gamma_0^2 - \Omega^2}$ real.
Therefore, Eqs.~(\ref{eq:lateDxp})-(\ref{eq:lateDpp}) can be applied to the overdamping case if the $\mathrm{Im}$ and $\mathrm{Re}$ terms are first expanded assuming that $\tilde{\Omega}$ is real, e.g. using $\mathrm{Im}[z]=(z-\bar{z})/(2\imath)$,  and then the analytical continuation $\tilde{\Omega} \to \imath \tilde{\gamma}$ is made.
}
\begin{align}
\hat{D}_{\!xp}(s) &=  - \frac{2\gamma_0 T}{\Lambda s} \mathcal{F}_s + \frac{\gamma_0}{s} \mbox{Im}\!\left[ \mathcal{I}_s \right] , \label{eq:lateDxp} \\ 
\hat{D}_{\!pp}(s) &=  \frac{2 \gamma_0 T}{s} \left(\!1 + \frac{s}{\Lambda}\right)\!\mathcal{F}_s + \frac{\gamma_0}{s} \mbox{Im}\!\left[ \! \left( \gamma_0 \!+\! \imath \tilde{\Omega} \right)\! \mathcal{I}_s \right] \label{eq:lateDpp} , 
\end{align}
in terms of the dimensionless quantities $\mathcal{F}_s$ and $\mathcal{I}_s$ defined
\begin{align}
\mathcal{F}_s &\equiv \left[ \left( 1 + \frac{\gamma_s}{\Lambda} \right)^{\!\!2} + \left( \frac{\tilde{\Omega}}{\Lambda} \right)^{\!\!2} \right]^{-1}  , \\
\mathcal{I}_s &\equiv \frac{2}{\pi} \frac{\imath + \frac{\gamma_s}{\tilde{\Omega}}}{ 1 - \left( \!\frac{\gamma_s \!+\! \imath \tilde{\Omega}}{\Lambda} \! \right)^{\!\!2} } \left\{ \!\mbox{H}\!\left( \frac{\Lambda}{2 \pi T} \right) \!-\! \mbox{H}\!\left( \!\frac{\gamma_s \!+\! \imath \tilde{\Omega} }{2 \pi T}\! \right) \right\} ,
\end{align}
and where $\gamma_s = \gamma_0 + s$.
Note that by making use of the final value theorem in Eq.~\eqref{eq:final_value}, we only need to discard the overall $1/s$ factor and replace $\gamma_s$ with $\gamma_0$ in Eqs.~\eqref{eq:lateDxp}-\eqref{eq:lateDpp} to obtain the late-time asymptotic values $D_{\!xp}(\infty)$ and $D_{\!pp}(\infty)$.
The $\mathrm{H}(z)$ functions are the harmonic number function discussed in \ref{sec:harmonic}.
These terms make up, among other things, the well known $\log(\Lambda/\Omega)$ divergence.
They behave asymptotically like logarithms but with $\mathrm{H}(0)=0$, making both their high and zero temperature limits trivial.
At high temperature, all of the harmonic number functions vanish, leaving only the second terms which are proportional to the temperature.
At zero temperature, all of the harmonic number functions can be equivalently evaluated as logarithms.

\begin{figure}[h!]
\centering
\includegraphics[width=0.5\textwidth]{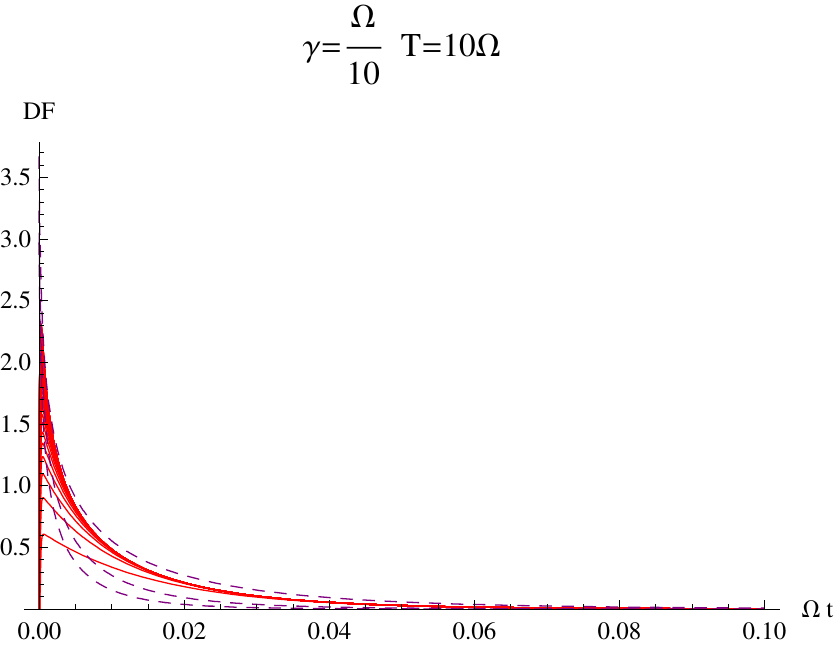}
\caption{Moderate temperature decay functions for \textcolor{red}{$\cdot$ a sequence of the first $50$ high temperature sums}, \textcolor{magenta}{$\cdots$ qualitative approximations for $k_i = \frac{1}{2},1,1\frac{1}{2}$} at $\Lambda = 10^5 \Omega$.
The high temperature sums are very slow to converge at the initial time.}
\label{fig:DFone}
\end{figure}

The diffusion coefficients can be expressed in the time domain as their asymptotic values plus damped oscillating differential operators acting on the same decay function $\mathrm{DF}(t)$ (although the sums over $k$ cannot in general be identified with any simply behaved special functions):
\begin{align}
D_{\!xp}(t) =&\; D_{\!xp}(\infty) \label{eq:Dxp2} \\
&- M \gamma_0 \left\{ \dot{G}_\mathrm{R}(t) + G_\mathrm{R}(t) \left( 2 \gamma_0 \!-\! \frac{d}{dt} \right) \right\} \mbox{DF}(t) , \nonumber \\
D_{\!pp}(t) =&\; D_{\!pp}(\infty) \label{eq:Dpp2} \\
&- M \gamma_0 \left\{ \dot{G}_\mathrm{R}(t) \left( \gamma_0 \!+\! \frac{d}{dt} \right) + G_\mathrm{R}(t) \, \Omega^2 \right\} \mbox{DF}(t) , \nonumber
\end{align}
with the thermal decay function
\begin{align}
\mbox{DF}(t) &= -\frac{\cot\left( \frac{\Lambda}{2T} \right) e^{-\Lambda t}}{\left( 1 + \frac{\gamma_0}{\Lambda} \right)^2 + \left( \frac{\tilde{\Omega}}{\Lambda} \right)^2} + \frac{2}{\pi} \mbox{TS}(t) , \label{eq:DF1}\\
\mbox{TS}(t) &= \sum_{k=1}^\infty \frac{ \left(\frac{\Lambda}{2 \pi T}\right)^2 }{\left(\frac{\Lambda}{2 \pi T}\right)^2-k^2} \frac{k \, e^{- 2 \pi T k t}}{ \left(k \!+\! \frac{\gamma_0}{2 \pi T}\right)^2 + \left(\frac{\tilde{\Omega}}{2 \pi T}\right)^2 } \label{eq:thermal_sum} .
\end{align}
For numerical evaluation purposes, it is useful to express this thermal sum in terms of Lerch transcendent functions:
\begin{align}
\mbox{TS}(t) =&\; \mbox{Re}\!\!\left[ \frac{1 - \imath \frac{\gamma_0}{\tilde{\Omega}} }{ 1 - \left( \frac{\gamma_0 + \imath \tilde{\Omega}}{\Lambda} \right)^2 }\, \varPhi_1\!\!\left( \frac{\gamma_0 \!+\! \imath \tilde{\Omega}}{2 \pi T} ; 2 \pi T t \right) \right] \nonumber \\
& - \mbox{Sy}_{\!\Lambda}\!\!\left[ \frac{\varPhi_1\!\left( \frac{\Lambda}{2 \pi T} ; 2 \pi T t \right)}{ \left( 1 - \frac{\gamma_0}{\Lambda} \right)^2 + \left( \frac{\tilde{\Omega}}{\Lambda} \right)^2 } \right] , \label{eq:DFLerch}
\end{align}
with the definitions of $\varPhi_1(z;\lambda)$, which is related to the \emph{Lerch transcendent} function by $\varPhi_1(z;\lambda) = \Phi \big(e^{-\lambda},1,z\big) - 1/z$, and of the symmetric part being
\begin{eqnarray}
\varPhi_1 ( z ; \lambda ) &=& \sum_{k=1}^\infty \frac{e^{-\lambda\, k}}{k+z} , \\
\mbox{Sy}_{\!z}\left[f(z)\right] &=& \frac{f(+z)+f(-z)}{2} \label{eq:DF5}.
\end{eqnarray}
The decay function is such that at the initial time it causes cancelation with the asymptotic values and the diffusion coefficients vanish.
In this (asymptotic) high temperature perspective, the decay function contains two terms.
The first decays at a cut-off dependent rate and can be expressed in closed form.
The second decays with primarily temperature dependent rates and cannot be expressed in closed form with intuitive functions.
It contains the initial time cancelation of the $\log(\Lambda/\Omega)$ divergence.
Although well convergent at moderate times, the sum's contribution to the regular diffusion coefficient is very slow to converge at the initial time, even for moderate temperatures; see Fig.~\ref{fig:DFone}.

While our expressions \eqref{eq:lateDxp}-\eqref{eq:lateDpp} can easily give us the zero temperature diffusion coefficients at asymptotically late time, they cannot easily give us the corresponding moderate time behavior in closed form.
Moreover, the zero temperature limit of \! $\coth(\omega/2T) \to \mathrm{sgn}(\omega)$
means that our diffusion coefficient integrals cannot be cast as closed contour integrals.
Nevertheless, the frequency integrals can be performed and the results expressed in terms of exponential integrals with predictable time scales.
At zero temperature (and in the high cut-off limit) we find the decay function to take the following form:
\begin{align}
&\lim_{T \to 0} \mbox{DF}(t) = \label{eq:DF_T0} \\
&\frac{2}{\pi} \frac{d}{dt} \left\{ \mbox{Re}\!\left[ \frac{\mbox{E}_1\!\!\left(\left[\gamma_0 \!+\! \imath \tilde{\Omega}\right]\!t\right) }{\imath \tilde{\Omega} \, e^{- \left(\gamma_0 + \imath \tilde{\Omega}\right) t } } \right] - \mbox{Sy}_{\!\Lambda}\!\left[ \frac{ \mbox{E}_1\!\left( \Lambda t \right) }{ \Lambda \, e^{-\Lambda t} } \right] \right\}  , \nonumber
\end{align}
where $\mathrm{E}_1(z)$ is the exponential integral, defined in \ref{sec:exp_int}, which behaves like $e^{-z}/z$ for large $z$.
It should be noted that unlike the asymptotic limits of the diffusion coefficients, the full time behavior is highly sensitive to the form of the cut-off regulator at low temperature.
For our smooth regulator, we find relatively smoothly evolving diffusion coefficients (similar to the result in Ref.~\cite{HPZ92} at $T=10\, \Omega$) all the way down to zero temperature.
In contrast, a sharp cut-off of the form $I(\omega) \propto \theta(\omega\!-\!\Lambda)$ would produce the same average behavior, but with a slowly decaying envelope modulating of considerable oscillations at the cut-off frequency.
\begin{figure}[h]
\centering
\includegraphics[width=0.5\textwidth]{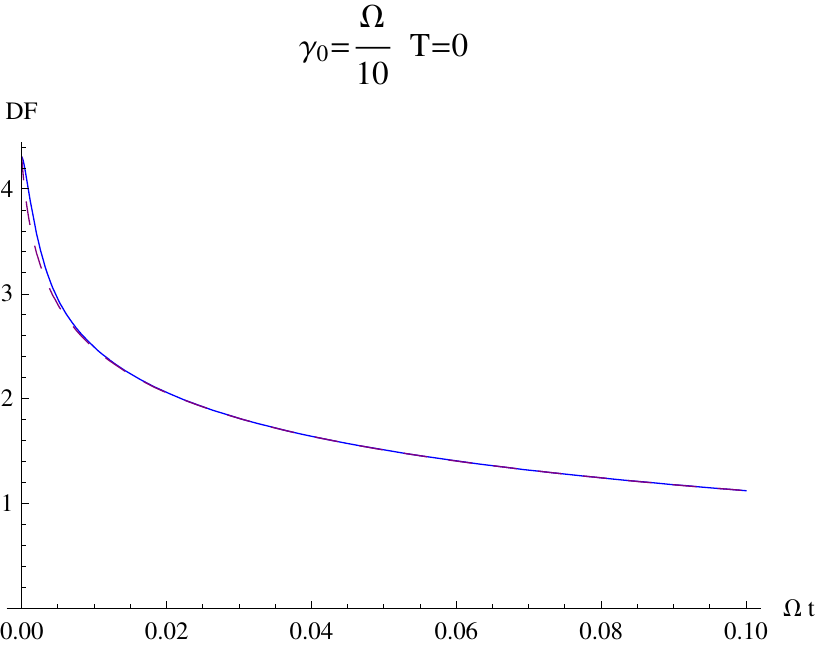}
\caption{Zero temperature decay functions for \textcolor{blue}{$\cdot$ zero temperature}, \textcolor{magenta}{$\cdots$ qualitative approximation} at $\Lambda = 10^3 \Omega$.
The slopes differ near the initial time (within the cut-off time scale).}
\label{fig:DFzero}
\end{figure}

Analogous functions appear when we approximate the thermal sum in \eqref{eq:thermal_sum}
[together with the first term on the right-hand side of \eqref{eq:DF1}, which cancels any spurious poles at $\Lambda = 2 \pi T k$]
as an integral with a comparably soft cut-off:
\begin{align}
\sum_{k=1}^{\infty} \frac{ \left( \frac{\Lambda}{2 \pi T} \right)^2 } { \left( \frac{\Lambda}{2 \pi T} \right)^2 - k^2 } f(k) &\approx \int_{k_i}^{\infty} \!\!\! dk \frac{ \left( \frac{\Lambda}{2 \pi T} \right)^2 } { \left( \frac{\Lambda}{2 \pi T} \right)^2 + k^2 } f(k) \, ,
\end{align}
where $k_i \approx 1$.
Still in the high cut-off limit, we find this qualitative approximation of the decay function to be
\begin{align}
\mbox{DF}(t) \approx&\; \frac{2}{\pi} \frac{d}{dt} \mbox{Re}\!\left[ \frac{\mbox{E}_1\!\!\left(\left[2 \pi T k_i \!+\! \gamma_0 \!+\! \imath \tilde{\Omega}\right]\!t\right) }{ \imath \tilde{\Omega} \, e^{- \left(\gamma_0 + \imath \tilde{\Omega}\right) t } } \right] \nonumber \\
&  - \frac{2}{\pi} \frac{d}{dt} \mbox{Sy}_{\!\Lambda}\!\left[ \frac{ \mbox{E}_1\!\left(\left[ 2 \pi T k_i \!+\! \imath \Lambda \right]t\right) }{ \imath \Lambda \, e^{-\imath \Lambda t} } \right] \label{eq:DF_T},
\end{align}
where we have discarded all finite terms at the initial time which decay at cut-off rates, as our approximation ultimately ruins the behavior of DF$(t)$ there.
Thus, when using this approximate decay function, the time-dependent, decaying part of the diffusion coefficients must be ``clamped'' at the initial time.
%
At moderate times, our approximation reveals the exact same form of exponential integral behavior as in the zero temperature limit.
But the temperature enters in such a way that the exponential decay inherent in $\mathrm{E}_1$ is not balanced out with a $e^{-2 \pi T k_i t}$ factor.
Therefore, temperature is an inherently stronger relaxation scale here [although there are additional $e^{-\gamma_0 t}$ factors from $G_\mathrm{R}(t)$ functions in the full diffusion coefficients].
\clearpage

\bibliography{bib}{}

\begin{thebibliography}{46}
\expandafter\ifx\csname natexlab\endcsname\relax\def\natexlab#1{#1}\fi
\expandafter\ifx\csname bibnamefont\endcsname\relax
  \def\bibnamefont#1{#1}\fi
\expandafter\ifx\csname bibfnamefont\endcsname\relax
  \def\bibfnamefont#1{#1}\fi
\expandafter\ifx\csname citenamefont\endcsname\relax
  \def\citenamefont#1{#1}\fi
\expandafter\ifx\csname url\endcsname\relax
  \def\url#1{\texttt{#1}}\fi
\expandafter\ifx\csname urlprefix\endcsname\relax\def\urlprefix{URL }\fi
\providecommand{\bibinfo}[2]{#2}
\providecommand{\eprint}[2][]{\url{#2}}

\bibitem[{\citenamefont{Breuer and Petruccione}(2002)}]{Breuer02}
\bibinfo{author}{\bibfnamefont{H.~P.} \bibnamefont{Breuer}} \bibnamefont{and}
  \bibinfo{author}{\bibfnamefont{F.}~\bibnamefont{Petruccione}},
  \emph{\bibinfo{title}{The Theory of Open Quantum Systems}}
  (\bibinfo{publisher}{Oxford University Press}, \bibinfo{address}{Oxford},
  \bibinfo{year}{2002}).

\bibitem[{\citenamefont{Calzetta and Hu}(2008)}]{CalzettaHu08}
\bibinfo{author}{\bibfnamefont{E.~A.} \bibnamefont{Calzetta}} \bibnamefont{and}
  \bibinfo{author}{\bibfnamefont{B.~L.} \bibnamefont{Hu}},
  \emph{\bibinfo{title}{Nonequilibrium Quantum Field Theory}}
  (\bibinfo{publisher}{Cambridge University Press},
  \bibinfo{address}{Cambridge}, \bibinfo{year}{2008}).

\bibitem[{\citenamefont{Giulini et~al.}(1996)\citenamefont{Giulini, Joos,
  Kiefer, Kupsch, Stamatescu, and Zeh}}]{Giulini96}
\bibinfo{author}{\bibfnamefont{D.}~\bibnamefont{Giulini}},
  \bibinfo{author}{\bibfnamefont{E.}~\bibnamefont{Joos}},
  \bibinfo{author}{\bibfnamefont{C.}~\bibnamefont{Kiefer}},
  \bibinfo{author}{\bibfnamefont{J.}~\bibnamefont{Kupsch}},
  \bibinfo{author}{\bibfnamefont{L.~O.} \bibnamefont{Stamatescu}},
  \bibnamefont{and} \bibinfo{author}{\bibfnamefont{H.~D.} \bibnamefont{Zeh}},
  \emph{\bibinfo{title}{Decoherence and the Appearance of a Classical World in
  Quantum Theory}} (\bibinfo{publisher}{Springer-Verlag},
  \bibinfo{address}{Berlin}, \bibinfo{year}{1996}).

\bibitem[{\citenamefont{Zurek}(2003)}]{Zurek03}
\bibinfo{author}{\bibfnamefont{W.~H.} \bibnamefont{Zurek}},
  \bibinfo{journal}{Rev. Mod. Phys.} \textbf{\bibinfo{volume}{75}},
  \bibinfo{pages}{715} (\bibinfo{year}{2003}).

\bibitem[{\citenamefont{Caldeira and
  Leggett}(1983{\natexlab{a}})}]{CaldeiraLeggett83b}
\bibinfo{author}{\bibfnamefont{A.~O.} \bibnamefont{Caldeira}} \bibnamefont{and}
  \bibinfo{author}{\bibfnamefont{A.~J.} \bibnamefont{Leggett}},
  \bibinfo{journal}{Ann. Phys. (N.Y.)} \textbf{\bibinfo{volume}{149}},
  \bibinfo{pages}{374} (\bibinfo{year}{1983}{\natexlab{a}}).

\bibitem[{\citenamefont{Leggett et~al.}(1987)\citenamefont{Leggett,
  Chakravarty, Dorsey, Fisher, Garg, and Zwerger}}]{Leggett87}
\bibinfo{author}{\bibfnamefont{A.~J.} \bibnamefont{Leggett}},
  \bibinfo{author}{\bibfnamefont{S.}~\bibnamefont{Chakravarty}},
  \bibinfo{author}{\bibfnamefont{A.~T.} \bibnamefont{Dorsey}},
  \bibinfo{author}{\bibfnamefont{M.~P.~A.} \bibnamefont{Fisher}},
  \bibinfo{author}{\bibfnamefont{A.}~\bibnamefont{Garg}}, \bibnamefont{and}
  \bibinfo{author}{\bibfnamefont{W.}~\bibnamefont{Zwerger}},
  \bibinfo{journal}{Rev. Mod. Phys.} \textbf{\bibinfo{volume}{59}},
  \bibinfo{pages}{1} (\bibinfo{year}{1987}).

\bibitem[{\citenamefont{Weiss}(1999)}]{Weiss99}
\bibinfo{author}{\bibfnamefont{U.}~\bibnamefont{Weiss}},
  \emph{\bibinfo{title}{Quantum dissipative systems}}
  (\bibinfo{publisher}{World Scientific}, \bibinfo{address}{Singapore},
  \bibinfo{year}{1999}).

\bibitem[{\citenamefont{Scully and Zubairy}(2007)}]{Scully07}
\bibinfo{author}{\bibfnamefont{M.~O.} \bibnamefont{Scully}} \bibnamefont{and}
  \bibinfo{author}{\bibfnamefont{M.~S.} \bibnamefont{Zubairy}},
  \emph{\bibinfo{title}{Quantum Optics}} (\bibinfo{publisher}{Cambridge
  University Press}, \bibinfo{address}{Cambridge}, \bibinfo{year}{2007}).

\bibitem[{\citenamefont{Meystre}(2001)}]{Meystre01}
\bibinfo{author}{\bibfnamefont{P.}~\bibnamefont{Meystre}},
  \emph{\bibinfo{title}{Atom Optics}} (\bibinfo{publisher}{Springer-Verlag},
  \bibinfo{address}{Berlin}, \bibinfo{year}{2001}).

\bibitem[{\citenamefont{Wisman and Milburn}(2010)}]{Wiseman10}
\bibinfo{author}{\bibfnamefont{H.}~\bibnamefont{Wisman}} \bibnamefont{and}
  \bibinfo{author}{\bibfnamefont{G.}~\bibnamefont{Milburn}},
  \emph{\bibinfo{title}{Quantum Measurement and Control}}
  (\bibinfo{publisher}{Cambridge University Press},
  \bibinfo{address}{Cambridge}, \bibinfo{year}{2010}).

\bibitem[{\citenamefont{Naik et~al.}(2006)}]{Naik06}
\bibinfo{author}{\bibfnamefont{A.}~\bibnamefont{Naik}} \bibnamefont{et~al.},
  \bibinfo{journal}{Nature} \textbf{\bibinfo{volume}{443}},
  \bibinfo{pages}{193} (\bibinfo{year}{2006}).

\bibitem[{\citenamefont{LaHaye et~al.}(2004)\citenamefont{LaHaye, Buu,
  Camarota, and Schwab}}]{Lahaye04}
\bibinfo{author}{\bibfnamefont{M.~D.} \bibnamefont{LaHaye}},
  \bibinfo{author}{\bibfnamefont{O.}~\bibnamefont{Buu}},
  \bibinfo{author}{\bibfnamefont{B.}~\bibnamefont{Camarota}}, \bibnamefont{and}
  \bibinfo{author}{\bibfnamefont{K.~C.} \bibnamefont{Schwab}},
  \bibinfo{journal}{Science} \textbf{\bibinfo{volume}{304}},
  \bibinfo{pages}{74} (\bibinfo{year}{2004}).

\bibitem[{\citenamefont{Feynman and Vernon}(1963)}]{Feynman63}
\bibinfo{author}{\bibfnamefont{R.~P.} \bibnamefont{Feynman}} \bibnamefont{and}
  \bibinfo{author}{\bibfnamefont{F.~L.} \bibnamefont{Vernon}},
  \bibinfo{journal}{Ann. Phys. (N.Y.)} \textbf{\bibinfo{volume}{24}},
  \bibinfo{pages}{118} (\bibinfo{year}{1963}).

\bibitem[{\citenamefont{Caldeira and
  Leggett}(1983{\natexlab{b}})}]{CaldeiraLeggett83}
\bibinfo{author}{\bibfnamefont{A.~O.} \bibnamefont{Caldeira}} \bibnamefont{and}
  \bibinfo{author}{\bibfnamefont{A.~J.} \bibnamefont{Leggett}},
  \bibinfo{journal}{Physica A} \textbf{\bibinfo{volume}{121}},
  \bibinfo{pages}{587} (\bibinfo{year}{1983}{\natexlab{b}}).

\bibitem[{\citenamefont{Caldeira et~al.}(1989)\citenamefont{Caldeira, Cerdeira,
  and Ramaswamy}}]{Caldeira89}
\bibinfo{author}{\bibfnamefont{A.~O.} \bibnamefont{Caldeira}},
  \bibinfo{author}{\bibfnamefont{H.~A.} \bibnamefont{Cerdeira}},
  \bibnamefont{and}
  \bibinfo{author}{\bibfnamefont{R.}~\bibnamefont{Ramaswamy}},
  \bibinfo{journal}{Phys. Rev. A} \textbf{\bibinfo{volume}{40}},
  \bibinfo{pages}{3438} (\bibinfo{year}{1989}).

\bibitem[{\citenamefont{Unruh and Zurek}(1989)}]{UnruhZurek89}
\bibinfo{author}{\bibfnamefont{W.~G.} \bibnamefont{Unruh}} \bibnamefont{and}
  \bibinfo{author}{\bibfnamefont{W.~H.} \bibnamefont{Zurek}},
  \bibinfo{journal}{Phys. Rev. D} \textbf{\bibinfo{volume}{40}},
  \bibinfo{pages}{1071} (\bibinfo{year}{1989}).

\bibitem[{\citenamefont{Hu et~al.}(1992)\citenamefont{Hu, Paz, and
  Zhang}}]{HPZ92}
\bibinfo{author}{\bibfnamefont{B.~L.} \bibnamefont{Hu}},
  \bibinfo{author}{\bibfnamefont{J.~P.} \bibnamefont{Paz}}, \bibnamefont{and}
  \bibinfo{author}{\bibfnamefont{Y.}~\bibnamefont{Zhang}},
  \bibinfo{journal}{Phys. Rev. D} \textbf{\bibinfo{volume}{45}},
  \bibinfo{pages}{2843} (\bibinfo{year}{1992}).

\bibitem[{\citenamefont{Halliwell and Yu}(1996)}]{HalliwellYu96}
\bibinfo{author}{\bibfnamefont{J.~J.} \bibnamefont{Halliwell}}
  \bibnamefont{and} \bibinfo{author}{\bibfnamefont{T.}~\bibnamefont{Yu}},
  \bibinfo{journal}{Phys. Rev. D} \textbf{\bibinfo{volume}{53}},
  \bibinfo{pages}{2012} (\bibinfo{year}{1996}).

\bibitem[{\citenamefont{Calzetta et~al.}(2003)\citenamefont{Calzetta, Roura,
  and Verdaguer}}]{CRV03}
\bibinfo{author}{\bibfnamefont{E.}~\bibnamefont{Calzetta}},
  \bibinfo{author}{\bibfnamefont{A.}~\bibnamefont{Roura}}, \bibnamefont{and}
  \bibinfo{author}{\bibfnamefont{E.}~\bibnamefont{Verdaguer}},
  \bibinfo{journal}{Physica A} \textbf{\bibinfo{volume}{319}},
  \bibinfo{pages}{188} (\bibinfo{year}{2003}), \eprint{arXiv:quant-ph/0011097}.

\bibitem[{\citenamefont{Calzetta et~al.}(2001)\citenamefont{Calzetta, Roura,
  and Verdaguer}}]{CRV01}
\bibinfo{author}{\bibfnamefont{E.}~\bibnamefont{Calzetta}},
  \bibinfo{author}{\bibfnamefont{A.}~\bibnamefont{Roura}}, \bibnamefont{and}
  \bibinfo{author}{\bibfnamefont{E.}~\bibnamefont{Verdaguer}},
  \bibinfo{journal}{Int. J. Theor. Phys.} \textbf{\bibinfo{volume}{40}},
  \bibinfo{pages}{2317} (\bibinfo{year}{2001}).

\bibitem[{\citenamefont{Ford and O'Connell}(2001)}]{FordOconnell01}
\bibinfo{author}{\bibfnamefont{G.~W.} \bibnamefont{Ford}} \bibnamefont{and}
  \bibinfo{author}{\bibfnamefont{R.~F.} \bibnamefont{O'Connell}},
  \bibinfo{journal}{Phys. Rev. D} \textbf{\bibinfo{volume}{64}},
  \bibinfo{pages}{105020} (\bibinfo{year}{2001}).

\bibitem[{\citenamefont{Ford et~al.}(1988)\citenamefont{Ford, Lewis, and
  O'Connell}}]{FordOconnell88}
\bibinfo{author}{\bibfnamefont{G.~W.} \bibnamefont{Ford}},
  \bibinfo{author}{\bibfnamefont{J.~T.} \bibnamefont{Lewis}}, \bibnamefont{and}
  \bibinfo{author}{\bibfnamefont{R.~F.} \bibnamefont{O'Connell}},
  \bibinfo{journal}{Phys. Rev. A} \textbf{\bibinfo{volume}{37}},
  \bibinfo{pages}{4419} (\bibinfo{year}{1988}).

\bibitem[{\citenamefont{Chou et~al.}(2008)\citenamefont{Chou, Hu, and
  Yu}}]{Chou08}
\bibinfo{author}{\bibfnamefont{C.-H.} \bibnamefont{Chou}},
  \bibinfo{author}{\bibfnamefont{B.~L.} \bibnamefont{Hu}}, \bibnamefont{and}
  \bibinfo{author}{\bibfnamefont{T.}~\bibnamefont{Yu}},
  \bibinfo{journal}{Physica A} \textbf{\bibinfo{volume}{387}},
  \bibinfo{pages}{432} (\bibinfo{year}{2008}).

\bibitem[{\citenamefont{Anastopoulos and Halliwell}(1995)}]{Anastopoulos95}
\bibinfo{author}{\bibfnamefont{C.}~\bibnamefont{Anastopoulos}}
  \bibnamefont{and} \bibinfo{author}{\bibfnamefont{J.~J.}
  \bibnamefont{Halliwell}}, \bibinfo{journal}{Phys. Rev. D}
  \textbf{\bibinfo{volume}{51}}, \bibinfo{pages}{6870} (\bibinfo{year}{1995}).

\bibitem[{\citenamefont{Hu and Zhang}(1993)}]{HuZhang93}
\bibinfo{author}{\bibfnamefont{B.~L.} \bibnamefont{Hu}} \bibnamefont{and}
  \bibinfo{author}{\bibfnamefont{Y.}~\bibnamefont{Zhang}},
  \bibinfo{journal}{Mod. Phys. Lett. A} \textbf{\bibinfo{volume}{8}},
  \bibinfo{pages}{3575} (\bibinfo{year}{1993}).

\bibitem[{\citenamefont{Hu and Zhang}(1995)}]{HuZhang95}
\bibinfo{author}{\bibfnamefont{B.~L.} \bibnamefont{Hu}} \bibnamefont{and}
  \bibinfo{author}{\bibfnamefont{Y.}~\bibnamefont{Zhang}},
  \bibinfo{journal}{Int. J. Mod. Phys. A} \textbf{\bibinfo{volume}{10}},
  \bibinfo{pages}{4537} (\bibinfo{year}{1995}).

\bibitem[{\citenamefont{Hillery et~al.}(1984)\citenamefont{Hillery, O'Connell,
  Scully, and Wigner}}]{HilleryOconnell86}
\bibinfo{author}{\bibfnamefont{M.}~\bibnamefont{Hillery}},
  \bibinfo{author}{\bibfnamefont{R.~F.} \bibnamefont{O'Connell}},
  \bibinfo{author}{\bibfnamefont{M.~O.} \bibnamefont{Scully}},
  \bibnamefont{and} \bibinfo{author}{\bibfnamefont{E.~P.}
  \bibnamefont{Wigner}}, \bibinfo{journal}{Phys. Rep.}
  \textbf{\bibinfo{volume}{106}}, \bibinfo{pages}{121} (\bibinfo{year}{1984}).

\bibitem[{\citenamefont{Gardiner and Collett}(1985)}]{Gardiner85}
\bibinfo{author}{\bibfnamefont{C.~W.} \bibnamefont{Gardiner}} \bibnamefont{and}
  \bibinfo{author}{\bibfnamefont{M.~J.} \bibnamefont{Collett}},
  \bibinfo{journal}{Phys. Rev. A} \textbf{\bibinfo{volume}{31}},
  \bibinfo{pages}{3761} (\bibinfo{year}{1985}).

\bibitem[{\citenamefont{Roura and Verdaguer}(1999)}]{roura99}
\bibinfo{author}{\bibfnamefont{A.}~\bibnamefont{Roura}} \bibnamefont{and}
  \bibinfo{author}{\bibfnamefont{E.}~\bibnamefont{Verdaguer}},
  \bibinfo{journal}{Phys. Rev. D} \textbf{\bibinfo{volume}{60}},
  \bibinfo{pages}{107503} (\bibinfo{year}{1999}).

\bibitem[{\citenamefont{Hu and Matacz}(1994)}]{HuMatacz94}
\bibinfo{author}{\bibfnamefont{B.~L.} \bibnamefont{Hu}} \bibnamefont{and}
  \bibinfo{author}{\bibfnamefont{A.}~\bibnamefont{Matacz}},
  \bibinfo{journal}{Phys. Rev. D} \textbf{\bibinfo{volume}{49}},
  \bibinfo{pages}{6612} (\bibinfo{year}{1994}).

\bibitem[{\citenamefont{W{\l}odarz}(2003)}]{Wlodarz03}
\bibinfo{author}{\bibfnamefont{J.~J.} \bibnamefont{W{\l}odarz}},
  \bibinfo{journal}{Int. J. Theor. Phys.} \textbf{\bibinfo{volume}{42}},
  \bibinfo{pages}{1075} (\bibinfo{year}{2003}).

\bibitem[{\citenamefont{Zurek}(1991)}]{Zurek91}
\bibinfo{author}{\bibfnamefont{W.~H.} \bibnamefont{Zurek}},
  \bibinfo{journal}{Physics Today} \textbf{\bibinfo{volume}{44}},
  \bibinfo{pages}{36} (\bibinfo{year}{1991}).

\bibitem[{\citenamefont{Paz et~al.}(1993)\citenamefont{Paz, Habib, and
  Zurek}}]{Paz93}
\bibinfo{author}{\bibfnamefont{J.~P.} \bibnamefont{Paz}},
  \bibinfo{author}{\bibfnamefont{S.}~\bibnamefont{Habib}}, \bibnamefont{and}
  \bibinfo{author}{\bibfnamefont{W.~H.} \bibnamefont{Zurek}},
  \bibinfo{journal}{Phys. Rev. D} \textbf{\bibinfo{volume}{47}},
  \bibinfo{pages}{488} (\bibinfo{year}{1993}).

\bibitem[{\citenamefont{Anglin et~al.}(1997)\citenamefont{Anglin, Paz, and
  Zurek}}]{Anglin97}
\bibinfo{author}{\bibfnamefont{J.~R.} \bibnamefont{Anglin}},
  \bibinfo{author}{\bibfnamefont{J.~P.} \bibnamefont{Paz}}, \bibnamefont{and}
  \bibinfo{author}{\bibfnamefont{W.~H.} \bibnamefont{Zurek}},
  \bibinfo{journal}{Phys. Rev. A} \textbf{\bibinfo{volume}{55}},
  \bibinfo{pages}{4041} (\bibinfo{year}{1997}).

\bibitem[{\citenamefont{Lombardo and Villar}(2005)}]{Lombardo05}
\bibinfo{author}{\bibfnamefont{F.~C.} \bibnamefont{Lombardo}} \bibnamefont{and}
  \bibinfo{author}{\bibfnamefont{P.~I.} \bibnamefont{Villar}},
  \bibinfo{journal}{Phys. Lett. A} \textbf{\bibinfo{volume}{336}},
  \bibinfo{pages}{16} (\bibinfo{year}{2005}).

\bibitem[{\citenamefont{Hakim and Ambegaokar}(1985)}]{Hakim85}
\bibinfo{author}{\bibfnamefont{V.}~\bibnamefont{Hakim}} \bibnamefont{and}
  \bibinfo{author}{\bibfnamefont{V.}~\bibnamefont{Ambegaokar}},
  \bibinfo{journal}{Phys. Rev. A} \textbf{\bibinfo{volume}{32}},
  \bibinfo{pages}{423} (\bibinfo{year}{1985}).

\bibitem[{\citenamefont{Kimble et~al.}(2002)}]{Kimble02}
\bibinfo{author}{\bibfnamefont{H.~J.} \bibnamefont{Kimble}}
  \bibnamefont{et~al.}, \bibinfo{journal}{Phys. Rev. D}
  \textbf{\bibinfo{volume}{65}}, \bibinfo{pages}{022002}
  (\bibinfo{year}{2002}).

\bibitem[{\citenamefont{Buonanno and Chen}(2001)}]{Buonanno01}
\bibinfo{author}{\bibfnamefont{A.}~\bibnamefont{Buonanno}} \bibnamefont{and}
  \bibinfo{author}{\bibfnamefont{Y.}~\bibnamefont{Chen}},
  \bibinfo{journal}{Phys. Rev. D} \textbf{\bibinfo{volume}{64}},
  \bibinfo{pages}{042006} (\bibinfo{year}{2001}).

\bibitem[{\citenamefont{Paz and Roncaglia}(2008)}]{Paz08}
\bibinfo{author}{\bibfnamefont{J.~P.} \bibnamefont{Paz}} \bibnamefont{and}
  \bibinfo{author}{\bibfnamefont{C.~A.} \bibnamefont{Roncaglia}},
  \bibinfo{journal}{Phys. Rev. Lett.} \textbf{\bibinfo{volume}{100}},
  \bibinfo{pages}{220401} (\bibinfo{year}{2008}).

\bibitem[{\citenamefont{Paz and Roncaglia}(2009)}]{Paz09}
\bibinfo{author}{\bibfnamefont{J.~P.} \bibnamefont{Paz}} \bibnamefont{and}
  \bibinfo{author}{\bibfnamefont{C.~A.} \bibnamefont{Roncaglia}},
  \bibinfo{journal}{Phys. Rev. A} \textbf{\bibinfo{volume}{79}},
  \bibinfo{pages}{032102} (\bibinfo{year}{2009}).

\bibitem[{\citenamefont{Fleming et~al.}()\citenamefont{Fleming, Roura, and
  Hu}}]{NQBM}
\bibinfo{author}{\bibfnamefont{C.~H.} \bibnamefont{Fleming}},
  \bibinfo{author}{\bibfnamefont{A.}~\bibnamefont{Roura}}, \bibnamefont{and}
  \bibinfo{author}{\bibfnamefont{B.~L.} \bibnamefont{Hu}},
  \bibinfo{note}{\emph{in preparation}}.

\bibitem[{\citenamefont{Hu et~al.}(2004)\citenamefont{Hu, Roura, and
  Verdaguer}}]{HRV04}
\bibinfo{author}{\bibfnamefont{B.~L.} \bibnamefont{Hu}},
  \bibinfo{author}{\bibfnamefont{A.}~\bibnamefont{Roura}}, \bibnamefont{and}
  \bibinfo{author}{\bibfnamefont{E.}~\bibnamefont{Verdaguer}},
  \bibinfo{journal}{Phys. Rev. D} \textbf{\bibinfo{volume}{73}},
  \bibinfo{pages}{044002} (\bibinfo{year}{2004}).

\bibitem[{\citenamefont{Grabert et~al.}(1988)\citenamefont{Grabert, Schramm,
  and Ingold}}]{Grabert88}
\bibinfo{author}{\bibfnamefont{H.}~\bibnamefont{Grabert}},
  \bibinfo{author}{\bibfnamefont{P.}~\bibnamefont{Schramm}}, \bibnamefont{and}
  \bibinfo{author}{\bibfnamefont{G.~L.} \bibnamefont{Ingold}},
  \bibinfo{journal}{Phys. Rep.} \textbf{\bibinfo{volume}{168}},
  \bibinfo{pages}{115} (\bibinfo{year}{1988}).

\bibitem[{\citenamefont{Romero and Paz}(1997)}]{Romero97}
\bibinfo{author}{\bibfnamefont{L.~D.} \bibnamefont{Romero}} \bibnamefont{and}
  \bibinfo{author}{\bibfnamefont{J.~P.} \bibnamefont{Paz}},
  \bibinfo{journal}{Phys. Rev. A} \textbf{\bibinfo{volume}{55}},
  \bibinfo{pages}{4070} (\bibinfo{year}{1997}).

\bibitem[{\citenamefont{Robertson}(1934)}]{Robertson34}
\bibinfo{author}{\bibfnamefont{H.~P.} \bibnamefont{Robertson}},
  \bibinfo{journal}{Phys. Rev.} \textbf{\bibinfo{volume}{46}},
  \bibinfo{pages}{794} (\bibinfo{year}{1934}).

\bibitem[{\citenamefont{Trifonov}(2002)}]{Trifonov02}
\bibinfo{author}{\bibfnamefont{D.~A.} \bibnamefont{Trifonov}},
  \bibinfo{journal}{Eur. Phys. J. B} \textbf{\bibinfo{volume}{29}},
  \bibinfo{pages}{349} (\bibinfo{year}{2002}).

\end{thebibliography}
\bibliographystyle{apsrev}

\end{document}